\documentclass[review]{elsarticle}

\usepackage[pdftex,
        colorlinks=true,
        urlcolor=darkblue,               % \href{...}{...}
        anchorcolor=darkblue,
        filecolor=green,                   % \href*{...}
        linkcolor=darkblue,              % \ref{...} and \pageref{...}
        menucolor=darkblue,
        citecolor=darkblue,
        pagebackref,
        backref=false,
        pdfpagemode=true,
        bookmarks=true,
        bookmarksopen=true,
        ]{hyperref}
\usepackage[
uncertainty-mode = separate, % pm sign for uncertainties
exponent-product = \cdot, % dot instead of x in products with exponents
]{siunitx}
\usepackage{booktabs}

\usepackage{upgreek} 
\usepackage[utf8]{inputenc}
\usepackage[british]{babel}

\journal{Journal of Magnetism and Magnetic Materials}

\biboptions{numbers,sort&compress}

\begin{document}

\begin{frontmatter}

\title{Magnetoimpedance properties of CoNbZr, multilayer CoNbZr/Au and multilayer NiFe/Au thin films}

\author[1]{Indujan Sivanesarajah\corref{cor1}}
 \ead{sivanesa@outlook.de}
\author[2]{Leon Abelmann}
\author[1]{Uwe Hartmann}

\cortext[cor1]{Corresponding author}
\address[1]{Institute of Experimental Physics, Saarland University, D-66041 Saarbr\"{u}cken, Germany}
\address[2]{Department of Microelectronics, Delft University of Technology, 2600 AA Delft, The Netherlands}

\begin{abstract}
Thin-film magnetic sensors using the giant magnetoimpedance (GMI) effect show great promise for sensitive low-field magnetic measurements. Optimising sensor performance requires a thorough understanding of the properties of various soft magnetic materials. This study examines the electric, magnetic, and GMI properties of sputtered single-layer amorphous CoNbZr, multilayer amorphous CoNbZr/Au, and crystalline NiFe/Au thin films. GMI measurements reveal distinct ferromagnetic resonance (FMR) frequencies: \qty{1.4}{GHz} for CoNbZr, \qty{0.7}{GHz} for CoNbZr/Au, and \qty{0.5}{GHz} for NiFe/Au. Au interlayers improve the GMI response, increasing the GMI ratio by \qty{50}{\percent} and reducing FMR frequency compared to single-layer CoNbZr. The highest GMI ratio of \qty{300}{\percent} occurs in a \qtyproduct{20x5000}{\micro m} CoNbZr/Au strip at \qty{1.8}{GHz} under \qty{2}{mT}, while NiFe/Au exhibits \qty{280}{\percent} at \qty{4}{mT}. These differences are linked to variations in in-plane demagnetising factors and saturation magnetisations, emphasising the role of material and geometry in GMI sensor performance.
\end{abstract}


\end{frontmatter}

%\linenumbers

\section{Introduction}

Magnetic sensors are indispensable in various fields, playing a pivotal role in detecting and measuring magnetic fields. They are crucial in biomedical diagnostics for imaging and tracking biological processes \cite{Lin2017,Murzin2020,Elzenheimer2022}, in industrial non-destructive testing for detecting magnetic contamination and defects \cite{Zhang2016,Nakai2021,Eslamlou2023}, and in applications such as navigation \cite{Caruso1997,El-Sheimy2020,Saltus2023}, environmental monitoring \cite{Grimes1999,Hojjati-Najafabadi2022,Culik2023}, and the automotive sector \cite{Treutler2001,Granig2007,Nonomura2020}.

The development of sensors for detecting weak magnetic fields, particularly in the pT/nT range, represents a crucial area of research. These sensors are essential for a wide range of applications, including magnetoencephalography for measuring brain activity \cite{Boto2016,Kanno2022,Brookes2022} and geophysical exploration \cite{Robinson2008,Dentith2014,Stolz2022}. The advancement of highly sensitive sensor technologies has resulted in various types of sensors, including fluxgate sensors \cite{Aschenbrenner1936}, superconducting quantum interference devices (SQUIDs) \cite{Josephson1962,Anderson1963,Jaklevic1964}, Hall sensors \cite{Hall1879,Uzlu2019}, and magnetoresistive sensors, such as anisotropic magnetoresistance (AMR) \cite{Thomson1857}, giant magnetoresistance (GMR) \cite{Baibich1988, Binasch1989}, and tunnel magnetoresistance (TMR) \cite{Julliere1975}. Additionally, sensors based on nitrogen-vacancy centres in diamonds \cite{Maze2008,Balasubramanian2008,Taylor2008} and those utilising the giant magnetoimpedance (GMI) effect \cite{Panina1994, Beach1994} have shown significant promise. Although each technology offers distinct advantages and faces specific challenges, the ideal sensing mechanism for long-term applications and market leadership has yet to be determined.

Among various magnetic field sensors, GMI sensors are particularly notable for their exceptional sensitivity \cite{Mohri2015, Khan2021}. Theoretical studies indicate a potential relative impedance change of up to \qty{3000}{\percent} \cite{Betzholz2013}, while experimental results have demonstrated values as high as \qty{1733}{\percent} at \qty{100}{kHz} \cite{Xiao2000}. However, the relatively large size of the investigated sensor, measuring \qtyproduct{3x15}{mm}, poses challenges for miniaturisation. Despite this limitation, these promising findings highlight the significant potential of GMI sensors for highly precise magnetic field measurements.

Soft magnetic materials, whether crystalline (e.g., NiFe \cite{Garcia-Arribas2016}) or amorphous (e.g., CoSiB \cite{Morikawa1996}, FeSiB \cite{Yu2000}, CoNbZr \cite{Yokoyama2019}), are crucial for GMI sensors. These materials are characterised by high permeability and lack of magnetocrystalline anisotropy \cite{Garcia-Arribas2017}. Notably, NiFe and CoNbZr have been extensively researched.

Research indicates that the soft magnetic properties of NiFe deteriorate above a critical thickness of \qty{150}{nm} - \qty{200}{nm} due to stripe domain formation from columnar growth, inducing out-of-plane anisotropy \cite{Saito1964,Silva2017,Garcia-Arribas2017}. To mitigate this, sputtering NiFe layers below this critical thickness with additional thin interlayers (e.g., Ti, Cu, Ag, Au) between the ferromagnetic layers is essential to disrupt columnar growth and enhance effective ferromagnetic thickness \cite{DeCos2008,Correa2010,Kurlyandskaya2010,Svalov2012,Vaskovskii2013}.

Unlike NiFe, CoNbZr retains its soft magnetic properties even at thicknesses above \qty{1}{\micro m} \cite{Yokoyama2019}. Studies have explored various factors affecting its magnetoimpedance behaviour, including easy axis direction \cite{Kikuchi2014}, GMI element shape \cite{Kikuchi2015}, driving power \cite{Kikuchi2015a}, parallel and meander configurations \cite{Kikuchi2020}, joule heating \cite{Kikuchi2022}, and control of anisotropy direction \cite{Kikuchi2023}.

The effect of multilayer amorphous/metallic interlayer systems on magnetoimpedance behaviour has been extensively investigated \cite{Correa2007,Correa2008,Vaskovskii2013}. However, despite this substantial body of research, the potential of CoNbZr/metallic interlayer systems remains largely unexplored. While NiFe/Au systems are widely regarded as the standard in the field, CoNbZr has been proposed as a promising alternative due to its favourable properties. To address this gap, the present study systematically examines the magnetoimpedance performance of CoNbZr/Au multilayers, comparing them to both single-layer CoNbZr and the established NiFe/Au systems. This comparative analysis focuses on the electric, magnetic, and magnetoimpedance properties, aiming to assess the viability of CoNbZr/Au as a competitive alternative to the standard NiFe/Au system.

\section{Theory}

This section presents the theoretical background necessary to understand the experimental results. First, the principles of magnetoimpedance and ferromagnetic resonance are discussed, followed by an overview of theoretical models used in this study. Finally, the influence of the demagnetising field on the system is considered.

\subsection{Magnetoimpedance}

When an alternating current $i = i_0 \exp(i\omega t)$, with amplitude $i_0$ $\left([i_0] = \mathrm{A}\right)$ and angular frequency $\omega = 2\pi f$ $\left([f] = \mathrm{Hz}\right)$, flows through an electrical conductor, its behaviour can be described by the impedance $Z =R + iX$ $\left([Z] = \mathrm{\Omega}\right)$, where $R$ $\left([R] = \mathrm{\Omega}\right)$ is the resistance and $X$ $\left([X] = \mathrm{\Omega}\right)$ the reactance. For ferromagnetic conductors, $X = \omega L$, with $L$ $\left([L] = \mathrm{H}\right)$, being the inductance. The alternating current within these conductors is subject to the skin effect, characterised by an exponential decay of current density from the surface due to eddy currents. This decay is quantified by the skin penetration depth:

\begin{equation}
\delta = \sqrt{\frac{2\rho}{\omega \mu}}, [\delta] = \mathrm{m},
\label{penetration depth}
\end{equation}

where $\rho$ $\left([\rho] = \mathrm{\Omega \cdot m}\right)$ is the electric resistivity, and $\mu = \mu_0\mu_\mathrm{t}$ $\left([\mu] = \mathrm{H/m}\right)$ is the permeability of the material, referring to the response to the alternating magnetic field generated by the alternating current. Here, $\mu_0 = 4\pi \cdot 10^{-7} \ \mathrm{H/m}$ is the vacuum permeability and $\mu_\mathrm{t}$ is the transverse permeability. 

For a film whose lateral dimensions $\left( l, w \right)$ are much larger than its thickness $2t$, the impedance is given by \cite{Chen1999,Phan2008}:

\begin{equation}
Z = R_\mathrm{dc} \ ikt \coth (ikt),
\label{Z_Film}
\end{equation}

where $R_\mathrm{dc} = \rho l/A$ $\left([R_\mathrm{dc}] = \mathrm{\Omega}\right)$ is the dc resistance, with $l$ $\left([l] = \mathrm{m}\right)$ being the length, $A = wt$ $\left([A] = \mathrm{m^2}\right)$ being the cross-sectional area, with $w$ $\left([w] = \mathrm{m}\right)$ being the width and $t$ $\left([t] = \mathrm{m}\right)$ being the thickness of the film, and $k=(1+i)/\delta$ $\left([k] = \mathrm{1/m}\right)$ is the propagation constant. 

If the alternating current decays exponentially from the surface, Eq. \ref{Z_Film} simplifies to \cite{Buschow2003}: 

 \begin{equation}
Z \approx R_\mathrm{dc} \ k \frac{A}{l} = (1+i) \sqrt{\pi f \rho \mu}.
\label{Z_Film_approx}
\end{equation}

In ferromagnetic materials, both the skin depth $\delta$ and impedance $Z$ vary with the external magnetic field $H$ $\left([H] = \mathrm{A/m}\right)$, influenced by the frequency and field dependence of the transverse permeability $\mu_\mathrm{t} = \mu_\mathrm{t}(\omega,H)$. This magnetic field dependence leads to the giant magnetoimpedance (GMI) effect, expressed by the GMI ratio \cite{Buschow2003}:
 
\begin{equation}
\eta = \frac{|Z(B)|-|Z(B_\mathrm{ref})|}{|Z(B_\mathrm{ref})|},
\label{GMI}
\end{equation}

where $|Z|^2 = R^2 + X^2$ $\left([Z] = \mathrm{\Omega}\right)$ is the absolute value of the impedance, $B=\mu_0H$ $\left([B] = \mathrm{T}\right)$ is the magnetic flux density, and $B_\mathrm{ref}=\mu_0H_\mathrm{ref}$ is the reference flux density, corresponding to the maximum achievable flux density in the experiment at which the material is magnetically saturated.

By measuring $|Z|$ and its phase, $R$ and $X$ can be determined. Then, the real, imaginary and absolute values of $k, \delta$ and $\mu_\mathrm{t}$ can be estimated numerically using the Levenberg-Marquardt (L-M) method to solve Eq. \ref{Z_Film}. The L-M method is an iterative algorithm combining the advantages of gradient descent and the Gauss-Newton method, enabling efficient minimisation of non-linear least-squares problems \cite{Levenberg1944,Marquardt1963}. It adjusts parameters iteratively to minimise the error between the measured and modelled values of $Z$, balancing fast convergence with robustness to local minima. The transverse permeability $\mu_\mathrm{t}$ is corrected by subtracting the numerically calculated value with $\mu_\mathrm{t}(B_\mathrm{ref})$.

\begin{figure}[t]
	\centering
	\includegraphics[width=0.8\textwidth]{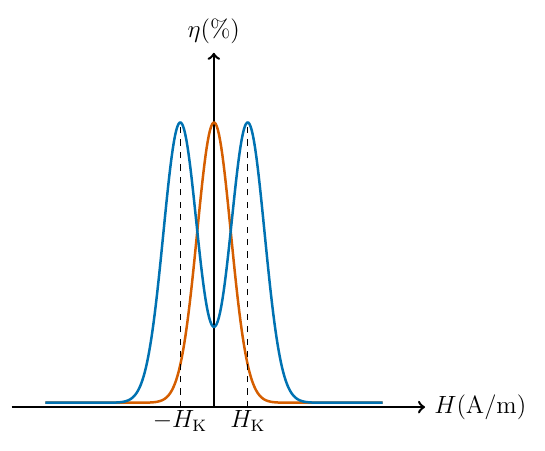}
	\caption[Schematic representation of the GMI curve.]
	{Schematic representation of the GMI cure as a function of the external field $H$. In the case of longitudinal anisotropy (red), the GMI ratio $\eta$ reaches its maximum at $H = 0$. In the transverse case (blue), the maximum shifts to the anisotropy field $\pm H_\mathrm{K}$, as described in \cite{Garcia-Arribas2017}.}
\label{fig:GMI_curve}
\end{figure}

The shape of a GMI curve is influenced by the direction of the alternating current, the external field, and the anisotropy axis of the material. Figure \ref{fig:GMI_curve} illustrates the GMI curve's behaviour as a function of the anisotropy axis when the current and field directions are collinear. In the case of longitudinal anisotropy (red), the transverse permeability is maximal at $H = 0$ and decreases with increasing field until the material is saturated in the field direction. Consequently, the GMI ratio is at its maximum at zero field and equals zero at the reference field. In contrast, with transverse anisotropy (blue), the transverse permeability is maximal at the anisotropy field

\begin{equation}
H_\mathrm{K} = \frac{2K_\mathrm{eff}}{\mu_0M_\mathrm{s}}, [H_\mathrm{K}] = \mathrm{A/m}.
\label{Hk}
\end{equation}

The sensitivity $s$ of the GMI effect is calculated using the slope of the GMI curve:

\begin{equation}
s = \frac{\mathrm{d}\eta}{\mathrm{d}H}, [s] = \mathrm{\%/mT}.
\label{Sensitivity}
\end{equation}

\subsection{Ferromagnetic resonance}

Ferromagnetic resonance (FMR) is a phenomenon in which the precessional motion of magnetic moments in a ferromagnetic material resonates with an applied alternating magnetic field. This resonance occurs at a specific frequency, $f_\mathrm{res}$ $\left([f_\mathrm{res}] = \mathrm{Hz}\right)$, determined by the material's magnetic properties and external conditions. The resonance condition is governed by the Kittel equation \cite{Kittel1948}:

\begin{equation}
f_\mathrm{res} = \frac{\gamma}{2\pi}\mu_0 \sqrt{(H_\mathrm{res} + H_\mathrm{K})(H_\mathrm{res} + H_\mathrm{K} + M_\mathrm{s})},
\label{Kittel}
\end{equation}

where $\gamma = \frac{g\mu_\mathrm{B}}{\hbar}$ $\left([\gamma] = \mathrm{rad/s\cdot T}\right)$ is the gyromagnetic ratio, with $g$ being the Land\'e g-factor, $\mu_\mathrm{B}$ $\left([\mu_\mathrm{B}] = \mathrm{J/T}\right)$ the Bohr magneton, and $\hbar$ $\left([\hbar] = \mathrm{Js}\right)$ the reduced Planck constant. $H_\mathrm{res}$ $\left([H_\mathrm{res}] = \mathrm{A/m}\right)$ represents the resonance field, while $M_\mathrm{s}$ $\left([M_\mathrm{s}] = \mathrm{A/m}\right)$ denotes the saturation magnetisation.

\subsection{Theoretical models}

This section outlines the theoretical models used to describe the resistivity and saturation magnetisation in multilayer systems. First, electric models are discussed, focusing on the parallel resistor model and dilution models, which are used to calculate resistivity in ferromagnetic and non-ferromagnetic layers. Then, magnetic models are introduced, where the dilution model is applied to estimate the saturation magnetisation in layered systems.

\subsubsection{Electric models}

The resistivity of ferromagnetic (FM) and non-ferromagnetic (NFM) layers within a multilayered system, consisting of $N$ FM layers and $N-1$ NFM layers, can be determined from the measured resistance using a parallel resistor model. The total measured resistance, $R$, is given by:

\begin{equation}
\frac{1}{R} = \frac{N-1}{R_\mathrm{NFM}} + \frac{N}{R_\mathrm{FM}},
\label{parallel_resistor}
\end{equation}

where $R_\mathrm{NFM}$ and $R_\mathrm{FM}$ are the resistances of the NFM and FM layers, respectively. From this relationship, the resistivities $\rho_\mathrm{NFM}$ and $\rho_\mathrm{FM}$ of the NFM and FM layers can be derived as:

\begin{equation}
\rho_\mathrm{NFM} = \frac{(N-1)t_\mathrm{NFM}w \, \rho_\mathrm{FM} R}{\rho_\mathrm{FM}l - Nt_\mathrm{FM}w R}, 
\label{Rho_NFM}
\end{equation}

\begin{equation}
\rho_\mathrm{FM} = \frac{Nt_\mathrm{FM}w \, \rho_\mathrm{NFM} R}{\rho_\mathrm{NFM}l - (N-1)t_\mathrm{NFM}w R},
\label{Rho_FM}
\end{equation}

where $\rho_\mathrm{NFM}$ and $\rho_\mathrm{FM}$ are the resistivities, and $t_\mathrm{NFM}$ and $t_\mathrm{FM}$ are the thicknesses of the NFM and FM layers, respectively. Here, $w$ represents the width, and $l$ represents the length of the layers.

An alternative approach involves using a dilution model, which considers the relative contributions of the FM and NFM layers to the overall resistivity of the multilayer structure. This model assumes that the NFM layers have a negligible impact on the total resistivity, allowing the resistivity of the composite to be approximated primarily by the FM layers. The generalised expression for the resistivity, $\rho_\mathrm{layer}'$, in both single-layer and multilayer films is:

\begin{equation}
\rho_\mathrm{layer}'= \frac{t_\mathrm{layer,tot}}{t_\mathrm{tot}} \rho_\mathrm{layer},
\label{Rho_Dil}
\end{equation}

where $t_\mathrm{layer,tot}$ is the total thickness of the FM or NFM layers being considered, $t_\mathrm{tot}$ is the total thickness of the entire film, and $\rho_\mathrm{layer}$ is the measured resistivity of the FM or NFM layer.

In the specific case where the dilution model is applied to calculate the resistivity of individual layers with $t_\mathrm{layer,tot} = t_\mathrm{tot}$, Eq. \ref{Rho_Dil} simplifies to $\rho_\mathrm{layer}' = \rho_\mathrm{layer}$. This indicates that for a single layer, its resistivity is equal to its intrinsic measured resistivity, as there are no other layers affecting the overall resistivity.

\subsubsection{Magnetic models}

The saturation magnetisation $M_\mathrm{s,FM}$ of the FM layer can be used to predict the saturation magnetisation in a layered system consisting of FM and NFM layers using a dilution model. Assuming that the NFM layers have a negligible influence on the saturation magnetisation, the saturation magnetisation $M'_\mathrm{s,layer}$ of the layered system can be calculated using the ratio of the total thickness $t_\mathrm{FM,tot}$ of the FM layers to the total thickness $t_\mathrm{tot}$f the entire film:

\begin{equation} 
M'_\mathrm{s,layer} = \frac{t_\mathrm{FM,tot}}{t_\mathrm{tot}}M_\mathrm{s,FM}. \label{Ms_Dil} 
\end{equation}

%For an FM layer that is composed of an alloy of an FM element (e.g., Co, Ni, Fe) with an atomic fraction $p_\mathrm{FM}$ and NFM elements, the magnetic saturation moment per FM atom $n_\mathrm{B}$ can be calculated as follows:
%
%\begin{equation} 
%n_\mathrm{B} = \frac{p_\mathrm{FM} M_\mathrm{s,FM}}{\mu_\mathrm{B} N_\mathrm{FM}} \label{nB} 
%\end{equation}
%
%where $\mu_\mathrm{B} \approx \qty{9,27e-24}{J/T}$ is the Bohr magneton, and the volumetric density
%
%\begin{equation} N_\mathrm{FM} = \frac{\rho_\mathrm{FM} N_\mathrm{A}}{M_\mathrm{FM}} 
%\label{N_FM} 
%\end{equation}
%
%of the FM atoms can be calculated from the density $\rho_\mathrm{FM}$, the molar mass $M_\mathrm{FM}$ of the FM atoms, and the Avogadro constant $N_\mathrm{A}$.

\subsection{Demagnetising field}

The demagnetising factor quantifies the reduction in the internal magnetic field due to the geometry of a ferromagnetic sample. For rectangular films, the demagnetising factors ($N_x$, $N_y$, $N_z$) depend on the aspect ratio of the sample dimensions (length $l$, width $w$, thickness $t$). These factors satisfy the condition $N_x + N_y + N_z = 1$ for a uniformly magnetised sample~\cite{Osborn1945,Aharoni1998}. 

In thin films, the demagnetising factor $N_z$ along the thickness is typically near unity. Conversely, $N_x$ and $N_y$ are much smaller but influence the in-plane anisotropy. The demagnetising field $H_\mathrm{d}$ $\left([H] = \mathrm{A/m}\right)$ is related to $N$ by \cite{Cullity2009}:

\begin{equation}
H_\mathrm{d} = -N M,
\label{Hd}
\end{equation}

where $M$ is the magnetisation, and $N$ is the relevant demagnetising factor for the considered direction. The corresponding shape anisotropy constant $K_\mathrm{s}$ is described as: 

\begin{equation}
K_\mathrm{s} = \frac{1}{2}\mu_0\left( N_x - N_y \right)M^2.
\label{Ks}
\end{equation}

Table~\ref{demag_factors} lists the demagnetisation factors and corresponding demagnetisation fields for the structures used in this work.

\begin{table}[!t]
\centering
\caption{In-plane demagnetising factors $N_x$ and $N_y$ (length and width direction, respectively) and fields $\mu_0|H_{\mathrm{d},x}|$ and $\mu_0|H_{\mathrm{d},x}|$ for a rectangular single-layer and multilayer film with varying lateral aspect ratios $l:w$ assuming $M = \qty{1}{MA/m}$ \cite{Aharoni1998}.}
\begin{tabular}{@{\extracolsep{\fill}}
l r@{:}l S S S S S 
@{ }}
{System} & {$l$}&{$w$} & {$N_x $} & {$N_y$} & {$\mu_0|H_{\mathrm{d},x}|$} & {$\mu_0|H_{\mathrm{d},y}|$} \\ 
{} & \multicolumn{2}{c}{} & {\num{e-4}} & {\num{e-4}} & {mT} & {mT} \\ 
\hline
 &1000&100& 19 & 198 & 2.39 & 24.88 \\ 
 &1000&50 & 17 & 352 & 2.14 & 44.23 \\ 
 &500&20&27 & 731& 3.39 & 91.86 \\ 
{Single-layer} &1000&20&14 & 734 & 1.76 & 92.24 \\ 
{Thickness:} \qty{1040}{nm} &2000&20&7 & 736& 0.88 & 92.49 \\ 
 &1000&10& 12& 1243 & 1.51 & 156.20 \\ 
 &1000&5& 9 & 2032 & 1.13 & 255.35 \\ 
 &5000&20&3 &737 & 0.38 & 92.61 \\ 
\hline
 &1000&100 & 20 & 212 & 2.51 & 26.64 \\ 
 &1000&50 & 18 & 377 & 2.26 & 47.38 \\ 
 &500&20&29 & 780 & 3.64 & 98.02 \\ 
{Multilayer} & 1000 &20 & 15 & 783 & 1.88 & 98.39 \\ 
{Thickness:} \qty{1130}{nm} & 2000 & 20 & 7 & 784 & 0.88 & 98.52 \\
&1000&10& 12& 1321 & 1.51 & 166.00 \\ 
&1000&5 & 10 & 2149 & 1.26 & 270.05 \\ 
&5000 & 20 & 3 & 786 & 0.38 & 98.77 \\ 
\end{tabular}
\label{demag_factors}
\end{table}

\section{Materials and methods}

This section provides an overview of the materials and methods employed in this study, covering the sample preparation, microstructural characterisation techniques, and experimental setup.

\subsection{Sample preparation}

\begin{figure}[t!]
\centering
\includegraphics[width=\textwidth]{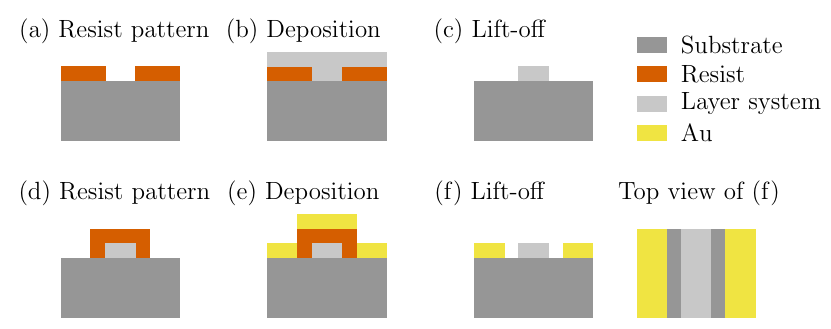} 
\caption{Schematic representation of the sample preparation process. The steps include patterning the photoresist (a,d), depositing the layer system (b) and gold (e), and removing the photoresist via lift-off (c,f).}
\label{fig:Process_flow}
\end{figure}

Figure \ref{fig:Process_flow} illustrates the schematic representation of the sample preparation process. Pre-cut Si-SiO$_2$ wafers with lateral dimensions of \qtyproduct{20x20}{mm} were used for structuring. The substrates were first cleaned with acetone, isopropanol, and deionised water, and an adhesion promoter (AR 300-80, Allresist) was spin-coated for \qty{1}{min} at \qty{4000}{rpm} to enhance the adhesion between the photoresist and the substrate. The substrates were baked on a hotplate for \qty{2}{min} at \qty{180}{\degreeCelsius}. Subsequently, the positive photoresist (AR-P 5320, Allresist) was spin-coated for \qty{1}{min} at \qty{4000}{rpm} and baked for \qty{4}{min} at \qty{105}{\degreeCelsius}.

\begin{figure}[t!]
\centering
\includegraphics[width=\linewidth]{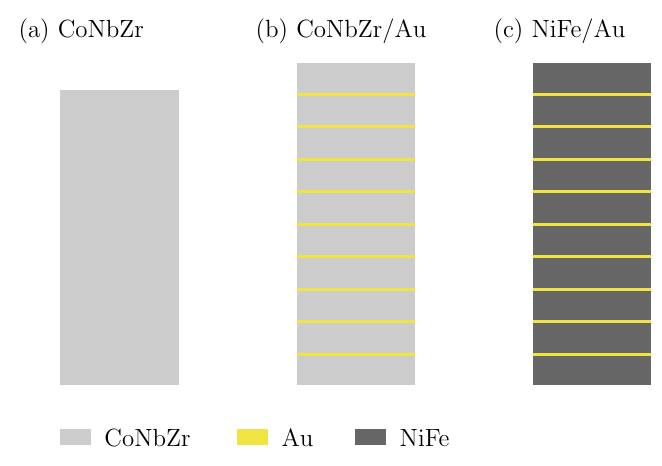} 
\caption{Schematic representation of the layer systems in cross-section: (a) a single-layer CoNbZr system with a film thickness of \qty{1040}{nm}; (b) a multilayer system consisting of 10 layers of CoNbZr and 9 layers of Au, with individual film thicknesses of \qty{104}{nm} and \qty{10}{nm}, respectively; and (c) a multilayer system composed of 10 layers of NiFe and 9 layers of Au, with the same respective thicknesses of \qty{104}{nm} and \qty{10}{nm}.}
\label{fig:layer_systems}
\end{figure}

\begin{table}[!b]
\centering
\caption{Parameters for sputter deposition on the system from VSW Technology Limited.}
\sisetup{
table-alignment-mode = format,
table-number-alignment = center
}
\begin{tabular} {@{\extracolsep{\fill}}
 c 
 S[table-format = 2.1(1.1)] 
 S[table-format = 3.0(1.0){***}] 
 S[table-format = 3.1(1.1)]
 @{ }}
{Target Composition} & {Pressure} & {Power} & {Rate} \\
{(at.\%)} & {\qty{e-3}{mbar}} & {W} & {nm/min} \\
\hline
{Co$_{85}$Nb$_{12}$Zr$_{3}$} & 1.7(1) & 100(2){\ RF} & 4.9(2) \\
{Ni$_{81}$Fe$_{19}$} & 1.7(1) & 85(2){\ RF} & 4.9(2) \\ 
{Au} & 5.0(1) & 50(1){\ DC} & 21.0(2) \\ 
\end{tabular}
\label{sputtering_parameters}
\end{table}

\begin{figure}[!b]
\includegraphics[width=\textwidth]{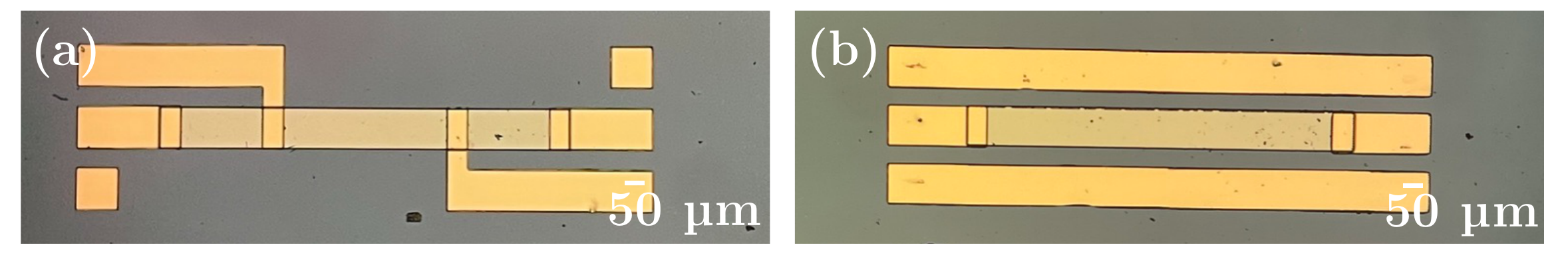}
\caption{Images of the patterns used for (a) resistance and (b) magnetoimpedance measurements.}
\label{fig:Patterns}
\end{figure}

The samples were patterned into stripes with widths ranging from \qty{5}{\micro m} to \qty{100}{\micro m} and lengths from \qty{0.5}{mm} to \qty{5}{mm} using laser lithography ($\upmu$PG 101, Heidelberg Instruments). Following development in developer (AR 300-26, Allresist) for \qty{3}{min}, rinsing with deionised water, and nitrogen drying, the patterned substrates were mounted on a rotating holder, maintaining a target-to-substrate distance of \qty{12}{cm}. Once the base pressure reached below \qty{1e-8}{mbar}, the layer systems, as defined in Fig. \ref{fig:layer_systems}, were deposited using the sputtering parameters listed in Tab. \ref{sputtering_parameters}. After the lift-off process, contact pads for the RF probes (FPC-GSG-150, Cascade Microtech) were patterned onto the stripes, followed by the sputtering of Au under the same conditions listed in Tab. \ref{sputtering_parameters} and another lift-off process. The fabricated samples are shown in Fig. \ref{fig:Patterns}, including the Au patterns used for resistance (see Fig. \ref{fig:Patterns}(a)) and magnetoimpedance (see Fig. \ref{fig:Patterns}(b)) measurements.

\subsection{Microstructural characterisation}

The electron diffraction patterns were acquired using a \textsc{JEOL ARM-200F} TEM equipped with a LaB$_6$ cathode and operated at \qty{200}{kV}. For this purpose, TEM slices were prepared by milling an area of \qtyproduct{15x5}{\um} out of a continuous film using a focused-ion beam (FIB) and an in situ lift-off with a micromanipulator \cite{Tomus2013}. The elemental composition of the TEM slices was determined using an EDAX energy dispersive X-ray spectroscopy (EDX) system. 

\subsection{Magnetic characterisation}

For the magnetic measurements, the in-plane hysteresis loops of one patterned stripe (\qtyproduct{0.1x1}{mm}) on \qtyproduct{8x8}{mm} samples were measured by a Vibrating Sample Magnetometer (PPMS-VSM, Quantum Design). The system was calibrated with a \qty{0.25}{g} Pd cylinder, which is assumed to have a magnetic moment of \qty{0.013}{mAm^2} at \qty{1}{T} field and \qty{298}{K} based on the literature value of the susceptibility of Pd (\qty{5.25e-2}{Am^2/kg}) \cite{Manuel1963}. The diamagnetic contribution of sample holder and Si substrate was subtracted as a linear background signal of \qty{0.1}{\micro Am^2/T}, obtained from a fit to the high-field branches of the in-plane hysteresis loops. Additionally, magneto-optic Kerr effect (MOKE) images were taken to observe the domain structures.

\subsection{Experimental setup}

\begin{figure}[t!]
\includegraphics[width=\textwidth]{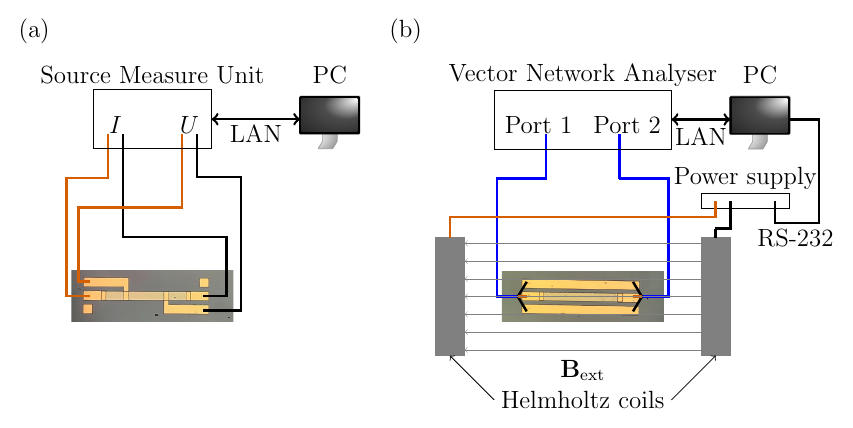}
\caption{Schematic setup used for (a) resistance and (b) magnetoimpedance measurements.}
\label{fig:Setup}
\end{figure}

Figure \ref{fig:Setup} show the setups used for resistance and magnetoimpedance measurements. In Fig. \ref{fig:Setup}(a), a Source Measure Unit (SMU; B2901A, Keysight) is used to determine the resistance using the four-point method. The samples with the lateral dimensions \qtyproduct{0.1x1}{mm} are contacted using two RF probes and are connected in series with the SMU. The voltage $U$ is measured over a defined length of \qty{0.45}{mm}. The SMU is controlled by a PC via LAN. 

\begin{figure}[t!]
\includegraphics[width=\textwidth]{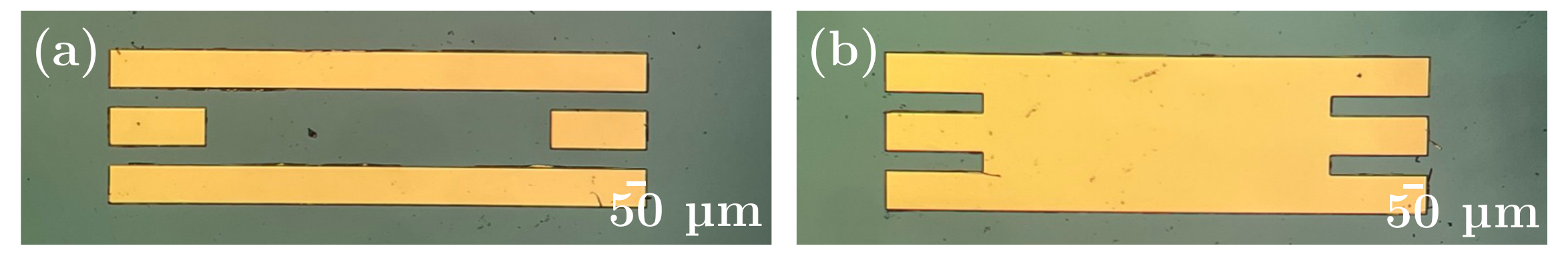}
\caption{Deembedding structures used for magnetoimpedance measurements containing (a) open and (b) short structure.}
\label{fig:Deembedding}
\end{figure}

For magnetoimpedance measurements, as shown in Fig. \ref{fig:Setup}(b), the samples are located in the center of a Helmholtz coil pair, which is operated with a unipolar current source (Hercules 2500 (250 V and 10 A),TET Electronics). A changeover switch is connected in between for polarity reversal. The current source is connected to the PC via an RS-232 USB converter. For the estimation of the GMI ratios, \qty{40}{mT} (corresponds to 10 A) is selected as the reference flux density. To determine the magnetoimpedance characteristics, the RF probes are connected to a vector network analyser (VNA; ZNL3, Rohde \& Schwarz) using a coaxial cable. A Thru-Open-Short-Match (TOSM) calibration on a standard impedance substrate (101-190C, Cascade Microtech) for each port is used to eliminate the influences of the VNA, the coaxial cables and the RF probes. 

To determine the intrinsic impedance of the samples, the de-embedding method \cite{Kolding2000} is used. In this method, reference measurements are carried out to determine parasitic losses. The reference structures required for this can be found in Fig. \ref{fig:Deembedding}. For the calibration, the frequency response of the open-circuit (see Fig. \ref{fig:Deembedding}(a)) and the short-circuit structure (see Fig. \ref{fig:Deembedding}(b)) is measured separately for each RF probe. For both the TOSM calibration and the deembedding method, the frequency response was measured from \qty{1}{MHz} to \qty{3}{GHz} in \qty{1}{MHz} steps with a bandwidth of \qty{1}{kHz} and an output power of \qty{-20}{dBm} (corresponds to \qty{10}{\micro W}) and stored in the VNA. 

\begin{figure}[!b]
\centering
\includegraphics[width=0.32\textwidth]{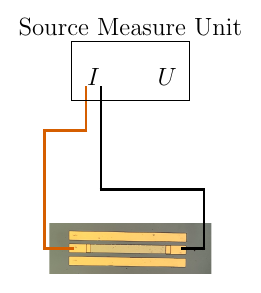}
\caption{Schematic setup used for annealing the GMI structures.}
\label{fig:Annealing_setup}
\end{figure}

The experimental setup for the heat treatment of the GMI structures is schematically illustrated in Fig. \ref{fig:Annealing_setup}. The structures were contacted using two RF probes and connected in series with the SMU via banana cables, which were linked to the RF probes through SMA-BNC adapters and BNC-banana connectors. The current $I$ passing through the structure resulted in Joule heating and generated an Oersted field, thereby inducing additional magnetic anisotropy in the sample. The heat treatment was performed in air by applying constant currents for a duration of \qty{1}{min}. This duration was selected based on previous studies, which showed that for similar structures, the temperature stabilised after 1 minute of current application \cite{Kikuchi2021}. To ensure consistent and reliable results, the same approach was followed in this work.

\section{Results and discussion}

This section presents the results and discusses the key findings of the study. It begins with an analysis of the microstructural properties of the samples, followed by an examination of their electrical, magnetic, and magnetoimpedance properties. Finally, the findings are discussed in relation to the observed experimental behaviours and mechanisms affecting the material properties.

\subsection{Microstructural properties} 

\begin{figure}[!b]
\centering
\includegraphics[width=\textwidth]{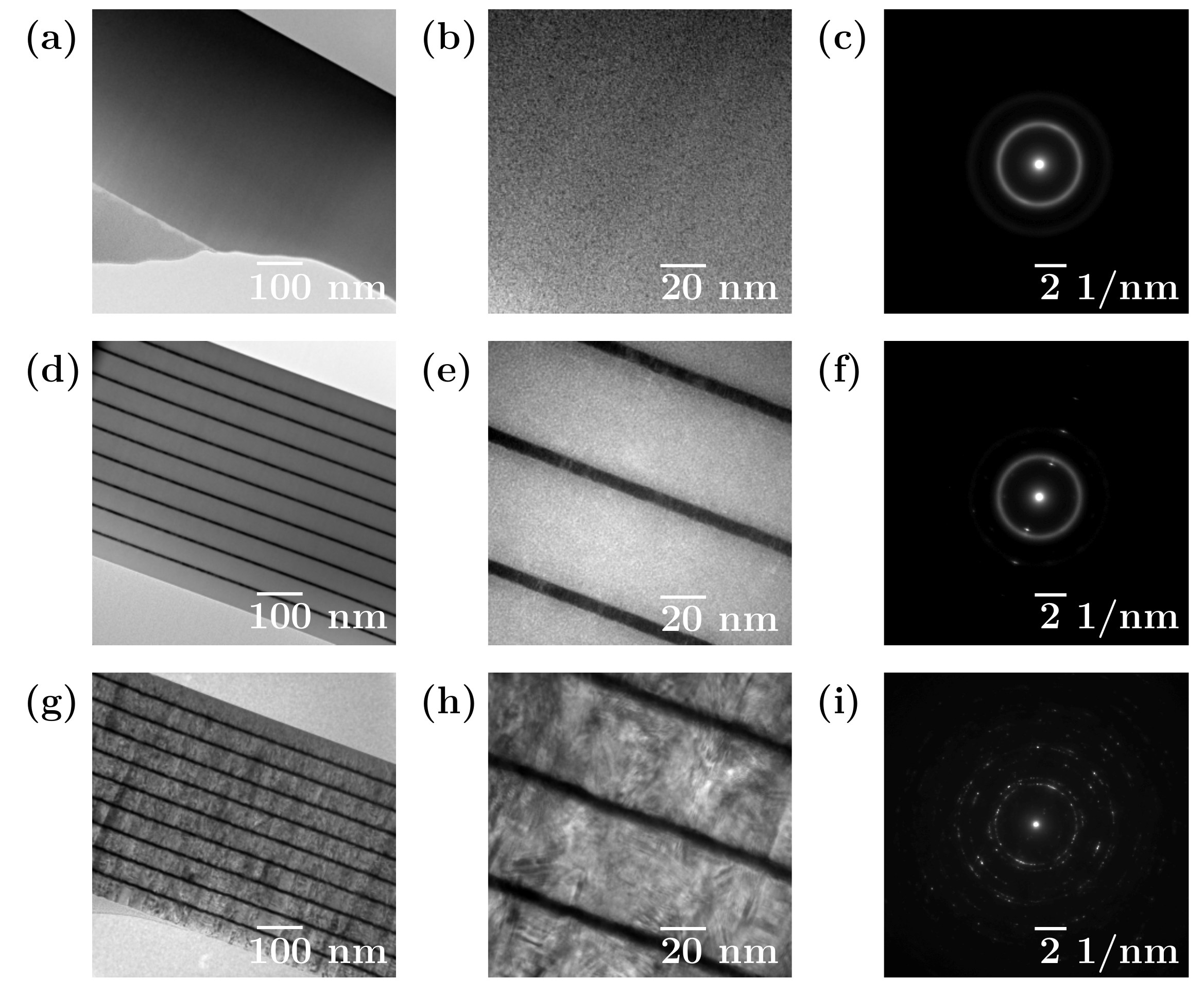} 
\caption{Cross-sectional TEM images and SAED patterns of (a-c) single-layer CoNbZr, (d-f) multilayer CoNbZr/Au, and (g-i) multilayer NiFe/Au films. (a-c): The uniform and featureless structure in the CoNbZr layer confirms its amorphous nature, supported by the diffuse ring in the SAED pattern. (d-f): The CoNbZr/Au multilayer shows well-defined alternating amorphous CoNbZr and crystalline Au layers, with distinct diffraction spots corresponding to Au lattice planes. (g-i): The NiFe/Au multilayer exhibits crystalline NiFe layers and Au layers, with corresponding SAED patterns confirming the crystallinity of both materials.}
\label{fig:TEM_SAED_comparison}
\end{figure}

The microstructure of single-layer CoNbZr, multilayer CoNbZr/Au, and multilayer NiFe/Au films were investigated using cross-sectional TEM and selected area electron diffraction (SAED). The results are shown in Fig. \ref{fig:TEM_SAED_comparison}. For further validation of the diffraction data, the distances between the diffraction spots and the rings were measured five times for each sample to improve statistical accuracy. Line profiles were extracted from the diffraction rings to identify the corresponding crystallographic planes. The average values were calculated, and the standard deviation was used to determine the error in d-spacing, ensuring robust and reliable measurements.

Single-layer CoNbZr (Figs. \ref{fig:TEM_SAED_comparison}(a) - (c)) exhibits a featureless, uniform cross-section (Figs. \ref{fig:TEM_SAED_comparison}(a) and \ref{fig:TEM_SAED_comparison}(b)), confirming the amorphous structure of the film further supported by the SAED pattern in Fig. \ref{fig:TEM_SAED_comparison}(c). The presence of a second, faint ring in Fig. \ref{fig:TEM_SAED_comparison}(c) is likely due to an instrumental artefact.

In the multilayer CoNbZr/Au system (Figs. \ref{fig:TEM_SAED_comparison}(d) - (f)), the TEM cross-sections (Figs. \ref{fig:TEM_SAED_comparison}(d) and \ref{fig:TEM_SAED_comparison}(e)) reveal well-defined, alternating layers of CoNbZr and Au. The CoNbZr layers maintain an amorphous structure, as indicated by the diffuse ring in the SAED pattern (Fig. \ref{fig:TEM_SAED_comparison}(f)). The additional discrete diffraction spots correspond to the crystalline Au layers consistent with the cross-sectional TEM in Fig. \ref{fig:TEM_SAED_comparison}(d), where crystals can be identified. The d-spacings of the observed diffraction spots were measured as \textit{d} = \qty{0.237(1)}{nm}, \textit{d} = \qty{0.118(1)}{nm}, and \textit{d} = \qty{0.080(1)}{nm}. To assign these to specific planes, the lattice constant of Au (\textit{a} = \qty{0.409}{nm} \cite{Ashcroft1976}) was used. The relation $ d
= a/ \sqrt{h^2+k^2+l^2}$ shows that the innermost spots correspond to the (111) plane, while the outer spots are assigned to the (222) and (333) planes, respectively. The sharp interfaces between the amorphous CoNbZr and crystalline Au suggest minimal interdiffusion and a well-controlled sputtering process.

For the multilayer NiFe/Au system (Figs. \ref{fig:TEM_SAED_comparison}(g) - (i)), the TEM cross-sections (Figs. \ref{fig:TEM_SAED_comparison}(g) and \ref{fig:TEM_SAED_comparison}(h)) reveal distinct, alternating layers of NiFe and Au. The NiFe layers exhibit grain-like features, indicating crystallinity, while the Au layers also remain crystalline. The SAED pattern in Fig. \ref{fig:TEM_SAED_comparison}(i) shows well-defined rings with discrete diffraction spots. Using the lattice constant for Ni$_{80}$Fe$_{20}$ (\textit{a} = \qty{0.355}{nm} \cite{Wakelin1953}), the d-spacings were calculated and matched to \qty{0.206(1)}{nm} for (111), \qty{0.178(1)}{nm} for (200), \qty{0.126(1)}{nm} for (220), and \qty{0.107(1)}{nm} for (222). Additionally, the fifth ring corresponds to the (333) plane of Au, confirming the crystalline nature of both NiFe and Au in the multilayer system.

Two faint features at approximately 1 and 7 o'clock are visible in the diffraction patterns (Figs. \ref{fig:TEM_SAED_comparison}(c), \ref{fig:TEM_SAED_comparison}(f), and \ref{fig:TEM_SAED_comparison}(i)). These features were observed in all three examined samples and are likely due to instrumental artefacts rather than sample-specific effects, as their positions remain unchanged with sample rotation or under varying measurement conditions. Importantly, this does not affect the interpretation of the crystalline structure.

\begin{table}[t!]
\centering
\caption{Elemental composition (at.\%) of the targets and sputtered films for CoNbZr and NiFe systems. The target compositions of CoNbZr and NiFe were optimised for nearly zero magnetostriction ($<$ \qty{1}{ppm}) \cite{Yamaguchi2015, Klokholm1981}. A significant increase in the Co/Nb ratio from 7.1 in the target to 13 in the sputtered film is observed.}
\begin{tabular} {@{\extracolsep{\fill}} c S S S | S S @{ }}
{Element} & {Co} & {Nb} & {Zr} & {Ni} & {Fe} \\
\hline
{Target (at.\%)} & 85(1) & 12(1) & 3(1) & 81(1) & 19(1) \\
{Film (at.\%)} & 91(1) & 7(1) & 2(1) & 80(1) & 20(1)\\ 
\end{tabular}
\label{EDX}
\end{table}

The EDX analyses in Tab. \ref{EDX} reveal a significant increase in the Co/Nb ratio from 7.1 in the target to 13 in the sputtered film. This variation is likely attributed to differences in the deposition rates of Co and Nb. These differences may arise from factors such as sputtering pressure, target-to-substrate distance, and the substantial disparities in atomic mass and radius between Co and Nb \cite{Neidhardt2008}.

\subsection{Electrical properties} 

\begin{table}[!b]
\centering
\caption[Comparison of the measured and calculated resistivities for the sputtered films.]
{Comparison of the measured and calculated resistivities ($\rho$) for the sputtered films. The measured resistivity of the multilayer amorphous CoNbZr/Au films is approximately three times higher than that of the multilayered NiFe/Au films, but only about half that of a single CoNbZr layer. The resistivities of Au and NiFe were determined using the parallel resistor model (see Eqs. \ref{parallel_resistor}, \ref{Rho_NFM} and \ref{Rho_FM}). The resistivities ($\rho'$) calculated using the dilution model (Eq. \ref{Rho_Dil}) indicate that the resistivity of CoNbZr/Au is \qty{55}{\percent} higher than the measured value, while the measured resistivity for NiFe/Au aligns closely with the predictions of the dilution model.}
\begin{tabular} {@{\extracolsep{\fill}}
 c S S S S S
 @{ }}
{Film} & {CoNbZr} & {CoNbZr/Au} & {NiFe/Au} & {Au} & {NiFe} \\
\hline
{$\rho$ (\qty{}{\micro\ohm\centi\meter})} & 119(4) & 71(3) & 26(1) & 12(4) & 29(20) \\
{$\rho'$ (\qty{}{\micro\ohm\centi\meter})} & 119(4) & 110(12) & 27(2) & 12(4) & 29(20)\\ 
\end{tabular}
\label{Rho}
\end{table}

Table \ref{Rho} presents a comparison of the measured and calculated resistivities for the sputtered films. The resistivity of the multilayer amorphous CoNbZr/Au films is approximately three times higher than that of the multilayered NiFe/Au films, yet only about half that of a single CoNbZr layer, which exhibits a resistivity of \qty{119(4)}{\micro\ohm\centi\meter}, similar to that of Co$_{85}$Nb$_{12}$Zr$_3$ \cite{Kikuchi2022}. The errors arise from uncertainties in the thickness ($t$), length ($l$), width ($w$), and the resistance ($R$) due to the slope uncertainty in the $U(I)$ curve. The resistivities of Au and NiFe were determined using a parallel resistor model (see Eqs. \ref{Rho_NFM} and \ref{Rho_FM}). First, the resistivity $\rho_\mathrm{CoNbZr}$ of CoNbZr is estimated and used in Eq. \ref{Rho_NFM} to calculate $\rho_\mathrm{Au}$. Then, $\rho_\mathrm{Au}$ is used in Eq. \ref{Rho_FM} to calculate $\rho_\mathrm{NiFe}$.

Due to the Au layer thickness being smaller than its mean free path (\qty{37.7}{nm} \cite{Gall2016}), the estimated resistivity for Au is six times higher than the bulk Au resistivity (\qty{2.214}{\micro\ohm\centi\meter} \cite{Gall2016}). Similarly, the estimated resistivity for NiFe is approximately \qty{30}{\percent} higher than the bulk value for NiFe (\qty{22}{\micro\ohm\centi\meter} \cite{Garcia-Arribas2016}). The errors result from inaccuracies in the lateral film dimensions, layer thicknesses, and resistivities, particularly for Au. Since the error in $\rho_\mathrm{Au}$, $\Delta\rho_\mathrm{Au} \approx \frac{1}{3} \rho_\mathrm{Au}$, is relatively large, it significantly impacts the uncertainty in the resistivity of NiFe/Au. This uncertainty is approximately \qty{70}{\percent} of $\rho_\mathrm{NiFe}$, bringing the measured value within the error limits of the literature value.

The resistivities ($\rho'$) calculated using the dilution model (Eq. \ref{Rho_Dil}) indicate that the resistivity of CoNbZr/Au is \qty{55}{\percent} higher than the measured value, while the measured resistivity for NiFe/Au aligns closely with the predictions of the dilution model.

\subsection{Magnetic properties} 

\begin{figure}[!b]
\centering
\includegraphics[width=\textwidth]{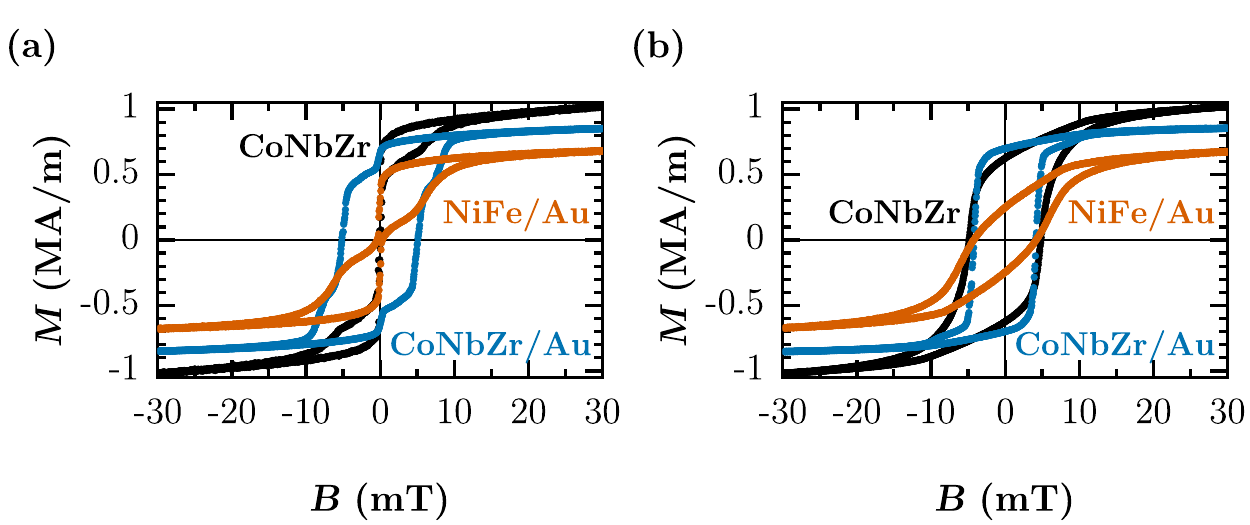}
\caption{In-plane hysteresis loops of the magnetisation $M$ for a field sweep of $B = \pm$ \qty{30}{mT} on patterned \qtyproduct{0.1x1}{mm} elements of single-layer CoNbZr, multilayer CoNbZr/Au, and multilayer NiFe/Au in (a) the length direction and (b) the width direction. The saturation magnetisation values, considering the Au layer in volume calculations, are \qty{1.01(5)}{MA/m} for CoNbZr, \qty{0.85(4)}{MA/m} for CoNbZr/Au, and \qty{0.68(3)}{MA/m} for NiFe/Au. In (a), CoNbZr exhibits a coercive field of \qty{0.09(1)}{mT}, which is 57 times smaller than that of CoNbZr/Au and 4 times smaller than that of NiFe/Au. Its maximum permeability reaches approximately 7000, significantly exceeding the values observed for both multilayers. The remanent-to-saturation magnetisation ratios are 0.17 for CoNbZr, 0.80 for CoNbZr/Au, and 0.41 for NiFe/Au. In (b), CoNbZr displays a coercive field of \qty{4.86(1)}{mT}, which is higher by \qty{17}{\percent} compared to CoNbZr/Au and by \qty{16}{\percent} compared to NiFe/Au. Its maximum permeability in this direction is 800, lower than in the length direction and also lower than in the multilayers. The remanent-to-saturation magnetisation ratios are 0.61 for CoNbZr, 0.82 for CoNbZr/Au, and 0.37 for NiFe/Au.}
\label{fig:VSM_single_bar}
\end{figure}

The measured in-plane hysteresis loops of the magnetisation $M$ for a field sweep of $B = \pm \qty{30}{mT}$ on patterned \qtyproduct{0.1x1}{mm} elements of single-layer CoNbZr, multilayer CoNbZr/Au, and NiFe/Au in both the length and width directions are depicted in Fig. \ref{fig:VSM_single_bar}. The saturation magnetisations were estimated by taking the magnetic moment at the maximum applied field of \qty{30}{mT} and normalising it to the volume of the sputtered stripes, accounting for the Au thickness in the calculation. This yielded values of \qty{1.01(5)}{MA/m} for CoNbZr (\qty{27}{\percent} higher than Co$_{85}$Nb$_{12}$Zr$_3$ from \cite{Kikuchi2022}), \qty{0.85(4)}{MA/m} for CoNbZr/Au, and \qty{0.68(3)}{MA/m} for NiFe/Au. It should be noted that the CoNbZr sample is nearly saturated at the maximum applied field of \qty{30}{mT}.

From Figs. \ref{fig:VSM_single_bar}(a) and \ref{fig:VSM_single_bar}(b), the coercive field (\(B_\mathrm{c}\)), remanent magnetisation (\(M_\mathrm{r}\)) to saturation magnetisation (\(M_\mathrm{s}\)) ratios, and maximum permeability (\(\mu_\mathrm{max}\)) were extracted and are summarised in Tab. \ref{VSM_summary}.

In the length direction, the coercive field for CoNbZr is \qty{0.09(1)}{mT}, which is 57 times lower than CoNbZr/Au (\qty{5.11(1)}{mT}) and 4 times lower than NiFe/Au (\qty{0.36(1)}{mT}). The maximum permeability for CoNbZr reaches approximately 7000, which is 6 times higher than CoNbZr/Au and 1.5 times higher than NiFe/Au. The remanent magnetisation to saturation magnetisation ratios are 0.17 for CoNbZr, 0.80 for CoNbZr/Au, and 0.41 for NiFe/Au. It should be noted that all samples exhibit distinct steps in their hysteresis loops in the length direction.

In the width direction, the coercive field for CoNbZr is \qty{4.86(1)}{mT}, which is \qty{17}{\percent} higher than CoNbZr/Au (\qty{4.16(1)}{mT}) and \qty{16}{\percent} higher than NiFe/Au (\qty{4.20(1)}{mT}). The maximum permeability for CoNbZr is around 800, which is 2.75 times lower than CoNbZr/Au and 2 times lower than NiFe/Au. The remanent magnetisation to saturation magnetisation ratios are 0.61 for CoNbZr, 0.82 for CoNbZr/Au, and 0.37 for NiFe/Au.

\begin{table}[t!]
\centering
\caption{Summary of magnetic properties for CoNbZr, CoNbZr/Au, and NiFe/Au thin films. The coercive field ($B_\mathrm{c}$), remanent magnetisation ($M_\mathrm{r}$) to saturation magnetisation ($M_\mathrm{s}$) ratios, and maximum permeability ($\mu_\mathrm{max}$) are presented for both in-plane (length) and out-of-plane (width) configurations.}
\sisetup{
table-alignment-mode = format,
table-number-alignment = center
}
\begin{tabular} {@{\extracolsep{\fill}}
 l l 
 S[table-format = 1.2(1)] 
 S[table-format = 1.2(1)]
 S[table-format = 1.2(1)] 
 @{ }}
{Sample} & {Parameter} & {CoNbZr} & {CoNbZr/Au} & {NiFe/Au} \\
\hline
{} & {$B_\mathrm{c}$ (\qty{}{mT})} & 0.09(1) & 5.11(1) & 0.36(1) \\
{In-plane (Length)} & {$M_\mathrm{r}/M_\mathrm{s}$} & 0.17(1) & 0.80(1) & 0.41(1) \\ 
{} & {$\mu_\mathrm{max}$(\num{e3})} & 7.0(1) & 1.2(1) & 4.6(1) \\ 
\hline
{} & {$B_\mathrm{c}$ (\qty{}{mT})} & 4.86(1) & 4.16(1) & 4.20(1) \\
{Out-of-plane (Width)} & {$M_\mathrm{r}/M_\mathrm{s}$} & 0.61(1) & 0.82(1) & 0.37(1) \\ 
{} & {$\mu_\mathrm{max}$(\num{e3})} & 0.8(1) & 2.2(1) & 1.7(1) \\ 
\end{tabular}
\label{VSM_summary}
\end{table}

The hysteresis loops indicate that for single-layer CoNbZr and multilayer NiFe/Au, the magnetic easy axis lies in the length direction, as evidenced by significantly smaller coercive fields compared to the width direction. This effect is attributed to the lower demagnetising factor in the length direction (approximately 0.002 \cite{Aharoni1998}), which is about 10 times smaller than in the width direction. For multilayer CoNbZr/Au, however, no clear magnetic easy axis is observed, as the hysteresis loops for both directions appear similar.

\begin{table}[!b]
\centering
\caption{Comparison of calculated saturation magnetisations considering the total thickness of the sputtered films and the magnetic materials, respectively. The estimated saturation magnetisations $M_\mathrm{S,t_\mathrm{tot}}$ considering the total sputtered thickness $t_\mathrm{tot}$ reveal that CoNbZr/Au is \qty{16}{\percent} lower than CoNbZr, and NiFe/Au is \qty{21}{\percent} lower than the literature value for NiFe (\qty{0,86}{MA/m} \cite{Nahrwold2010}). Additionally, employing a dilution model (see Eq. \ref{Ms_Dil}) reveals agreement for CoNbZr, CoNbZr/Au and NiFe/Au.}
\sisetup{
table-alignment-mode = format,
table-number-alignment = center
}
\begin{tabular} {@{\extracolsep{\fill}}
 l l 
 S[table-format = 1.2(1)]
 S[table-format = 1.2(1)]
 S[table-format = 1.2(1)] 
 @{ }}
\multicolumn{2}{r}{} & {CoNbZr} & {CoNbZr/Au} & {NiFe/Au} \\
\hline
$M_\mathrm{s,t_\mathrm{tot}}$ &(\qty{}{MA/m}) & 1.01(5) & 0.85(4) & 0.68(3) \\
$M_\mathrm{s,t_\mathrm{FM,tot}}$ &(\qty{}{MA/m}) & 1.01(5) & 0.92(5) & 0.74(4)\\ 
$M_\mathrm{s}'$ & (\qty{}{MA/m}) & 1.01(5) & 0.93(12) & 0.79(6)\\ 
\end{tabular}
\label{Ms_all}
\end{table}

The calculated saturation magnetisations considering the total thickness of the sputtered films and the magnetic materials, respectively are listed in Tab. \ref{Ms_all}. The error results from the uncertainties of the film dimensions as well as the measurement of the magnetic moment $m$. The estimated saturation magnetisations $M_\mathrm{S,t_\mathrm{tot}}$ considering the total sputtered thickness $t_\mathrm{tot}$ reveal that CoNbZr/Au is \qty{16}{\percent} lower than CoNbZr, and NiFe/Au is \qty{21}{\percent} lower than the literature value for NiFe (\qty{0,86}{MA/m} \cite{Nahrwold2010}). Additionally, employing a dilution model (see Eq. \ref{Ms_Dil}) reveals agreement for CoNbZr, CoNbZr/Au and NiFe/Au.

\begin{figure}[t!]
\centering
\includegraphics[width=\textwidth]{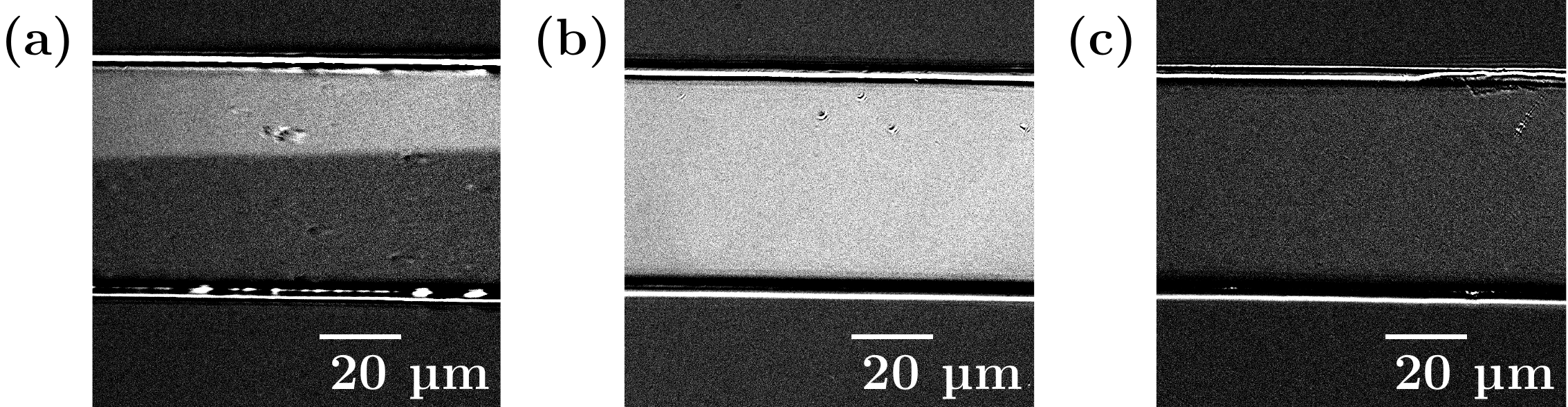} 
\caption{MOKE images of (a) single-layer CoNbZr, (b) multilayer CoNbZr/Au, and (c) multilayer NiFe/Au. While (a) exhibits a bipolar contrast, (b) and (c) display a unipolar contrast. The magnetic easy axis of the single-layer CoNbZr is oriented along the sample's length axis.}
\label{fig:MOKE}
\end{figure}

Figure \ref{fig:MOKE} presents the MOKE images of the patterned samples. In Fig. \ref{fig:MOKE}(a), the single-layer CoNbZr exhibits a bipolar contrast, indicating a magnetic domain structure with the easy axis oriented along the length of the sample. In contrast, Figs. \ref{fig:MOKE}(b) and \ref{fig:MOKE}(c) show a unipolar contrast, suggesting that the multilayered CoNbZr/Au and NiFe/Au films are in a single domain state. 

It should be noted that the MOKE technique is sensitive to the magnetisation of the surface and near-surface regions, as the penetration depth of light is limited to a few tens of nanometres \cite{Hubert2009}. Consequently, the observed contrast primarily reflects the magnetisation state of the topmost ferromagnetic layer. While the unipolar contrast observed in Figs. \ref{fig:MOKE}(b) and \ref{fig:MOKE}(c) suggests a single-domain state in the top layer, it does not exclude the possibility of different magnetisation states in deeper layers of the multilayer structures. This observation is consistent with Ref. \cite{Yokoyama2019}, which attributes the uniform domain state to magnetostatic coupling between the ferromagnetic layers in the multilayer structures.

\subsection{Magnetoimpedance properties} 

This section investigates the magnetoimpedance (MI) properties of the samples, examining the effects of key factors such as the applied magnetic field, frequency, annealing current, and lateral aspect ratio. The influence of these variables on the MI response is analysed to better understand the underlying mechanisms and to evaluate their impact on the material's overall behaviour.

\subsubsection{Influence of field and frequency}

\begin{figure}[t!]
\centering
\includegraphics[width=\textwidth]{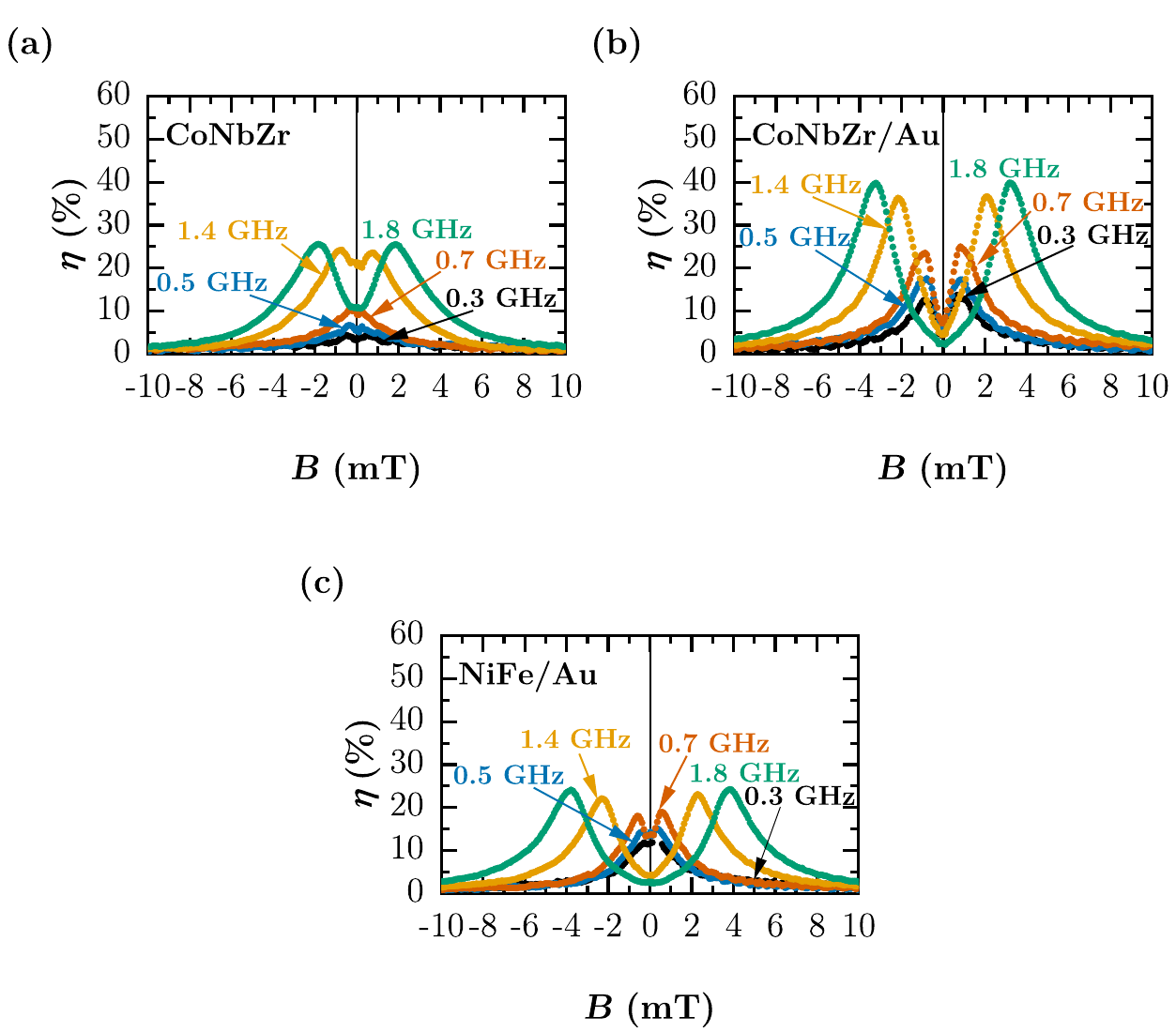} 
\caption{GMI curves of \qtyproduct{0.1x1}{mm} elements for (a) single-layer CoNbZr, (b) multilayer CoNbZr/Au, and (c) multilayer NiFe/Au, with a magnetic field sweep of $B = \pm \qty{10}{mT}$. The GMI ratio $\eta$ increases with frequency in all systems due to the rising impedance $Z$ (see Eq. \ref{Z_Film_approx}).
For single-layer CoNbZr, the response transitions from a single peak to minor double peaks above \qty{0.3}{GHz}, attributed to experimental misalignment, and broadens at frequencies exceeding \qty{1.4}{GHz} due to FMR. Multilayer CoNbZr/Au exhibits double peaks below \qty{0.7}{GHz}, transitioning to FMR above \qty{0.7}{GHz}. Compared to single-layer CoNbZr, it shows an enhanced GMI ratio and a reduced FMR frequency.
Multilayer NiFe/Au shows a single peak up to \qty{0.5}{GHz}, followed by splitting into double peaks in the FMR regime. Its GMI response in the FMR regime is similar to that of single-layer CoNbZr.}
\label{fig:GMI_comparison}
\end{figure}

The GMI ratios $\eta$ for \qtyproduct{0.1x1}{mm} elements as a function of frequency for single-layer CoNbZr, multilayer CoNbZr/Au, and multilayer NiFe/Au under a field sweep of $B = \pm$ \qty{10}{mT} are shown in Fig. \ref{fig:GMI_comparison}. Across all systems, the GMI ratio increases with frequency due to the rising impedance caused by the decreasing skin depth (Eqs. \ref{penetration depth} and \ref{Z_Film_approx}).

For single-layer CoNbZr (Fig. \ref{fig:GMI_comparison}(a)), the GMI response initially exhibits a single peak, consistent with the magnetic easy axis being aligned along the sample's length (Fig.\ref{fig:MOKE}(a)). However, minor double peaks emerge above \qty{0.3}{GHz}, potentially due to experimental misalignments. Beyond \qty{1.4}{GHz}, the broadening of the GMI curves and the shift of the double peaks to higher fields mark the onset of the FMR regime. This observation aligns with Endo et al. \cite{Endo2015}, who reported an FMR frequency of \qty{1.2}{GHz} for Co$_{85}$Nb$_{12}$Zr$_3$. For the Co$_{91}$-rich composition studied here, the FMR frequency is expected to be higher, approaching \qty{2.27}{GHz}, as reported for pure cobalt at zero field \cite{Bromer1978}.

The multilayer CoNbZr/Au system (Fig. \ref{fig:GMI_comparison}(b)) shows a more complex GMI response. A double-peak structure is observed below \qty{0.7}{GHz}, likely arising from the absence of a well-defined magnetic easy axis, as evidenced by the similar hysteresis loops in both directions (Fig \ref{fig:VSM_single_bar}). This lack of a clear anisotropy axis contributes to variability in the GMI behaviour, which may alternate between single and double peaks below the FMR frequency, depending on slight variations in the magnetisation configuration during measurement. Above \qty{0.7}{GHz}, the system transitions into the FMR regime, characterised by a broadening GMI response. The Au interlayers enhance the GMI ratio by approximately \qty{50}{\percent} and reduce the FMR frequency by nearly half compared to single-layer CoNbZr.

For the multilayer NiFe/Au system (Fig. \ref{fig:GMI_comparison}(c)), a single peak is observed up to \qty{0.5}{GHz}. At higher frequencies, this splits into a double peak due to FMR, consistent with previous studies on NiFe-based multilayers \cite{Barandiaran2006, DeCos2008}. The GMI response in the FMR regime closely resembles that of single-layer CoNbZr.

At \qty{1.8}{GHz}, the positions of the double peaks vary across systems: CoNbZr peaks at \qty{2}{mT}, CoNbZr/Au at \qty{3}{mT}, and NiFe/Au at \qty{4}{mT}. 

\begin{figure}[t!]
\centering
\includegraphics[width=\textwidth]{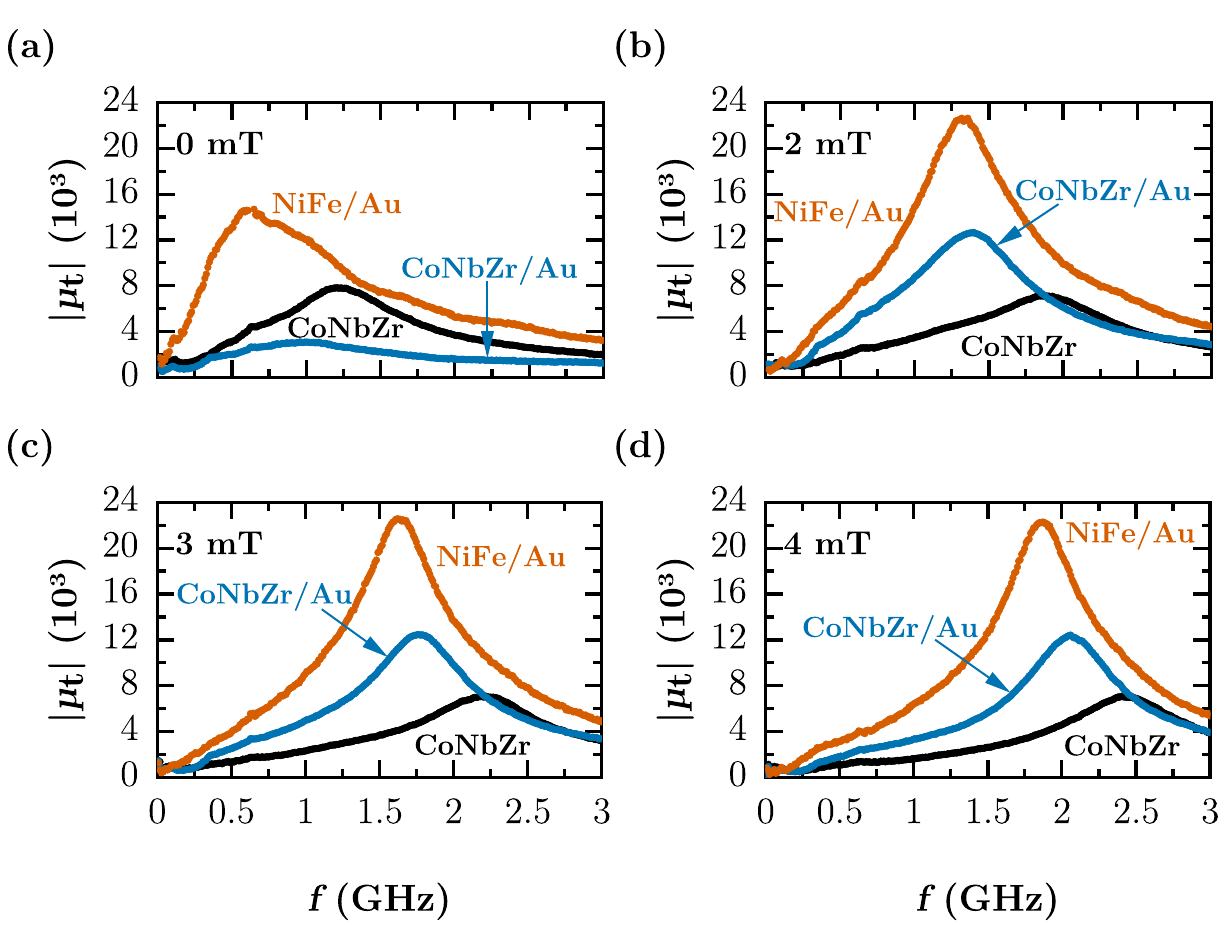} 
\caption {Calculated absolute value of the transverse permeability $|\mu_\mathrm{t}|$ of the sputtered \qtyproduct{0.1x1}{mm} elements as a function of frequency $f$ for different applied fields: (a) \qty{0}{mT}, (b) \qty{2}{mT}, (c) \qty{3}{mT}, and (d) \qty{4}{mT}. The calculations are based on the analytical formula from Eq. \ref{Z_Film}, assuming constant resistivities. At \qty{0}{mT}, CoNbZr exhibits a maximum permeability that is four times higher than CoNbZr/Au and \qty{50}{\percent} lower than NiFe/Au. As the field strength increases, the peak permeability shifts towards higher frequencies. At \qty{4}{mT}, the permeability of CoNbZr/Au rises to 12000, which is \qty{50}{\percent} higher than CoNbZr and approximately \qty{50}{\percent} lower than NiFe/Au, which reaches a maximum permeability of 22000.}
\label{fig:Permeability}
\end{figure}

Applying Eq. \ref{Z_Film} under the assumption of constant resistivities, the absolute value of the transverse permeability, $|\mu_\mathrm{t}|$, was calculated as a function of frequency $f$. The results for different applied fields are presented in Fig. \ref{fig:Permeability}. At \qty{0}{mT}, CoNbZr exhibits a maximum permeability four times higher than CoNbZr/Au and \qty{50}{\percent} lower than NiFe/Au, as shown in Fig. \ref{fig:Permeability}(a) . As the applied field increases, the peak permeability shifts to higher frequencies (see Figs. \ref{fig:Permeability}(b) - (d)). At \qty{4}{mT}, CoNbZr/Au reaches a permeability of 12000, approximately \qty{50}{\percent} higher than CoNbZr but still \qty{50}{\percent} lower than NiFe/Au, which peaks at 22000.

The enhancement of $|\mu_\mathrm{t}|$ in the multilayer CoNbZr compared to the single-layer system aligns with previous reports in Ref. \cite{Yokoyama2019} and can be attributed to the Au interlayer. This structure mitigates the pure skin effect and leverages the magneto-inductive effect, with the nonmagnetic Au spacer enhancing the GMI ratios at moderate frequencies \cite{Mohri1992, Hika1996, Morikawa1997, Garcia-Arribas2017}.

\begin{table}[!b]
\centering
\caption{Numerically calculated skin depth $\delta$ and transverse permeability $|\mu_\mathrm{t}|$ at \qty{1}{mT} from the GMI curves for CoNbZr and CoNbZr/Au at \qty{0.7}{GHz} and \qty{1.4}{GHz}.}
\begin{tabular} {@{\extracolsep{\fill}}
 		c S S S S 
 		@{}}
{System} & {$\delta$ (\qty{0.7}{GHz})} & {$\delta$ (\qty{1.4}{GHz})} & {$|\mu_\mathrm{t}|$ (\qty{0.7}{GHz})} & {$|\mu_\mathrm{t}|$ (\qty{1.4}{GHz})} \\
{} & {nm} & {nm} & {\num{e3}} & {\num{e3}} \\
\hline
{CoNbZr} & 145(1) & 101(1) & 3.6(1) & 7.2(1) \\
{CoNbZr/Au} & 102(1) & 82(1) & 8.8(1) & 7.7(1) \\
\end{tabular}
\label{skin_depth_1mT}
\end{table}

The reduced FMR frequency for CoNbZr/Au is related to its smaller skin depth. At \qty{0.7}{GHz}, the skin depth of CoNbZr/Au is \qty{102(1)}{nm}, significantly smaller than the \qty{145(1)}{nm} observed for single-layer CoNbZr, as shown in Tab. \ref{skin_depth_1mT}. This reduction in skin depth correlates with an enhanced transverse permeability, $|\mu_\mathrm{t}|$, of \num{8.8(1)e3} for CoNbZr/Au, compared to \num{3.6(1)e3} for CoNbZr.

At \qty{1.4}{GHz}, the skin depth for CoNbZr decreases to \qty{101(1)}{nm}, while for CoNbZr/Au, it further reduces to \qty{82(1)}{nm}. At this frequency, the transverse permeabilities of CoNbZr and CoNbZr/Au become comparable (\num{7.2(1)e3} and \num{7.7(1)e3}, respectively).

These findings suggest that the Au interlayer not only reduces the skin depth but also enhances magnetic permeability at lower frequencies, contributing to the improved GMI response of CoNbZr/Au. With CoNbZr approaching a skin depth of \qty{101(1)}{nm} at \qty{1.4}{GHz} and CoNbZr/Au reaching \qty{102(1)}{nm} at \qty{0.7}{GHz}, both systems approach the FMR regime, as observed in Figs. \ref{fig:GMI_comparison}(a) and \ref{fig:GMI_comparison}(b).

\begin{table}[t!]
\centering
\caption{Maximum GMI ratio, the magnetic field at which this maximum is reached, the slope of the GMI curve, transverse permeability, and impedance derived from the GMI curves (see Fig. \ref{fig:GMI_comparison}) at a fixed frequency of \qty{1.8}{GHz}.}
\begin{tabular} {@{\extracolsep{\fill}}
 l 
 S[table-format = 2.0(1.0)]
 S[table-format = 1.1(1.1)]
 S[table-format = 2.0(1.0)]
 S[table-format = 2.1(1.1)]
 S[table-format = 1.1(1.1)] 
 @{ }}
{System} & {$\eta_\mathrm{max}$} & {$B(\eta_\mathrm{max})$} & {$s_\mathrm{max}$} & {$|\mu_\mathrm{t}|$} & {$\sqrt{\pi f \rho \mu}$} \\
{} & {\%} & {mT} & {\%/mT} & {\num{e3}} & {$\Omega$} \\
\hline
{CoNbZr} & 26(1) & 2.0(1) & 14(1) & 7.0(1) & 7.7(1) \\
{CoNbZr/Au} & 40(1) & 3.0(1) & 26(1) & 12.4(1) & 7.9(1) \\ 
{NiFe/Au} & 24(1) & 4.0(1) & 15(1) & 21.5(1) & 6.3(1) \\ 
\end{tabular}
\label{GMI_values_0.1x1}
\end{table}

Table \ref{GMI_values_0.1x1} summarises the values derived from the GMI curves at \qty{1.8}{GHz}. Among the systems studied, CoNbZr/Au demonstrates the highest maximum GMI ratio (\qty{40}{\percent}) at \qty{3.0(1)}{mT} and the highest sensitivity (\qty{26(1)}{\percent/mT}), even though its transverse permeability is only half that of NiFe/Au. According to Eq.\ref{Z_Film_approx}, at a fixed frequency, the impedance is determined by the product of resistivity and transverse permeability. The superior GMI ratio of CoNbZr/Au can be attributed primarily to its significantly higher resistivity, which is approximately three times greater than that of NiFe/Au, as shown in Tab. \ref{Rho}.

Comparing CoNbZr/Au with single-layer CoNbZr, despite the reduced resistivity of CoNbZr/Au relative to single-layer CoNbZr (1.7 times lower), the enhanced transverse permeability of CoNbZr/Au (1.77 times higher) plays a crucial role in achieving an improved GMI ratio. On the other hand, comparing single-layer CoNbZr with NiFe/Au, although CoNbZr exhibits a transverse permeability that is three times lower than that of NiFe/Au, its 4.5-fold higher resistivity compensates effectively, resulting in a higher GMI ratio.

\subsubsection{Influence of annealing current}

\begin{figure}[t!]
\centering
\includegraphics[width=\textwidth]{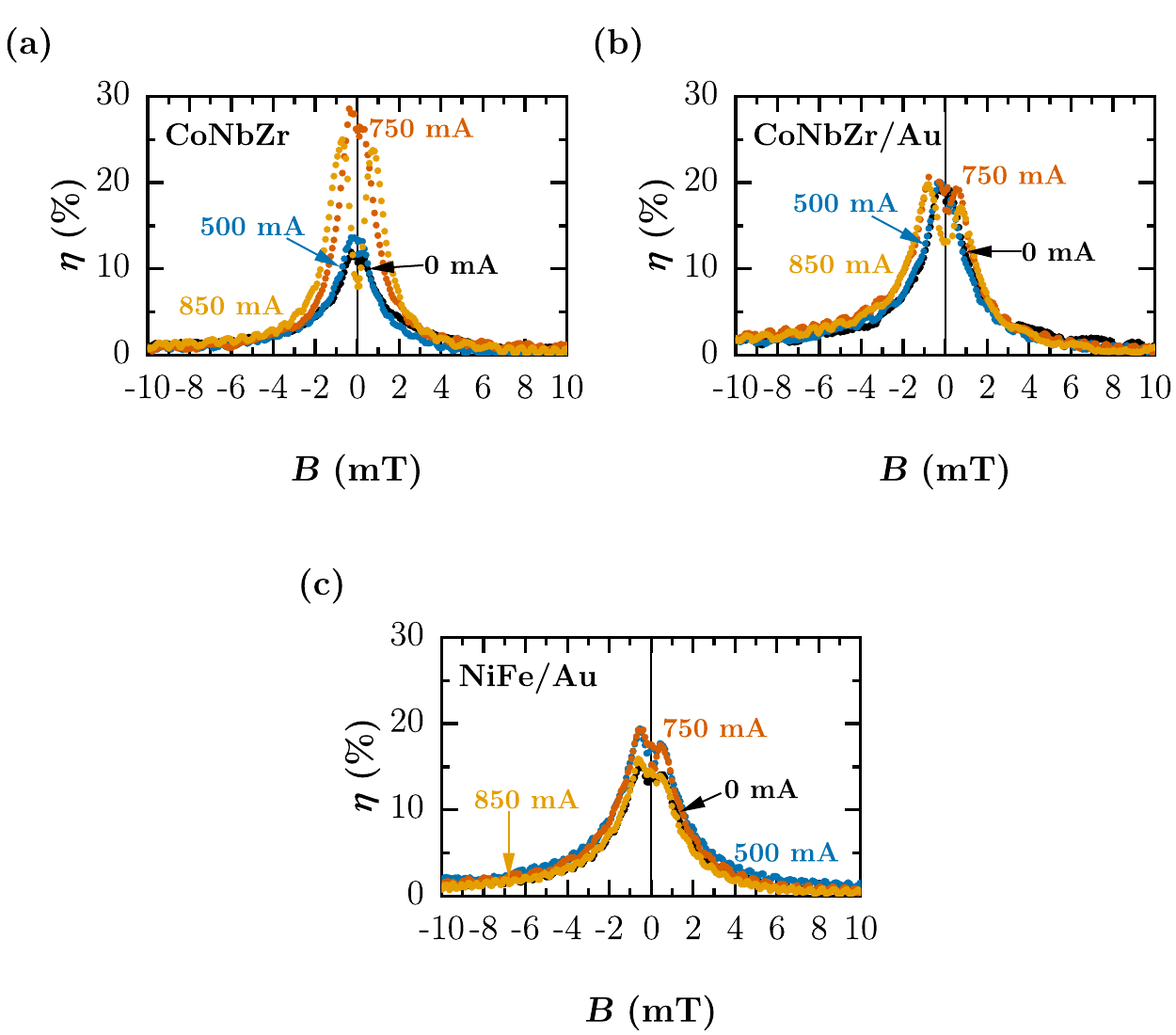} 
\caption{GMI curves for \qtyproduct{0.1x1}{mm} structures measured over a magnetic field range of $B=\pm$ \qty{10}{mT} as a function of annealing current at \qty{0.7}{GHz}. The samples were annealed at currents of \qty{0}{mA}, \qty{500}{mA}, \qty{750}{mA}, and \qty{850}{mA} for the following systems: (a) single-layer CoNbZr, (b) multilayer CoNbZr/Au, and (c) multilayer NiFe/Au. For CoNbZr, the GMI ratio $\eta$ increases from \qty{10}{\percent} at \qty{0}{mA} to nearly \qty{30}{\percent} at \qty{750}{mA}, with a pronounced double peak emerging at \qty{850}{mA}. In the case of CoNbZr/Au, the GMI ratio remains stable across all annealing currents, while the double peak becomes more pronounced at \qty{850}{mA}. For NiFe/Au, a double-peak curve is observed due to ferromagnetic resonance, with the GMI ratio varying from \qty{15}{\percent} at \qty{0}{mA} and \qty{850}{mA} to \qty{20}{\percent} at \qty{500}{mA} and \qty{750}{mA}.}
\label{fig:GMI_curr}
\end{figure}

Figure \ref{fig:GMI_curr} shows the GMI curves of \qtyproduct{0.1x1}{mm} structures measured over a magnetic field range of $B = \pm$ \qty{10}{mT} at \qty{0.7}{GHz}, as a function of annealing current $I$, for single-layer CoNbZr, multilayer CoNbZr/Au, and multilayer NiFe/Au. These measurements were conducted post-annealing.

For single-layer CoNbZr (Fig. \ref{fig:GMI_curr}(a)), the GMI ratio increases significantly, from \qty{10}{\percent} at $I$ = \qty{0}{mA} to approximately \qty{30}{\percent} at \textit{I} = \qty{750}{mA}. At \textit{I} = \qty{850}{mA}, a distinct double peak emerges in the GMI curve. Conversely, CoNbZr/Au (Fig. \ref{fig:GMI_curr}(b)) exhibits a constant GMI ratio across all currents, with a more pronounced double-peak feature at \textit{I} = \qty{850}{mA}. For NiFe/Au (Fig. \ref{fig:GMI_curr}(c)), the GMI ratio varies modestly, increasing from \qty{15}{\percent} at $I$ = \qty{0}{mA} and \textit{I} = \qty{850}{mA} to \qty{20}{\percent} at \textit{I} = \qty{500}{mA} and \textit{I} = \qty{750}{mA}. The double-peak feature observed in NiFe/Au is attributed to FMR.

%The current distribution within the multilayer structures explains these differences. Due to the parallel resistor effect, only a fraction of the total current flows through the ferromagnetic (FM) layers. For CoNbZr/Au, \qty{9}{\percent} of the current flows through the FM layer, compared to \qty{29}{\percent} for NiFe/Au, owing to the resistivity differences between the FM and Au layers (see Tab. \ref{Rho}).

%In CoNbZr/Au, the low current fraction is insufficient to induce domain structure changes, even at the maximum annealing current of \textit{I} = \qty{850}{mA}. Consequently, the GMI ratio remains constant. For NiFe/Au, the higher current fraction permits slight domain modifications, leading to a modest increase in the GMI ratio.

%Achieving domain changes comparable to single-layer CoNbZr would require an FM layer current of $I_\mathrm{FM} \approx$ \qty{0.5}{A} (see Fig. \ref{fig:CoNbZr_MOKE_500mA}). This corresponds to total currents of $ I \approx$ \qty{5.56}{A} for CoNbZr/Au and $ I \approx$ \qty{1.72}{A} for NiFe/Au. However, these currents exceed the capabilities of the Source Measure Unit used in this work and approach the breakdown voltage of the RF probe, posing risks to the experimental setup.

\begin{figure}[t!]
\centering
\includegraphics[width=\textwidth]{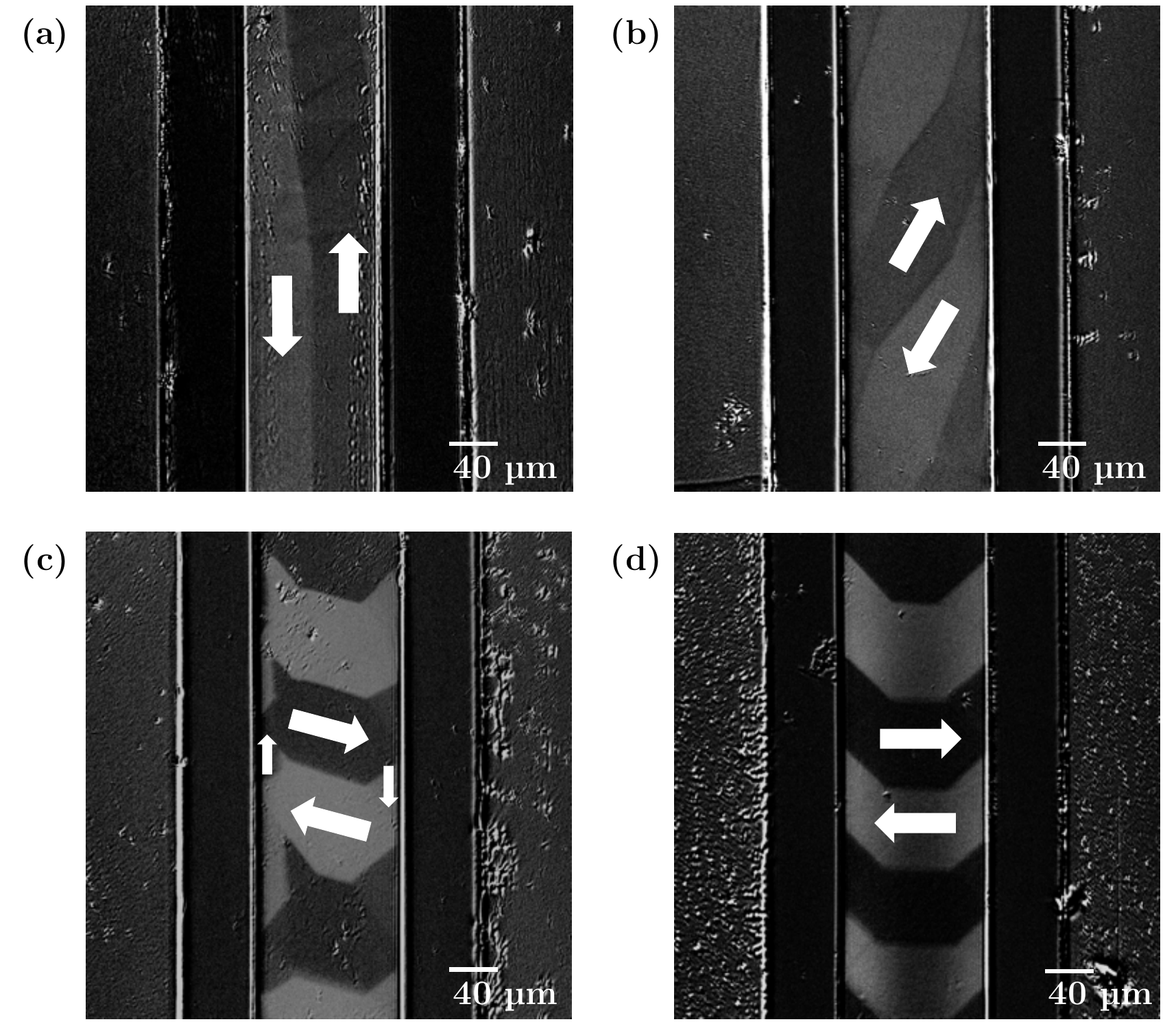} 
\caption{MOKE images of \qtyproduct{0.1x1}{mm} single-layer CoNbZr structures at (a) \qty{0}{mA}, (b) \qty{500}{mA}, (c) \qty{750}{mA}, and (d) \qty{850}{mA}. At \qty{0}{mA}, the magnetic domains are aligned along the length of the structure, as indicated by the bipolar contrast and the white arrows. At \qty{500}{mA}, the domains are rotated by approximately \qty{60}{\degree} relative to the width axis. At \qty{750}{mA}, the domains in the centre are rotated by about \qty{15}{\degree} relative to the width axis, while the edges remain parallel to the length direction. At \qty{850}{mA}, the domains in the centre are aligned parallel to the width axis, forming a horseshoe configuration. Also, a black band in the centre is visible.}
\label{fig:CoNbZr_MOKE_curr}
\end{figure}

To investigate the reasons behind the different GMI curves for CoNbZr, MOKE images of the \qtyproduct{0.1x1}{mm} structures were captured after annealing at various current levels, as illustrated in Fig. \ref{fig:CoNbZr_MOKE_curr}. At \qty{0}{mA} (see Fig. \ref{fig:CoNbZr_MOKE_curr}(a)), the domains are aligned along the longitudinal direction of the structure, as indicated by the bipolar contrast and the white arrows. This alignment explains the maximum GMI ratio observed at \qty{0}{mT} in Fig. \ref{fig:GMI_curr}(a).

At \qty{500}{mA}, the domains rotate by approximately \qty{60}{\degree} relative to the width axis (Fig. \ref{fig:CoNbZr_MOKE_curr}(b)), resulting in an increase in transverse permeability and, consequently, impedance (see Eq. \ref{Z_Film_approx}).

In Fig. \ref{fig:CoNbZr_MOKE_curr}(c), four distinct grayscale levels are visible, indicating four different magnetisation directions within the domains, as indicated by the white arrows. The magnetisation in the centre of the structure is rotated by about \qty{15}{\degree} relative to the width axis, while the edges remain parallel to the length direction. 

At \qty{850}{mA} (Fig. \ref{fig:CoNbZr_MOKE_curr}(d)), the domain structure exhibits a horseshoe-like configuration with two distinct grayscale levels and a prominent black band in the centre. These features differ significantly from the patterns observed at lower currents.

These variations in domain structure directly influence the stray field. At \qty{750}{mA}, the magnetic flux forms closed loops due to the magnetisation being rotated by about \qty{15}{\degree} relative to the width direction in the brightest and darkest areas of the grayscale, while it remains parallel in the intermediate contrast regions, effectively minimising the stray field. In contrast, the predominantly perpendicular magnetisation at \qty{850}{mA} gives rise to open magnetic flux at the ends of the domains, thereby generating stray fields.

The changes in domain structure at \qty{750}{mA} and \qty{850}{mA} are reflected in the GMI curves presented in Fig. \ref{fig:GMI_curr}(a). At \qty{750}{mA}, a double peak begins to emerge, as the maxima no longer occur solely at \qty{0}{mT}. By \qty{850}{mA}, the double peak is fully pronounced, a feature attributed to the transverse alignment of the easy magnetic axis. However, the presence of open magnetic flux in the structure results in a lower GMI ratio at \qty{850}{mA} compared to \qty{750}{mA}.

\subsubsection{Influence of lateral aspect ratio}

\begin{figure}[t!]
\centering
\includegraphics[width=\textwidth]{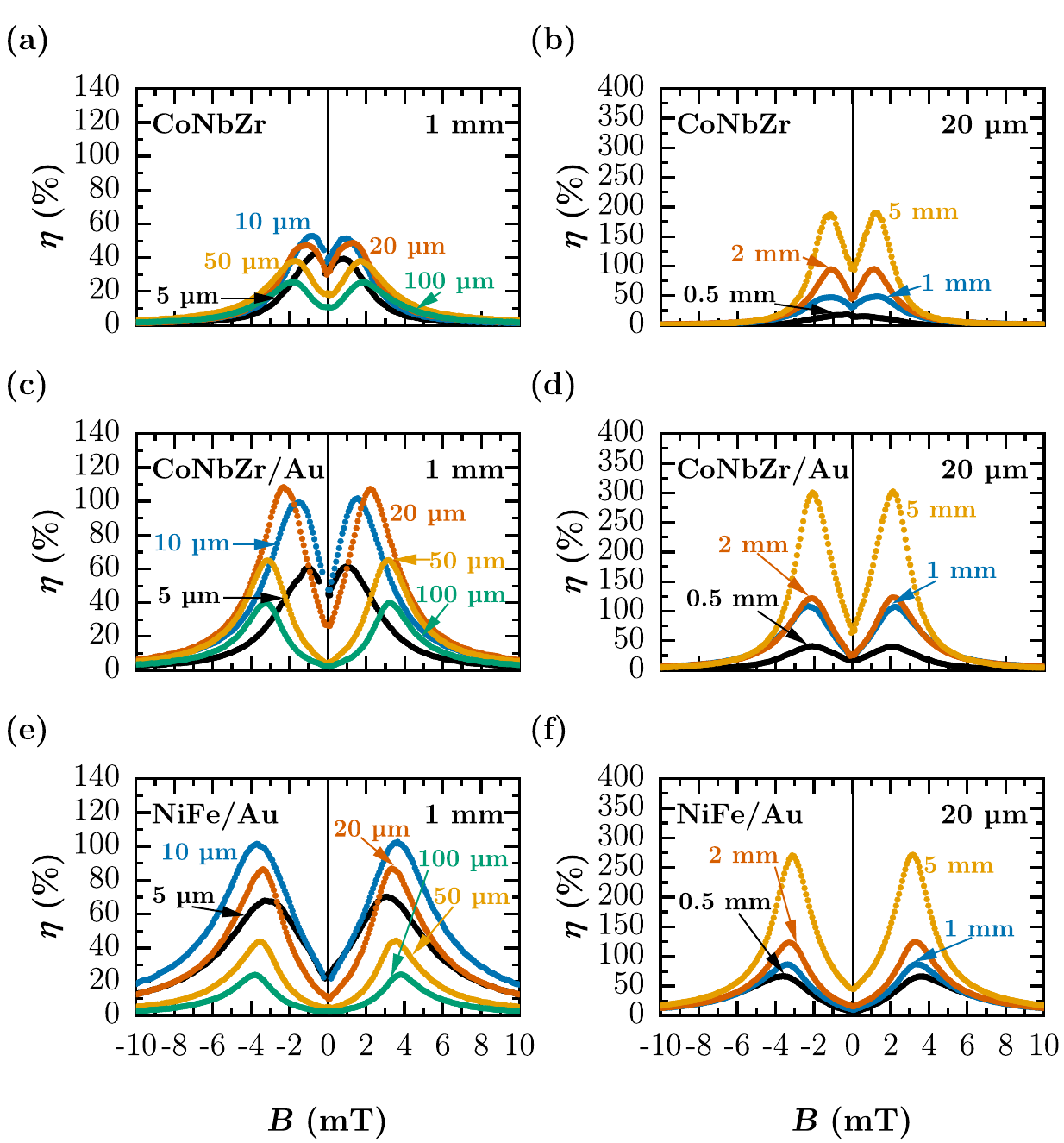} 
\caption{GMI curves measured as a function of aspect ratio over a magnetic field sweep of $B = \pm$ \qty{10}{mT} at \qty{1.8}{GHz} for (a,b) single-layer CoNbZr, (c,d) multilayer CoNbZr/Au, and (e,f) multilayer NiFe/Au. Maximum GMI ratios $\eta$ occur for widths of \qty{10}{\micro m} to \qty{20}{\micro m} at a fixed length of \qty{1}{mm}, with the field at peak maxima decreasing as width decreases. The Au interlayer enhances $\eta$ in CoNbZr/Au by \qty{50}{\percent}, while NiFe/Au exhibits slightly lower values. Increasing the length from \qty{0.5}{mm} to \qty{5}{mm} (at \qty{20}{\micro m} width) significantly enhances $\eta$ across all systems: by factors of 5 (\qty{180(1)}{\percent}) for CoNbZr, 7.5 (\qty{300(1)}{\percent}) for CoNbZr/Au, and 4 (\qty{280(1)}{\percent}) for NiFe/Au. For all systems, the peak position decreases slightly with increasing length, with NiFe/Au requiring a higher applied field (\qty{4}{mT}) to reach its maximum compared to CoNbZr/Au.}
\label{fig:GMI_AspectRatio}
\end{figure}

Figure \ref{fig:GMI_AspectRatio} presents the GMI response as a function of aspect ratio for three systems: (a,b) single-layer CoNbZr, (c,d) multilayer CoNbZr/Au, and (e,f) multilayer NiFe/Au. Measurements were performed at \qty{1.8}{GHz} over a magnetic field range of $B = \pm$ \qty{10}{mT}. Panels (a,c,e) show how the GMI ratio $\eta$ varies with width for elements of fixed length (\qty{1}{mm}), while panels (b,d,f) illustrate the length dependence for elements of fixed width (\qty{20}{\micro m}).

For the width-dependent measurements (a,c,e), the maximum GMI ratio is observed across all systems for widths between \qty{10}{\micro m} and \qty{20}{\micro m}. As the width decreases, the field corresponding to the double peak maxima shifts to lower values across all systems. The incorporation of an Au interlayer in the multilayer CoNbZr/Au system enhances the GMI ratio by \qty{50}{\percent} compared to single-layer CoNbZr. In contrast, the multilayer NiFe/Au system exhibits a slightly lower GMI ratio than CoNbZr/Au.

In panels (b,d,f), increasing the element length from \qty{0.5}{mm} to \qty{5}{mm} significantly enhances the GMI ratio across all systems. Single-layer CoNbZr (b) shows a fivefold increase, reaching a maximum value of \qty{180(1)}{\percent}. The multilayer CoNbZr/Au system (d) exhibits an even greater enhancement, with the GMI ratio increasing by a factor of 7.5 to \qty{300(1)}{\percent}. Similarly, the multilayer NiFe/Au system (f) exhibits a fourfold increase, reaching \qty{280(1)}{\percent}. The maximum GMI ratio for NiFe/Au occurs at an applied field of approximately \qty{4}{mT}, which is twice the field required for CoNbZr/Au. For all systems, increasing the element length causes a slight shift in the GMI peak position.

\begin{table}[!b]
\centering
\caption{Maximum sensitivity obtained from the GMI curves as a function of element width at \qty{1}{mm} width for a fixed frequency at \qty{1.8}{GHz}.}
\begin{tabular} {@{\extracolsep{\fill}}
r@{:}l 
S[table-format = 3.0(1.0)] 
S[table-format = 3.0(1.0)] 
S[table-format = 3.0(1.0)] 
 @{ }}
\multicolumn{2}{c}{Sensitivity} & {CoNbZr} & {CoNbZr/Au} & {NiFe/Au} \\
\multicolumn{2}{c}{(\%/mT)} \\
\hline
1000&100 & 14(1) & 26(1) & 15(1) \\
1000&50 & 21(1) & 37(1) & 24(1) \\ 
500&20 & 21(1) & 18(1) & 31(1) \\ 
1000&20 & 31(1) & 56(1) & 43(1) \\
2000&20 & 61(1) & 68(1) & 68(1) \\ 
1000&10 & 25(1) & 54(1) & 33(1) \\
1000&5 & 19(1) & 22(1) &22(1) \\ 
5000&20 & 135(1) & 199(1) & 146(1) \\ 
\end{tabular}
\label{sensitivity_values}
\end{table}

Table \ref{sensitivity_values} presents the maximum sensitivity values obtained from the GMI curves as a function of element width and length at a fixed frequency of \qty{1.8}{GHz}. For a fixed length of \qty{1}{mm}, the highest sensitivity is observed at a width of \qty{20}{\micro m}, reaching up to \qty{61(1)}{\percent/mT} for single-layer CoNbZr, \qty{68(1)}{\percent/mT} for multilayer CoNbZr/Au, and \qty{68(1)}{\percent/mT} for multilayer NiFe/Au. The incorporation of the Au interlayer enhances the sensitivity of multilayer CoNbZr/Au compared to single-layer CoNbZr.

As the element length increases while maintaining a fixed width of \qty{20}{\micro m}, the sensitivity exhibits further enhancements. Single-layer CoNbZr reaches a maximum sensitivity of \qty{135(1)}{\percent/mT}, an increase by a factor of 6.5. Similarly, multilayer CoNbZr/Au shows the highest enhancement, with sensitivity increasing by a factor of 11 to \qty{199(1)}{\percent/mT}. In comparison, multilayer NiFe/Au achieves a sensitivity increase by a factor of 5, reaching up to \qty{146(1)}{\percent/mT}.

\subsection{Discussion}

This section presents a detailed discussion of the key findings from the experiments, focusing on the magnetic anisotropy and reversal mechanisms, resonance field variations, thermal effects, magnetic domain reconfiguration, and the influence of lateral aspect ratio. Each of these factors is examined in relation to the observed behaviours and their impact on the overall properties of the materials under study.

\subsubsection{Magnetic anisotropy and reversal mechanisms}

For the multilayer CoNbZr/Au in Fig. \ref{fig:VSM_single_bar}, no distinct magnetic easy axis is observed, as the hysteresis loops for both the length and width directions appear similar. This behaviour is likely attributed to the additional interface anisotropy introduced at the CoNbZr/Au interfaces during deposition. The TEM images (Figs. \ref{fig:TEM_SAED_comparison}(d) - (f)) reveal well-defined, sharp interfaces between the amorphous CoNbZr and crystalline Au layers, suggesting minimal interdiffusion. The SAED patterns confirm the amorphous nature of CoNbZr and the crystalline nature of Au, with the latter showing diffraction spots corresponding to the (111), (222), and (333) planes. This suggests that Au grows as multiples of the (111) direction, resembling a single-crystalline structure within the multilayer system.

The pronounced interface anisotropy in CoNbZr/Au likely arises from the structural disparity between the amorphous CoNbZr and the crystalline Au layers. The sharp interfaces, combined with the alignment of Au along the (111) direction, may enhance interfacial coupling, which in turn influences the magnetic energy landscape. This effect competes with shape anisotropy, reorienting the magnetic easy axis and leading to the observed isotropy in the hysteresis loops.

In contrast, the NiFe/Au system exhibits distinct alternating layers of crystalline NiFe and Au, as shown in the TEM images (Figs. \ref{fig:TEM_SAED_comparison}(g) - (i)). The SAED patterns indicate that both materials retain their crystalline nature, with NiFe showing grain-like features. The interface anisotropy in this system is likely less pronounced due to the similar crystalline nature of the layers, resulting in a magnetic easy axis predominantly governed by shape anisotropy.

The step observed in the length-direction hysteresis loop of CoNbZr/Au and NiFe/Au in Fig. \ref{fig:VSM_single_bar}(a) suggests that at least one ferromagnetic layer reverses its magnetisation at a different field than the others. This behaviour can be attributed to interlayer interactions or a distribution in magnetic properties among the layers, potentially arising from slight variations in layer thickness or deposition conditions, as indicated by the TEM cross-sections. In the CoNbZr/Au system, the sharp interfaces and well-controlled sputtering process suggest minimal interdiffusion. Local variations in interface anisotropy, combined with inhomogeneities induced by the patterning process, such as edge roughness and sidewall deposition on the resist, may influence the reversal fields of individual layers. Similarly, in NiFe/Au, the granular structure of NiFe, as revealed in the TEM analysis, implies that grains with different sizes or orientations may reverse magnetisation at different fields, further contributing to the step-like behaviour.

A step is also present in the hysteresis loop of single-layer CoNbZr films. This can be attributed to inhomogeneities within the single layer, such as variations in magnetic properties or localised stress-induced anisotropies, as well as the same patterning-induced effects mentioned above. These inhomogeneities are consistent with the amorphous structure of CoNbZr observed in the TEM and SAED results, which may lead to regions within the film reversing magnetisation at different fields. Additionally, magnetoelastic effects arising from intrinsic stresses during sputtering could further contribute to the non-uniform reversal behaviour observed in the hysteresis loop.

\subsubsection{Resonance field variations}

The GMI curves in Fig. \ref{fig:GMI_comparison} at \qty{1.8}{GHz} show double-peak positions at \qty{2}{mT} for CoNbZr, \qty{3}{mT} for CoNbZr/Au, and \qty{4}{mT} for NiFe/Au. These differences are primarily attributed to variations in shape anisotropy $K_\mathrm{s}$ (Eq. \ref{Ks}), which is governed by the saturation magnetisation $M_\mathrm{s}$ for systems with constant geometry. CoNbZr, with the highest $M_\mathrm{s}$, aligns its magnetic easy axis more strongly with the sample length, requiring a lower field for resonance compared to NiFe/Au, which has the lowest $M_\mathrm{s}$ (see Tab. \ref{Ms_all}).

In the FMR regime, additional factors come into play. Interfacial exchange coupling in the multilayers (CoNbZr/Au and NiFe/Au) modifies the effective field by contributing an additional anisotropy term. The TEM images (Figs. \ref{fig:TEM_SAED_comparison}(d) - (i)) reveal sharp interfaces for CoNbZr/Au and granular structures in NiFe/Au, suggesting that interfacial strain or roughness may introduce localised anisotropies. Furthermore, intrinsic properties such as damping and gyromagnetic ratio differences among the systems may also shift the resonance field.

These combined effects \textendash{} shape anisotropy, interfacial coupling, and intrinsic material properties \textendash{} offer a comprehensive explanation for the observed resonance field variations. Future studies could explore the quantitative contribution of each factor using micromagnetic simulations or additional experimental techniques.

\subsubsection{Thermal effects and magnetic domain reconfiguration}

The pronounced double peak observed in the GMI curves of CoNbZr/Au at \qty{850}{mA} and \qty{0.7}{GHz} (Fig. \ref{fig:GMI_curr}(b)) can primarily be attributed to current-induced heating effects. This heating reduces the intrinsic stresses within the CoNbZr layer and the interfacial stresses between Au and CoNbZr. These thermal effects promote the relaxation of internal stresses, facilitating a more favourable alignment of magnetic domains, which enhances the resonance conditions.

Similarly, for NiFe/Au (Fig. \ref{fig:GMI_curr}(c)), the increase in the GMI ratio is also attributed to current-induced heating, which alleviates intrinsic and interfacial stresses, resulting in a magnetically softer structure. Additionally, NiFe/Au exhibits a higher transverse permeability, which further contributes to the increase in the GMI ratio. However, the drop in the GMI ratio at \qty{850}{mA} suggests that, beyond a certain threshold current, the transverse permeability may begin to decrease due to excessive heating or related effects.

The horseshoe-like domain structure observed at \qty{850}{mA} (Fig. \ref{fig:CoNbZr_MOKE_curr}(d)) could indicate a reconfiguration of magnetic domains under the influence of localised magnetic anisotropies or strain effects caused by annealing. The black band in the centre is particularly striking and may represent a region of reduced magnetisation or magnetic flux closure. However, these interpretations remain speculative and require further validation through micromagnetic simulations or additional experimental techniques. Future studies could investigate whether domain reconfiguration is influenced by edge effects or variations in the current distribution during annealing.

\subsubsection{Lateral aspect ratio}

The observed shift in the GMI peak position (see Fig. \ref{fig:GMI_AspectRatio}) for a fixed geometry is primarily attributed to variations in the saturation magnetisation $M_\mathrm{s}$, which directly affect the shape anisotropy $K_\mathrm{s}$ (Eq. \ref{Ks}), as previously discussed. Additionally, geometric factors such as the lateral aspect ratio influence the effective demagnetisation fields and must be considered for a comprehensive understanding. These effects are further modulated by interfacial exchange coupling, intrinsic material properties, and local anisotropies introduced by interfacial strain or roughness, as revealed in TEM images (Figs. \ref{fig:TEM_SAED_comparison}(d) - (i)).

Table \ref{demag_factors} shows that, irrespective of whether the sample is single-layer or multilayer, the demagnetisation field $H_{\mathrm{d},x}$ increases as the width decreases. This elevated field aligns the magnetic easy axis more strongly along the length direction, thereby enhancing microwave absorption at resonance and leading to a higher GMI ratio.

Conversely, the observed decline in the GMI ratio at smaller widths ($<$ \qty{10}{\micro m}) is attributed to edge effects, which become prominent in this regime. These effects stem from the distortion of magnetic domain structures near the sample edges, where stray fields are stronger. The resulting non-uniform magnetisation disrupts the resonance condition required for efficient microwave absorption, thereby reducing the maximum GMI ratio despite the increased demagnetising field. Similar observations on the significant role of edge effects in shaping GMI behaviour in thin films have been reported in a previous study \cite{Garcia-Arribas2008}.

In contrast, increasing the sample length improves GMI performance through two primary mechanisms, regardless of the sample being single-layer or multilayer. First, the reduction in $H_{\mathrm{d},x}$ (as shown in Tab. \ref{demag_factors}) facilitates easier resonance alignment along the length direction. Second, the increased sample volume allows a greater number of magnetic moments to oscillate. Consequently, a lower external field is required to achieve the resonance condition as the length increases, enhancing microwave absorption and resulting in improved GMI performance.

\section{Conclusion}

Our motivation to study CoNbZr was driven by its ability to remain amorphous even at high film thicknesses, a property that makes it highly suitable for applications in high-sensitivity magnetic sensors. Indeed, microstructural analysis confirms that CoNbZr films remain amorphous even in multilayer CoNbZr/Au systems, as verified by TEM diffraction patterns, in contrast to the crystalline structure of NiFe/Au films.

When sputtering CoNbZr, one should be aware that the target elemental composition may not transfer directly to the deposited thin film. EDX analyses reveal a significant increase in the Co/Nb ratio from 7.1 in the target to 13 in the film, while the Ni/Fe ratio in NiFe films remains consistent with the target.

Of primary interest for applications are the electrical properties of the films, which demonstrate notable differences. The resistivity of CoNbZr/Au multilayers is approximately three times higher than that of NiFe/Au films, while being half that of single-layer CoNbZr films. These differences highlight the impact of Au interlayers in modifying electrical characteristics.

Magnetic property analysis reveals further distinctions. In-plane hysteresis loops indicate saturation magnetisation values of \qty{1.01}{MA/m} for single-layer CoNbZr, \qty{0.85}{MA/m} for CoNbZr/Au, and \qty{0.68}{MA/m} for NiFe/Au. While CoNbZr/Au and NiFe/Au multilayers exhibit single-domain structures, single-layer CoNbZr exhibits a multi-domain structure.

The simple dilution model provides a useful framework for understanding the resistivity and saturation magnetisation results. While the model predicts resistivity accurately for NiFe/Au films, CoNbZr/Au films show resistivity values \qty{55}{\percent} higher than predicted. In terms of saturation magnetisation, all systems agree well with the model, validating its applicability.

The magnetoimpedance (GMI) properties of the films further emphasise the advantages of CoNbZr/Au for high-sensitivity applications. GMI measurements reveal ferromagnetic resonance (FMR) frequencies of \qty{1.4}{GHz} for CoNbZr, \qty{0.7}{GHz} for CoNbZr/Au, and \qty{0.5}{GHz} for NiFe/Au. Au interlayers enhance the GMI ratio by \qty{50}{\percent} and lower the FMR frequency by \qty{50}{\percent} compared to single-layer CoNbZr. GMI ratios increase significantly with element length, reaching maximum values at element widths of \qty{10}{\micro m} to \qty{20}{\micro m}. The highest GMI ratio of \qty{300}{\percent} is achieved in CoNbZr/Au at \qty{1.8}{GHz} and \qty{2}{mT} for a \qtyproduct{20x5000}{\micro m} strip, compared to \qty{280}{\percent} for NiFe/Au at \qty{4}{mT}.

These results highlight the significant potential of CoNbZr/Au multilayers in advancing magnetic sensor technology. Future work should focus on optimising magnetoimpedance properties through variation of interlayer and CoNbZr layer thicknesses, as well as exploring alternative interlayer materials such as Ti or Cu. Sputtering under an applied perpendicular magnetic field to enhance transverse permeability offers another promising direction, particularly in the skin-effect regime. However, in the FMR regime, care must be taken to avoid introducing additional anisotropy, as this could adversely affect the desired resonance behaviour.

Additionally, incorporating a thicker conductive non-ferromagnetic (NFM) layer within the multilayer stack, while retaining the thin interlayers, provides a promising strategy for enhancing sensor sensitivity. This configuration, which leverages the improved conductivity and current distribution from the thicker NFM layer, could enable higher GMI performance in next-generation magnetic sensors. Furthermore, the observed increase in the GMI ratio of CoNbZr with current annealing demonstrates its potential for further enhancing sensor performance, particularly in structures with high lateral aspect ratios. This scalability underscores the viability of current annealing and advanced multilayer designs for optimising CoNbZr-based magnetic sensors, paving the way for their integration into advanced sensing technologies.

\section*{Acknowledgements}

The project was conducted at Saarland University as part of the BMBF-funded collaborative research initiative "ForMikro-spinGMI".

The authors thank Christoph Pauly and Christian Sch\"{a}fer from the Chair of Functional Materials for preparing the TEM lamellae by FIB milling.  
We further thank J\"{o}rg Schmauch from INM Saarbrücken for his support in collecting the TEM, SEM, and EDX data presented in this paper.  
Additional thanks go to Bj\"{o}rn Heinz and Philipp Schwenke (RPTU Kaiserslautern, Germany) for performing the VSM measurements, to Carsten Brill from KIST Europe for his assistance with film deposition, and to Gregor B\"{u}ttel for his valuable contributions throughout the project.

\section*{Declaration of generative AI and AI-assisted technologies in the writing process}

During the preparation of this work, the authors used ChatGPT to improve the readability, language, grammar, spelling, and style of the manuscript. After using this tool, the authors reviewed and edited the content as needed and take full responsibility for the content of the publication.

% If all references are compiled correctly, comment \bibliography and
% outcomment \input{Paper.bbl} so that we don't have to upload the
% bibtex file to ArXiV or a journal
%\bibliography{ThesisReferences}

\begin{thebibliography}{88}%
\makeatletter
\providecommand \@ifxundefined [1]{%
 \@ifx{#1\undefined}
}%
\providecommand \@ifnum [1]{%
 \ifnum #1\expandafter \@firstoftwo
 \else \expandafter \@secondoftwo
 \fi
}%
\providecommand \@ifx [1]{%
 \ifx #1\expandafter \@firstoftwo
 \else \expandafter \@secondoftwo
 \fi
}%
\providecommand \natexlab [1]{#1}%
\providecommand \enquote  [1]{``#1''}%
\providecommand \bibnamefont  [1]{#1}%
\providecommand \bibfnamefont [1]{#1}%
\providecommand \citenamefont [1]{#1}%
\providecommand \href@noop [0]{\@secondoftwo}%
\providecommand \href [0]{\begingroup \@sanitize@url \@href}%
\providecommand \@href[1]{\@@startlink{#1}\@@href}%
\providecommand \@@href[1]{\endgroup#1\@@endlink}%
\providecommand \@sanitize@url [0]{\catcode `\\12\catcode `\$12\catcode
  `\&12\catcode `\#12\catcode `\^12\catcode `\_12\catcode `\%12\relax}%
\providecommand \@@startlink[1]{}%
\providecommand \@@endlink[0]{}%
\providecommand \url  [0]{\begingroup\@sanitize@url \@url }%
\providecommand \@url [1]{\endgroup\@href {#1}{\urlprefix }}%
\providecommand \urlprefix  [0]{URL }%
\providecommand \Eprint [0]{\href }%
\providecommand \doibase [0]{https://doi.org/}%
\providecommand \selectlanguage [0]{\@gobble}%
\providecommand \bibinfo  [0]{\@secondoftwo}%
\providecommand \bibfield  [0]{\@secondoftwo}%
\providecommand \translation [1]{[#1]}%
\providecommand \BibitemOpen [0]{}%
\providecommand \bibitemStop [0]{}%
\providecommand \bibitemNoStop [0]{.\EOS\space}%
\providecommand \EOS [0]{\spacefactor3000\relax}%
\providecommand \BibitemShut  [1]{\csname bibitem#1\endcsname}%
\let\auto@bib@innerbib\@empty
%</preamble>
\bibitem [{\citenamefont {Lin}\ \emph {et~al.}(2017)\citenamefont {Lin},
  \citenamefont {Makarov},\ and\ \citenamefont {Schmidt}}]{Lin2017}%
  \BibitemOpen
  \bibfield  {author} {\bibinfo {author} {\bibfnamefont {G.}~\bibnamefont
  {Lin}}, \bibinfo {author} {\bibfnamefont {D.}~\bibnamefont {Makarov}},\ and\
  \bibinfo {author} {\bibfnamefont {O.~G.}\ \bibnamefont {Schmidt}},\ }\href
  {https://doi.org/10.1039/c7lc00026j} {\bibfield  {journal} {\bibinfo
  {journal} {Lab on a Chip}\ }\textbf {\bibinfo {volume} {17}},\ \bibinfo
  {pages} {1884} (\bibinfo {year} {2017})}\BibitemShut {NoStop}%
\bibitem [{\citenamefont {Murzin}\ \emph {et~al.}(2020)\citenamefont {Murzin},
  \citenamefont {Mapps}, \citenamefont {Levada}, \citenamefont {Belyaev},
  \citenamefont {Omelyanchik}, \citenamefont {Panina},\ and\ \citenamefont
  {Rodionova}}]{Murzin2020}%
  \BibitemOpen
  \bibfield  {author} {\bibinfo {author} {\bibfnamefont {D.}~\bibnamefont
  {Murzin}}, \bibinfo {author} {\bibfnamefont {D.~J.}\ \bibnamefont {Mapps}},
  \bibinfo {author} {\bibfnamefont {K.}~\bibnamefont {Levada}}, \bibinfo
  {author} {\bibfnamefont {V.}~\bibnamefont {Belyaev}}, \bibinfo {author}
  {\bibfnamefont {A.}~\bibnamefont {Omelyanchik}}, \bibinfo {author}
  {\bibfnamefont {L.}~\bibnamefont {Panina}},\ and\ \bibinfo {author}
  {\bibfnamefont {V.}~\bibnamefont {Rodionova}},\ }\bibfield  {journal}
  {\bibinfo  {journal} {Sensors (Switzerland)}\ }\textbf {\bibinfo {volume}
  {20}},\ \href {https://doi.org/10.3390/s20061569} {10.3390/s20061569}
  (\bibinfo {year} {2020})\BibitemShut {NoStop}%
\bibitem [{\citenamefont {Elzenheimer}\ \emph {et~al.}(2022)\citenamefont
  {Elzenheimer}, \citenamefont {Bald}, \citenamefont {Engelhardt},
  \citenamefont {Hoffmann}, \citenamefont {Hayes}, \citenamefont {Arbustini},
  \citenamefont {Bahr}, \citenamefont {Quandt}, \citenamefont {Höft},\ and\
  \citenamefont {Schmidt}}]{Elzenheimer2022}%
  \BibitemOpen
  \bibfield  {author} {\bibinfo {author} {\bibfnamefont {E.}~\bibnamefont
  {Elzenheimer}}, \bibinfo {author} {\bibfnamefont {C.}~\bibnamefont {Bald}},
  \bibinfo {author} {\bibfnamefont {E.}~\bibnamefont {Engelhardt}}, \bibinfo
  {author} {\bibfnamefont {J.}~\bibnamefont {Hoffmann}}, \bibinfo {author}
  {\bibfnamefont {P.}~\bibnamefont {Hayes}}, \bibinfo {author} {\bibfnamefont
  {J.}~\bibnamefont {Arbustini}}, \bibinfo {author} {\bibfnamefont
  {A.}~\bibnamefont {Bahr}}, \bibinfo {author} {\bibfnamefont {E.}~\bibnamefont
  {Quandt}}, \bibinfo {author} {\bibfnamefont {M.}~\bibnamefont {Höft}},\ and\
  \bibinfo {author} {\bibfnamefont {G.}~\bibnamefont {Schmidt}},\ }\href
  {https://doi.org/10.3390/s22031018} {\bibfield  {journal} {\bibinfo
  {journal} {Sensors}\ }\textbf {\bibinfo {volume} {22}},\ \bibinfo {pages} {1}
  (\bibinfo {year} {2022})}\BibitemShut {NoStop}%
\bibitem [{\citenamefont {Zhang}\ \emph {et~al.}(2016)\citenamefont {Zhang},
  \citenamefont {Liao}, \citenamefont {Zhao}, \citenamefont {Zhou},
  \citenamefont {Yang},\ and\ \citenamefont {Xia}}]{Zhang2016}%
  \BibitemOpen
  \bibfield  {author} {\bibinfo {author} {\bibfnamefont {H.}~\bibnamefont
  {Zhang}}, \bibinfo {author} {\bibfnamefont {L.}~\bibnamefont {Liao}},
  \bibinfo {author} {\bibfnamefont {R.}~\bibnamefont {Zhao}}, \bibinfo {author}
  {\bibfnamefont {J.}~\bibnamefont {Zhou}}, \bibinfo {author} {\bibfnamefont
  {M.}~\bibnamefont {Yang}},\ and\ \bibinfo {author} {\bibfnamefont
  {R.}~\bibnamefont {Xia}},\ }\bibfield  {journal} {\bibinfo  {journal}
  {Sensors (Switzerland)}\ }\textbf {\bibinfo {volume} {16}},\ \href
  {https://doi.org/10.3390/s16091439} {10.3390/s16091439} (\bibinfo {year}
  {2016})\BibitemShut {NoStop}%
\bibitem [{\citenamefont {Nakai}(2021)}]{Nakai2021}%
  \BibitemOpen
  \bibfield  {author} {\bibinfo {author} {\bibfnamefont {T.}~\bibnamefont
  {Nakai}},\ }\bibfield  {journal} {\bibinfo  {journal} {Sensors}\ }\textbf
  {\bibinfo {volume} {21}},\ \href {https://doi.org/10.3390/s21124063}
  {10.3390/s21124063} (\bibinfo {year} {2021})\BibitemShut {NoStop}%
\bibitem [{\citenamefont {Eslamlou}\ \emph {et~al.}(2023)\citenamefont
  {Eslamlou}, \citenamefont {Ghaderiaram}, \citenamefont {Schlangen},\ and\
  \citenamefont {Fotouhi}}]{Eslamlou2023}%
  \BibitemOpen
  \bibfield  {author} {\bibinfo {author} {\bibfnamefont {A.~D.}\ \bibnamefont
  {Eslamlou}}, \bibinfo {author} {\bibfnamefont {A.}~\bibnamefont
  {Ghaderiaram}}, \bibinfo {author} {\bibfnamefont {E.}~\bibnamefont
  {Schlangen}},\ and\ \bibinfo {author} {\bibfnamefont {M.}~\bibnamefont
  {Fotouhi}},\ }\href {https://doi.org/10.1016/j.conbuildmat.2023.132460}
  {\bibfield  {journal} {\bibinfo  {journal} {Construction and Building
  Materials}\ }\textbf {\bibinfo {volume} {397}},\ \bibinfo {pages} {132460}
  (\bibinfo {year} {2023})}\BibitemShut {NoStop}%
\bibitem [{\citenamefont {Caruso}(1997)}]{Caruso1997}%
  \BibitemOpen
  \bibfield  {author} {\bibinfo {author} {\bibfnamefont {M.~J.}\ \bibnamefont
  {Caruso}},\ }\bibfield  {journal} {\bibinfo  {journal} {SAE Technical
  Papers}\ }\href {https://doi.org/10.4271/970602} {10.4271/970602} (\bibinfo
  {year} {1997})\BibitemShut {NoStop}%
\bibitem [{\citenamefont {El-Sheimy}\ and\ \citenamefont
  {Youssef}(2020)}]{El-Sheimy2020}%
  \BibitemOpen
  \bibfield  {author} {\bibinfo {author} {\bibfnamefont {N.}~\bibnamefont
  {El-Sheimy}}\ and\ \bibinfo {author} {\bibfnamefont {A.}~\bibnamefont
  {Youssef}},\ }\href {https://doi.org/10.1186/s43020-019-0001-5} {\bibfield
  {journal} {\bibinfo  {journal} {Satellite Navigation}\ }\textbf {\bibinfo
  {volume} {1}},\ \bibinfo {pages} {1} (\bibinfo {year} {2020})}\BibitemShut
  {NoStop}%
\bibitem [{\citenamefont {Saltus}\ \emph {et~al.}(2023)\citenamefont {Saltus},
  \citenamefont {Alken}, \citenamefont {Balmes}, \citenamefont {Boneh},
  \citenamefont {Califf}, \citenamefont {Chulliat}, \citenamefont {Meyer},\
  and\ \citenamefont {Nair}}]{Saltus2023}%
  \BibitemOpen
  \bibfield  {author} {\bibinfo {author} {\bibfnamefont {R.}~\bibnamefont
  {Saltus}}, \bibinfo {author} {\bibfnamefont {P.}~\bibnamefont {Alken}},
  \bibinfo {author} {\bibfnamefont {A.}~\bibnamefont {Balmes}}, \bibinfo
  {author} {\bibfnamefont {N.}~\bibnamefont {Boneh}}, \bibinfo {author}
  {\bibfnamefont {S.}~\bibnamefont {Califf}}, \bibinfo {author} {\bibfnamefont
  {A.}~\bibnamefont {Chulliat}}, \bibinfo {author} {\bibfnamefont
  {B.}~\bibnamefont {Meyer}},\ and\ \bibinfo {author} {\bibfnamefont
  {M.}~\bibnamefont {Nair}},\ }\href
  {https://doi.org/10.1109/PLANS53410.2023.10140025} {\bibfield  {journal}
  {\bibinfo  {journal} {2023 IEEE/ION Position, Location and Navigation
  Symposium, PLANS 2023}\ ,\ \bibinfo {pages} {805}} (\bibinfo {year}
  {2023})}\BibitemShut {NoStop}%
\bibitem [{\citenamefont {Grimes}\ \emph {et~al.}(1999)\citenamefont {Grimes},
  \citenamefont {Ong}, \citenamefont {Loiselle}, \citenamefont {Stoyanov},
  \citenamefont {Kouzoudis}, \citenamefont {Liu}, \citenamefont {Tong},\ and\
  \citenamefont {Tefiku}}]{Grimes1999}%
  \BibitemOpen
  \bibfield  {author} {\bibinfo {author} {\bibfnamefont {C.~A.}\ \bibnamefont
  {Grimes}}, \bibinfo {author} {\bibfnamefont {K.}~\bibnamefont {Ong}},
  \bibinfo {author} {\bibfnamefont {K.}~\bibnamefont {Loiselle}}, \bibinfo
  {author} {\bibfnamefont {P.~G.}\ \bibnamefont {Stoyanov}}, \bibinfo {author}
  {\bibfnamefont {D.}~\bibnamefont {Kouzoudis}}, \bibinfo {author}
  {\bibfnamefont {Y.}~\bibnamefont {Liu}}, \bibinfo {author} {\bibfnamefont
  {C.}~\bibnamefont {Tong}},\ and\ \bibinfo {author} {\bibfnamefont
  {F.}~\bibnamefont {Tefiku}},\ }\href
  {http://iopscience.iop.org/article/10.1088/0964-1726/8/5/314/pdf} {\bibfield
  {journal} {\bibinfo  {journal} {Smart Materials and Structures}\ }\textbf
  {\bibinfo {volume} {8}},\ \bibinfo {pages} {639} (\bibinfo {year}
  {1999})}\BibitemShut {NoStop}%
\bibitem [{\citenamefont {Hojjati-Najafabadi}\ \emph
  {et~al.}(2022)\citenamefont {Hojjati-Najafabadi}, \citenamefont
  {Mansoorianfar}, \citenamefont {Liang}, \citenamefont {Shahin},\ and\
  \citenamefont {Karimi-Maleh}}]{Hojjati-Najafabadi2022}%
  \BibitemOpen
  \bibfield  {author} {\bibinfo {author} {\bibfnamefont {A.}~\bibnamefont
  {Hojjati-Najafabadi}}, \bibinfo {author} {\bibfnamefont {M.}~\bibnamefont
  {Mansoorianfar}}, \bibinfo {author} {\bibfnamefont {T.}~\bibnamefont
  {Liang}}, \bibinfo {author} {\bibfnamefont {K.}~\bibnamefont {Shahin}},\ and\
  \bibinfo {author} {\bibfnamefont {H.}~\bibnamefont {Karimi-Maleh}},\
  }\bibfield  {journal} {\bibinfo  {journal} {Science of the Total
  Environment}\ }\textbf {\bibinfo {volume} {824}},\ \href
  {https://doi.org/10.1016/j.scitotenv.2022.153844}
  {10.1016/j.scitotenv.2022.153844} (\bibinfo {year} {2022})\BibitemShut
  {NoStop}%
\bibitem [{\citenamefont {$\mathrm{\check{C}}$ulik}\ \emph
  {et~al.}(2023)\citenamefont {$\mathrm{\check{C}}$ulik}, \citenamefont
  {$\mathrm{\check{S}}$tefancová},\ and\ \citenamefont {Hrudkay}}]{Culik2023}%
  \BibitemOpen
  \bibfield  {author} {\bibinfo {author} {\bibfnamefont {K.}~\bibnamefont
  {$\mathrm{\check{C}}$ulik}}, \bibinfo {author} {\bibfnamefont
  {V.}~\bibnamefont {$\mathrm{\check{S}}$tefancová}},\ and\ \bibinfo {author}
  {\bibfnamefont {K.}~\bibnamefont {Hrudkay}},\ }\href
  {https://doi.org/10.3390/s23125740} {\bibfield  {journal} {\bibinfo
  {journal} {Sensors}\ }\textbf {\bibinfo {volume} {23}},\ \bibinfo {pages}
  {5740} (\bibinfo {year} {2023})}\BibitemShut {NoStop}%
\bibitem [{\citenamefont {Treutler}(2001)}]{Treutler2001}%
  \BibitemOpen
  \bibfield  {author} {\bibinfo {author} {\bibfnamefont {C.~P.~O.}\
  \bibnamefont {Treutler}},\ }\href
  {https://doi.org/10.1016/S0924-4247(01)00621-5} {\bibfield  {journal}
  {\bibinfo  {journal} {Sensors and Actuators, A: Physical}\ }\textbf {\bibinfo
  {volume} {91}},\ \bibinfo {pages} {2} (\bibinfo {year} {2001})}\BibitemShut
  {NoStop}%
\bibitem [{\citenamefont {Granig}\ \emph {et~al.}(2007)\citenamefont {Granig},
  \citenamefont {Hartmann},\ and\ \citenamefont {Köppl}}]{Granig2007}%
  \BibitemOpen
  \bibfield  {author} {\bibinfo {author} {\bibfnamefont {W.}~\bibnamefont
  {Granig}}, \bibinfo {author} {\bibfnamefont {S.}~\bibnamefont {Hartmann}},\
  and\ \bibinfo {author} {\bibfnamefont {B.}~\bibnamefont {Köppl}},\ }\href
  {https://doi.org/10.4271/2007-01-0397} {\bibfield  {journal} {\bibinfo
  {journal} {SAE Technical Papers}\ }\textbf {\bibinfo {volume} {2007}},\
  \bibinfo {pages} {776} (\bibinfo {year} {2007})}\BibitemShut {NoStop}%
\bibitem [{\citenamefont {Nonomura}(2020)}]{Nonomura2020}%
  \BibitemOpen
  \bibfield  {author} {\bibinfo {author} {\bibfnamefont {Y.}~\bibnamefont
  {Nonomura}},\ }\href {https://doi.org/10.1002/tee.23142} {\bibfield
  {journal} {\bibinfo  {journal} {IEEJ Transactions on Electrical and
  Electronic Engineering}\ }\textbf {\bibinfo {volume} {15}},\ \bibinfo {pages}
  {984} (\bibinfo {year} {2020})}\BibitemShut {NoStop}%
\bibitem [{\citenamefont {Boto}\ \emph {et~al.}(2016)\citenamefont {Boto},
  \citenamefont {Bowtell}, \citenamefont {Krüger}, \citenamefont {Fromhold},
  \citenamefont {Morris}, \citenamefont {Meyer}, \citenamefont {Barnes},\ and\
  \citenamefont {Brookes}}]{Boto2016}%
  \BibitemOpen
  \bibfield  {author} {\bibinfo {author} {\bibfnamefont {E.}~\bibnamefont
  {Boto}}, \bibinfo {author} {\bibfnamefont {R.}~\bibnamefont {Bowtell}},
  \bibinfo {author} {\bibfnamefont {P.}~\bibnamefont {Krüger}}, \bibinfo
  {author} {\bibfnamefont {T.~M.}\ \bibnamefont {Fromhold}}, \bibinfo {author}
  {\bibfnamefont {P.~G.}\ \bibnamefont {Morris}}, \bibinfo {author}
  {\bibfnamefont {S.~S.}\ \bibnamefont {Meyer}}, \bibinfo {author}
  {\bibfnamefont {G.~R.}\ \bibnamefont {Barnes}},\ and\ \bibinfo {author}
  {\bibfnamefont {M.~J.}\ \bibnamefont {Brookes}},\ }\href
  {https://doi.org/10.1371/journal.pone.0157655} {\bibfield  {journal}
  {\bibinfo  {journal} {PLoS ONE}\ }\textbf {\bibinfo {volume} {11}},\ \bibinfo
  {pages} {1} (\bibinfo {year} {2016})}\BibitemShut {NoStop}%
\bibitem [{\citenamefont {Kanno}\ \emph {et~al.}(2022)\citenamefont {Kanno},
  \citenamefont {Nakasato}, \citenamefont {Oogane}, \citenamefont {Fujiwara},
  \citenamefont {Nakano}, \citenamefont {Arimoto}, \citenamefont {Matsuzaki},\
  and\ \citenamefont {Ando}}]{Kanno2022}%
  \BibitemOpen
  \bibfield  {author} {\bibinfo {author} {\bibfnamefont {A.}~\bibnamefont
  {Kanno}}, \bibinfo {author} {\bibfnamefont {N.}~\bibnamefont {Nakasato}},
  \bibinfo {author} {\bibfnamefont {M.}~\bibnamefont {Oogane}}, \bibinfo
  {author} {\bibfnamefont {K.}~\bibnamefont {Fujiwara}}, \bibinfo {author}
  {\bibfnamefont {T.}~\bibnamefont {Nakano}}, \bibinfo {author} {\bibfnamefont
  {T.}~\bibnamefont {Arimoto}}, \bibinfo {author} {\bibfnamefont
  {H.}~\bibnamefont {Matsuzaki}},\ and\ \bibinfo {author} {\bibfnamefont
  {Y.}~\bibnamefont {Ando}},\ }\href
  {https://doi.org/10.1038/s41598-022-10155-6} {\bibfield  {journal} {\bibinfo
  {journal} {Scientific Reports}\ }\textbf {\bibinfo {volume} {12}},\ \bibinfo
  {pages} {1} (\bibinfo {year} {2022})}\BibitemShut {NoStop}%
\bibitem [{\citenamefont {Brookes}\ \emph {et~al.}(2022)\citenamefont
  {Brookes}, \citenamefont {Leggett}, \citenamefont {Rea}, \citenamefont
  {Hill}, \citenamefont {Holmes}, \citenamefont {Boto},\ and\ \citenamefont
  {Bowtell}}]{Brookes2022}%
  \BibitemOpen
  \bibfield  {author} {\bibinfo {author} {\bibfnamefont {M.~J.}\ \bibnamefont
  {Brookes}}, \bibinfo {author} {\bibfnamefont {J.}~\bibnamefont {Leggett}},
  \bibinfo {author} {\bibfnamefont {M.}~\bibnamefont {Rea}}, \bibinfo {author}
  {\bibfnamefont {R.~M.}\ \bibnamefont {Hill}}, \bibinfo {author}
  {\bibfnamefont {N.}~\bibnamefont {Holmes}}, \bibinfo {author} {\bibfnamefont
  {E.}~\bibnamefont {Boto}},\ and\ \bibinfo {author} {\bibfnamefont
  {R.}~\bibnamefont {Bowtell}},\ }\href
  {https://doi.org/10.1016/j.tins.2022.05.008} {\bibfield  {journal} {\bibinfo
  {journal} {Trends in Neurosciences}\ }\textbf {\bibinfo {volume} {45}},\
  \bibinfo {pages} {621} (\bibinfo {year} {2022})}\BibitemShut {NoStop}%
\bibitem [{\citenamefont {Robinson}\ \emph {et~al.}(2008)\citenamefont
  {Robinson}, \citenamefont {Binley}, \citenamefont {Crook}, \citenamefont
  {Day-Lewis}, \citenamefont {Ferré}, \citenamefont {Grauch}, \citenamefont
  {Knight}, \citenamefont {Knoll}, \citenamefont {Lakshmi}, \citenamefont
  {Miller}, \citenamefont {Nyquist}, \citenamefont {Pellerin}, \citenamefont
  {Singha},\ and\ \citenamefont {Slater}}]{Robinson2008}%
  \BibitemOpen
  \bibfield  {author} {\bibinfo {author} {\bibfnamefont {D.~A.}\ \bibnamefont
  {Robinson}}, \bibinfo {author} {\bibfnamefont {A.}~\bibnamefont {Binley}},
  \bibinfo {author} {\bibfnamefont {N.}~\bibnamefont {Crook}}, \bibinfo
  {author} {\bibfnamefont {F.~D.}\ \bibnamefont {Day-Lewis}}, \bibinfo {author}
  {\bibfnamefont {T.~P.~A.}\ \bibnamefont {Ferré}}, \bibinfo {author}
  {\bibfnamefont {V.~J.~S.}\ \bibnamefont {Grauch}}, \bibinfo {author}
  {\bibfnamefont {R.}~\bibnamefont {Knight}}, \bibinfo {author} {\bibfnamefont
  {M.}~\bibnamefont {Knoll}}, \bibinfo {author} {\bibfnamefont
  {V.}~\bibnamefont {Lakshmi}}, \bibinfo {author} {\bibfnamefont
  {R.}~\bibnamefont {Miller}}, \bibinfo {author} {\bibfnamefont
  {J.}~\bibnamefont {Nyquist}}, \bibinfo {author} {\bibfnamefont
  {L.}~\bibnamefont {Pellerin}}, \bibinfo {author} {\bibfnamefont
  {K.}~\bibnamefont {Singha}},\ and\ \bibinfo {author} {\bibfnamefont
  {L.}~\bibnamefont {Slater}},\ }\href {https://doi.org/10.1002/hyp.6963}
  {\bibfield  {journal} {\bibinfo  {journal} {Hydrological Processes}\ }\textbf
  {\bibinfo {volume} {22}},\ \bibinfo {pages} {3604} (\bibinfo {year}
  {2008})}\BibitemShut {NoStop}%
\bibitem [{\citenamefont {Dentith}\ and\ \citenamefont
  {Mudge}(2014)}]{Dentith2014}%
  \BibitemOpen
  \bibfield  {author} {\bibinfo {author} {\bibfnamefont {M.}~\bibnamefont
  {Dentith}}\ and\ \bibinfo {author} {\bibfnamefont {S.}~\bibnamefont
  {Mudge}},\ }\href {https://doi.org/10.1017/cbo9781139024358} {\emph {\bibinfo
  {title} {Geophysics for the mineral exploration geoscientist}}}\ (\bibinfo
  {publisher} {Cambridge University Press},\ \bibinfo {year}
  {2014})\BibitemShut {NoStop}%
\bibitem [{\citenamefont {Stolz}\ \emph {et~al.}(2022)\citenamefont {Stolz},
  \citenamefont {Schiffler}, \citenamefont {Becken}, \citenamefont {Thiede},
  \citenamefont {Schneider}, \citenamefont {Chubak}, \citenamefont {Marsden},
  \citenamefont {Bergshjorth}, \citenamefont {Schaefer},\ and\ \citenamefont
  {Terblanche}}]{Stolz2022}%
  \BibitemOpen
  \bibfield  {author} {\bibinfo {author} {\bibfnamefont {R.}~\bibnamefont
  {Stolz}}, \bibinfo {author} {\bibfnamefont {M.}~\bibnamefont {Schiffler}},
  \bibinfo {author} {\bibfnamefont {M.}~\bibnamefont {Becken}}, \bibinfo
  {author} {\bibfnamefont {A.}~\bibnamefont {Thiede}}, \bibinfo {author}
  {\bibfnamefont {M.}~\bibnamefont {Schneider}}, \bibinfo {author}
  {\bibfnamefont {G.}~\bibnamefont {Chubak}}, \bibinfo {author} {\bibfnamefont
  {P.}~\bibnamefont {Marsden}}, \bibinfo {author} {\bibfnamefont {A.~B.}\
  \bibnamefont {Bergshjorth}}, \bibinfo {author} {\bibfnamefont
  {M.}~\bibnamefont {Schaefer}},\ and\ \bibinfo {author} {\bibfnamefont
  {O.}~\bibnamefont {Terblanche}},\ }\href
  {https://doi.org/10.1007/s13563-022-00333-3} {\bibfield  {journal} {\bibinfo
  {journal} {Mineral Economics}\ }\textbf {\bibinfo {volume} {35}},\ \bibinfo
  {pages} {467} (\bibinfo {year} {2022})}\BibitemShut {NoStop}%
\bibitem [{\citenamefont {Aschenbrenner}\ and\ \citenamefont
  {Goubau}(1936)}]{Aschenbrenner1936}%
  \BibitemOpen
  \bibfield  {author} {\bibinfo {author} {\bibfnamefont {H.}~\bibnamefont
  {Aschenbrenner}}\ and\ \bibinfo {author} {\bibfnamefont {G.}~\bibnamefont
  {Goubau}},\ }\href@noop {} {\bibfield  {journal} {\bibinfo  {journal}
  {Hochfrequenztechnik und Elektroakustik}\ }\textbf {\bibinfo {volume} {47}},\
  \bibinfo {pages} {177} (\bibinfo {year} {1936})}\BibitemShut {NoStop}%
\bibitem [{\citenamefont {Josephson}(1962)}]{Josephson1962}%
  \BibitemOpen
  \bibfield  {author} {\bibinfo {author} {\bibfnamefont {B.~D.}\ \bibnamefont
  {Josephson}},\ }\href {https://doi.org/10.1016/0031-9163(62)91369-0}
  {\bibfield  {journal} {\bibinfo  {journal} {Physics Letters}\ }\textbf
  {\bibinfo {volume} {1}},\ \bibinfo {pages} {251} (\bibinfo {year}
  {1962})}\BibitemShut {NoStop}%
\bibitem [{\citenamefont {Anderson}\ and\ \citenamefont
  {Rowell}(1963)}]{Anderson1963}%
  \BibitemOpen
  \bibfield  {author} {\bibinfo {author} {\bibfnamefont {P.~W.}\ \bibnamefont
  {Anderson}}\ and\ \bibinfo {author} {\bibfnamefont {J.~M.}\ \bibnamefont
  {Rowell}},\ }\href {https://doi.org/10.1103/PhysRevLett.10.230} {\bibfield
  {journal} {\bibinfo  {journal} {Physical Review Letters}\ }\textbf {\bibinfo
  {volume} {10}},\ \bibinfo {pages} {230} (\bibinfo {year} {1963})}\BibitemShut
  {NoStop}%
\bibitem [{\citenamefont {Jaklevic}\ \emph {et~al.}(1964)\citenamefont
  {Jaklevic}, \citenamefont {Lambe}, \citenamefont {Silver},\ and\
  \citenamefont {Mercereau}}]{Jaklevic1964}%
  \BibitemOpen
  \bibfield  {author} {\bibinfo {author} {\bibfnamefont {R.~C.}\ \bibnamefont
  {Jaklevic}}, \bibinfo {author} {\bibfnamefont {J.}~\bibnamefont {Lambe}},
  \bibinfo {author} {\bibfnamefont {A.~H.}\ \bibnamefont {Silver}},\ and\
  \bibinfo {author} {\bibfnamefont {J.~E.}\ \bibnamefont {Mercereau}},\ }\href
  {https://doi.org/10.1103/PhysRevLett.12.159} {\bibfield  {journal} {\bibinfo
  {journal} {Physical Review Letters}\ }\textbf {\bibinfo {volume} {12}},\
  \bibinfo {pages} {159} (\bibinfo {year} {1964})}\BibitemShut {NoStop}%
\bibitem [{\citenamefont {Hall}(1879)}]{Hall1879}%
  \BibitemOpen
  \bibfield  {author} {\bibinfo {author} {\bibfnamefont {E.~H.}\ \bibnamefont
  {Hall}},\ }\href@noop {} {\bibfield  {journal} {\bibinfo  {journal} {American
  Journal of Mathematics}\ }\textbf {\bibinfo {volume} {2}},\ \bibinfo {pages}
  {287} (\bibinfo {year} {1879})}\BibitemShut {NoStop}%
\bibitem [{\citenamefont {Uzlu}\ \emph {et~al.}(2019)\citenamefont {Uzlu},
  \citenamefont {Wang}, \citenamefont {Lukas}, \citenamefont {Otto},
  \citenamefont {Lemme},\ and\ \citenamefont {Neumaier}}]{Uzlu2019}%
  \BibitemOpen
  \bibfield  {author} {\bibinfo {author} {\bibfnamefont {B.}~\bibnamefont
  {Uzlu}}, \bibinfo {author} {\bibfnamefont {Z.}~\bibnamefont {Wang}}, \bibinfo
  {author} {\bibfnamefont {S.}~\bibnamefont {Lukas}}, \bibinfo {author}
  {\bibfnamefont {M.}~\bibnamefont {Otto}}, \bibinfo {author} {\bibfnamefont
  {M.~C.}\ \bibnamefont {Lemme}},\ and\ \bibinfo {author} {\bibfnamefont
  {D.}~\bibnamefont {Neumaier}},\ }\bibfield  {journal} {\bibinfo  {journal}
  {Scientific Reports}\ }\textbf {\bibinfo {volume} {9}},\ \href
  {https://doi.org/10.1038/s41598-019-54489-0} {10.1038/s41598-019-54489-0}
  (\bibinfo {year} {2019})\BibitemShut {NoStop}%
\bibitem [{\citenamefont {Thomson}(1857)}]{Thomson1857}%
  \BibitemOpen
  \bibfield  {author} {\bibinfo {author} {\bibfnamefont {W.}~\bibnamefont
  {Thomson}},\ }\href {https://doi.org/10.1098/rspl.1856.0144} {\bibfield
  {journal} {\bibinfo  {journal} {Proceedings of the Royal Society of London}\
  }\textbf {\bibinfo {volume} {8}},\ \bibinfo {pages} {546} (\bibinfo {year}
  {1857})}\BibitemShut {NoStop}%
\bibitem [{\citenamefont {Baibich}\ \emph {et~al.}(1988)\citenamefont
  {Baibich}, \citenamefont {Broto}, \citenamefont {Fert}, \citenamefont {Dau},
  \citenamefont {Petroff}, \citenamefont {Eitenne}, \citenamefont {Creuzet},
  \citenamefont {Friederich},\ and\ \citenamefont {Chazelas}}]{Baibich1988}%
  \BibitemOpen
  \bibfield  {author} {\bibinfo {author} {\bibfnamefont {M.~N.}\ \bibnamefont
  {Baibich}}, \bibinfo {author} {\bibfnamefont {J.~M.}\ \bibnamefont {Broto}},
  \bibinfo {author} {\bibfnamefont {A.}~\bibnamefont {Fert}}, \bibinfo {author}
  {\bibfnamefont {F.~N.~V.}\ \bibnamefont {Dau}}, \bibinfo {author}
  {\bibfnamefont {F.}~\bibnamefont {Petroff}}, \bibinfo {author} {\bibfnamefont
  {P.}~\bibnamefont {Eitenne}}, \bibinfo {author} {\bibfnamefont
  {G.}~\bibnamefont {Creuzet}}, \bibinfo {author} {\bibfnamefont
  {A.}~\bibnamefont {Friederich}},\ and\ \bibinfo {author} {\bibfnamefont
  {J.}~\bibnamefont {Chazelas}},\ }\href
  {https://doi.org/10.1103/PhysRevLett.61.2472} {\bibfield  {journal} {\bibinfo
   {journal} {Physical Review Letters}\ }\textbf {\bibinfo {volume} {61}},\
  \bibinfo {pages} {2472} (\bibinfo {year} {1988})}\BibitemShut {NoStop}%
\bibitem [{\citenamefont {Binasch}\ \emph {et~al.}(1989)\citenamefont
  {Binasch}, \citenamefont {Grünberg}, \citenamefont {Saurenbach},\ and\
  \citenamefont {Zinn}}]{Binasch1989}%
  \BibitemOpen
  \bibfield  {author} {\bibinfo {author} {\bibfnamefont {G.}~\bibnamefont
  {Binasch}}, \bibinfo {author} {\bibfnamefont {P.}~\bibnamefont {Grünberg}},
  \bibinfo {author} {\bibfnamefont {F.}~\bibnamefont {Saurenbach}},\ and\
  \bibinfo {author} {\bibfnamefont {W.}~\bibnamefont {Zinn}},\ }\href
  {https://doi.org/10.1103/PhysRevB.39.4828} {\bibfield  {journal} {\bibinfo
  {journal} {Physical Review B}\ }\textbf {\bibinfo {volume} {39}},\ \bibinfo
  {pages} {4828} (\bibinfo {year} {1989})}\BibitemShut {NoStop}%
\bibitem [{\citenamefont {Julliere}(1975)}]{Julliere1975}%
  \BibitemOpen
  \bibfield  {author} {\bibinfo {author} {\bibfnamefont {M.}~\bibnamefont
  {Julliere}},\ }\href {https://doi.org/10.1016/0375-9601(75)90174-7}
  {\bibfield  {journal} {\bibinfo  {journal} {Physics Letters A}\ }\textbf
  {\bibinfo {volume} {54}},\ \bibinfo {pages} {225} (\bibinfo {year}
  {1975})}\BibitemShut {NoStop}%
\bibitem [{\citenamefont {Maze}\ \emph {et~al.}(2008)\citenamefont {Maze},
  \citenamefont {Stanwix}, \citenamefont {Hodges}, \citenamefont {Hong},
  \citenamefont {Taylor}, \citenamefont {Cappellaro}, \citenamefont {Jiang},
  \citenamefont {Dutt}, \citenamefont {Togan}, \citenamefont {Zibrov},
  \citenamefont {Yacoby}, \citenamefont {Walsworth},\ and\ \citenamefont
  {Lukin}}]{Maze2008}%
  \BibitemOpen
  \bibfield  {author} {\bibinfo {author} {\bibfnamefont {J.~R.}\ \bibnamefont
  {Maze}}, \bibinfo {author} {\bibfnamefont {P.~L.}\ \bibnamefont {Stanwix}},
  \bibinfo {author} {\bibfnamefont {J.~S.}\ \bibnamefont {Hodges}}, \bibinfo
  {author} {\bibfnamefont {S.}~\bibnamefont {Hong}}, \bibinfo {author}
  {\bibfnamefont {J.~M.}\ \bibnamefont {Taylor}}, \bibinfo {author}
  {\bibfnamefont {P.}~\bibnamefont {Cappellaro}}, \bibinfo {author}
  {\bibfnamefont {L.}~\bibnamefont {Jiang}}, \bibinfo {author} {\bibfnamefont
  {M.~V.}\ \bibnamefont {Dutt}}, \bibinfo {author} {\bibfnamefont
  {E.}~\bibnamefont {Togan}}, \bibinfo {author} {\bibfnamefont {A.~S.}\
  \bibnamefont {Zibrov}}, \bibinfo {author} {\bibfnamefont {A.}~\bibnamefont
  {Yacoby}}, \bibinfo {author} {\bibfnamefont {R.~L.}\ \bibnamefont
  {Walsworth}},\ and\ \bibinfo {author} {\bibfnamefont {M.~D.}\ \bibnamefont
  {Lukin}},\ }\href {https://doi.org/10.1038/nature07279} {\bibfield  {journal}
  {\bibinfo  {journal} {Nature}\ }\textbf {\bibinfo {volume} {455}},\ \bibinfo
  {pages} {644} (\bibinfo {year} {2008})}\BibitemShut {NoStop}%
\bibitem [{\citenamefont {Balasubramanian}\ \emph {et~al.}(2008)\citenamefont
  {Balasubramanian}, \citenamefont {Chan}, \citenamefont {Kolesov},
  \citenamefont {Al-Hmoud}, \citenamefont {Tisler}, \citenamefont {Shin},
  \citenamefont {Kim}, \citenamefont {Wojcik}, \citenamefont {Hemmer},
  \citenamefont {Krueger}, \citenamefont {Hanke}, \citenamefont
  {Leitenstorfer}, \citenamefont {Bratschitsch}, \citenamefont {Jelezko},\ and\
  \citenamefont {Wrachtrup}}]{Balasubramanian2008}%
  \BibitemOpen
  \bibfield  {author} {\bibinfo {author} {\bibfnamefont {G.}~\bibnamefont
  {Balasubramanian}}, \bibinfo {author} {\bibfnamefont {I.~Y.}\ \bibnamefont
  {Chan}}, \bibinfo {author} {\bibfnamefont {R.}~\bibnamefont {Kolesov}},
  \bibinfo {author} {\bibfnamefont {M.}~\bibnamefont {Al-Hmoud}}, \bibinfo
  {author} {\bibfnamefont {J.}~\bibnamefont {Tisler}}, \bibinfo {author}
  {\bibfnamefont {C.}~\bibnamefont {Shin}}, \bibinfo {author} {\bibfnamefont
  {C.}~\bibnamefont {Kim}}, \bibinfo {author} {\bibfnamefont {A.}~\bibnamefont
  {Wojcik}}, \bibinfo {author} {\bibfnamefont {P.~R.}\ \bibnamefont {Hemmer}},
  \bibinfo {author} {\bibfnamefont {A.}~\bibnamefont {Krueger}}, \bibinfo
  {author} {\bibfnamefont {T.}~\bibnamefont {Hanke}}, \bibinfo {author}
  {\bibfnamefont {A.}~\bibnamefont {Leitenstorfer}}, \bibinfo {author}
  {\bibfnamefont {R.}~\bibnamefont {Bratschitsch}}, \bibinfo {author}
  {\bibfnamefont {F.}~\bibnamefont {Jelezko}},\ and\ \bibinfo {author}
  {\bibfnamefont {J.}~\bibnamefont {Wrachtrup}},\ }\href
  {https://doi.org/10.1038/nature07278} {\bibfield  {journal} {\bibinfo
  {journal} {Nature}\ }\textbf {\bibinfo {volume} {455}},\ \bibinfo {pages}
  {648} (\bibinfo {year} {2008})}\BibitemShut {NoStop}%
\bibitem [{\citenamefont {Taylor}\ \emph {et~al.}(2008)\citenamefont {Taylor},
  \citenamefont {Cappellaro}, \citenamefont {Childress}, \citenamefont {Jiang},
  \citenamefont {Budker}, \citenamefont {Hemmer}, \citenamefont {Yacoby},
  \citenamefont {Walsworth},\ and\ \citenamefont {Lukin}}]{Taylor2008}%
  \BibitemOpen
  \bibfield  {author} {\bibinfo {author} {\bibfnamefont {J.~M.}\ \bibnamefont
  {Taylor}}, \bibinfo {author} {\bibfnamefont {P.}~\bibnamefont {Cappellaro}},
  \bibinfo {author} {\bibfnamefont {L.}~\bibnamefont {Childress}}, \bibinfo
  {author} {\bibfnamefont {L.}~\bibnamefont {Jiang}}, \bibinfo {author}
  {\bibfnamefont {D.}~\bibnamefont {Budker}}, \bibinfo {author} {\bibfnamefont
  {P.~R.}\ \bibnamefont {Hemmer}}, \bibinfo {author} {\bibfnamefont
  {A.}~\bibnamefont {Yacoby}}, \bibinfo {author} {\bibfnamefont
  {R.}~\bibnamefont {Walsworth}},\ and\ \bibinfo {author} {\bibfnamefont
  {M.~D.}\ \bibnamefont {Lukin}},\ }\href {https://doi.org/10.1038/nphys1075}
  {\bibfield  {journal} {\bibinfo  {journal} {Nature Physics}\ }\textbf
  {\bibinfo {volume} {4}},\ \bibinfo {pages} {810} (\bibinfo {year}
  {2008})}\BibitemShut {NoStop}%
\bibitem [{\citenamefont {Panina}\ and\ \citenamefont
  {Mohri}(1994)}]{Panina1994}%
  \BibitemOpen
  \bibfield  {author} {\bibinfo {author} {\bibfnamefont {L.~V.}\ \bibnamefont
  {Panina}}\ and\ \bibinfo {author} {\bibfnamefont {K.}~\bibnamefont {Mohri}},\
  }\href {https://doi.org/10.1063/1.112104} {\bibfield  {journal} {\bibinfo
  {journal} {Applied Physics Letters}\ }\textbf {\bibinfo {volume} {65}},\
  \bibinfo {pages} {1189} (\bibinfo {year} {1994})}\BibitemShut {NoStop}%
\bibitem [{\citenamefont {Beach}\ and\ \citenamefont
  {Berkowitz}(1994)}]{Beach1994}%
  \BibitemOpen
  \bibfield  {author} {\bibinfo {author} {\bibfnamefont {R.~S.}\ \bibnamefont
  {Beach}}\ and\ \bibinfo {author} {\bibfnamefont {A.~E.}\ \bibnamefont
  {Berkowitz}},\ }\href {https://doi.org/10.1063/1.111170} {\bibfield
  {journal} {\bibinfo  {journal} {Applied Physics Letters}\ }\textbf {\bibinfo
  {volume} {64}},\ \bibinfo {pages} {3652} (\bibinfo {year}
  {1994})}\BibitemShut {NoStop}%
\bibitem [{\citenamefont {Mohri}\ \emph {et~al.}(2015)\citenamefont {Mohri},
  \citenamefont {Uchiyama}, \citenamefont {Panina}, \citenamefont {Yamamoto},\
  and\ \citenamefont {Bushida}}]{Mohri2015}%
  \BibitemOpen
  \bibfield  {author} {\bibinfo {author} {\bibfnamefont {K.}~\bibnamefont
  {Mohri}}, \bibinfo {author} {\bibfnamefont {T.}~\bibnamefont {Uchiyama}},
  \bibinfo {author} {\bibfnamefont {L.~V.}\ \bibnamefont {Panina}}, \bibinfo
  {author} {\bibfnamefont {M.}~\bibnamefont {Yamamoto}},\ and\ \bibinfo
  {author} {\bibfnamefont {K.}~\bibnamefont {Bushida}},\ }\href
  {https://doi.org/10.1155/2015/718069} {\bibfield  {journal} {\bibinfo
  {journal} {Journal of Sensors}\ }\textbf {\bibinfo {volume} {2015}},\
  \bibinfo {pages} {1} (\bibinfo {year} {2015})}\BibitemShut {NoStop}%
\bibitem [{\citenamefont {Khan}\ \emph {et~al.}(2021)\citenamefont {Khan},
  \citenamefont {Sun}, \citenamefont {Li}, \citenamefont {Przybysz},\ and\
  \citenamefont {Kosel}}]{Khan2021}%
  \BibitemOpen
  \bibfield  {author} {\bibinfo {author} {\bibfnamefont {M.~A.}\ \bibnamefont
  {Khan}}, \bibinfo {author} {\bibfnamefont {J.}~\bibnamefont {Sun}}, \bibinfo
  {author} {\bibfnamefont {B.}~\bibnamefont {Li}}, \bibinfo {author}
  {\bibfnamefont {A.}~\bibnamefont {Przybysz}},\ and\ \bibinfo {author}
  {\bibfnamefont {J.}~\bibnamefont {Kosel}},\ }\href
  {https://doi.org/10.1088/2631-8695/ac0838} {\bibfield  {journal} {\bibinfo
  {journal} {Engineering Research Express}\ }\textbf {\bibinfo {volume} {3}},\
  \bibinfo {pages} {022005} (\bibinfo {year} {2021})}\BibitemShut {NoStop}%
\bibitem [{\citenamefont {Betzholz}\ \emph {et~al.}(2013)\citenamefont
  {Betzholz}, \citenamefont {Gao}, \citenamefont {Zhao},\ and\ \citenamefont
  {Hartmann}}]{Betzholz2013}%
  \BibitemOpen
  \bibfield  {author} {\bibinfo {author} {\bibfnamefont {R.}~\bibnamefont
  {Betzholz}}, \bibinfo {author} {\bibfnamefont {H.}~\bibnamefont {Gao}},
  \bibinfo {author} {\bibfnamefont {Z.}~\bibnamefont {Zhao}},\ and\ \bibinfo
  {author} {\bibfnamefont {U.}~\bibnamefont {Hartmann}},\ }\bibfield  {journal}
  {\bibinfo  {journal} {Europhysics Letters}\ }\textbf {\bibinfo {volume}
  {101}},\ \href {https://doi.org/10.1209/0295-5075/101/17005}
  {10.1209/0295-5075/101/17005} (\bibinfo {year} {2013})\BibitemShut {NoStop}%
\bibitem [{\citenamefont {q.~Xiao}\ \emph {et~al.}(2000)\citenamefont
  {q.~Xiao}, \citenamefont {h.~Liu}, \citenamefont {h.~Yan}, \citenamefont
  {y.~Dai}, \citenamefont {Zhang},\ and\ \citenamefont {m.~Mei}}]{Xiao2000}%
  \BibitemOpen
  \bibfield  {author} {\bibinfo {author} {\bibfnamefont {S.}~\bibnamefont
  {q.~Xiao}}, \bibinfo {author} {\bibfnamefont {Y.}~\bibnamefont {h.~Liu}},
  \bibinfo {author} {\bibfnamefont {S.}~\bibnamefont {h.~Yan}}, \bibinfo
  {author} {\bibfnamefont {Y.}~\bibnamefont {y.~Dai}}, \bibinfo {author}
  {\bibfnamefont {L.}~\bibnamefont {Zhang}},\ and\ \bibinfo {author}
  {\bibfnamefont {L.}~\bibnamefont {m.~Mei}},\ }\href@noop {} {\bibfield
  {journal} {\bibinfo  {journal} {Physical Review B - Condensed Matter and
  Materials Physics}\ }\textbf {\bibinfo {volume} {61}},\ \bibinfo {pages}
  {5734} (\bibinfo {year} {2000})}\BibitemShut {NoStop}%
\bibitem [{\citenamefont {García-Arribas}\ \emph {et~al.}(2016)\citenamefont
  {García-Arribas}, \citenamefont {Fernández}, \citenamefont {Svalov},
  \citenamefont {Kurlyandskaya},\ and\ \citenamefont
  {Barandiarán}}]{Garcia-Arribas2016}%
  \BibitemOpen
  \bibfield  {author} {\bibinfo {author} {\bibfnamefont {A.}~\bibnamefont
  {García-Arribas}}, \bibinfo {author} {\bibfnamefont {E.}~\bibnamefont
  {Fernández}}, \bibinfo {author} {\bibfnamefont {A.}~\bibnamefont {Svalov}},
  \bibinfo {author} {\bibfnamefont {G.}~\bibnamefont {Kurlyandskaya}},\ and\
  \bibinfo {author} {\bibfnamefont {J.}~\bibnamefont {Barandiarán}},\ }\href
  {https://doi.org/10.1016/j.jmmm.2015.07.107} {\bibfield  {journal} {\bibinfo
  {journal} {Journal of Magnetism and Magnetic Materials}\ }\textbf {\bibinfo
  {volume} {400}},\ \bibinfo {pages} {321} (\bibinfo {year}
  {2016})}\BibitemShut {NoStop}%
\bibitem [{\citenamefont {Morikawa}\ \emph {et~al.}(1996)\citenamefont
  {Morikawa}, \citenamefont {Nishibe}, \citenamefont {Yamadera}, \citenamefont
  {Nonomura}, \citenamefont {Takeuchi}, \citenamefont {Sakata},\ and\
  \citenamefont {Taga}}]{Morikawa1996}%
  \BibitemOpen
  \bibfield  {author} {\bibinfo {author} {\bibfnamefont {T.}~\bibnamefont
  {Morikawa}}, \bibinfo {author} {\bibfnamefont {Y.}~\bibnamefont {Nishibe}},
  \bibinfo {author} {\bibfnamefont {H.}~\bibnamefont {Yamadera}}, \bibinfo
  {author} {\bibfnamefont {Y.}~\bibnamefont {Nonomura}}, \bibinfo {author}
  {\bibfnamefont {M.}~\bibnamefont {Takeuchi}}, \bibinfo {author}
  {\bibfnamefont {J.}~\bibnamefont {Sakata}},\ and\ \bibinfo {author}
  {\bibfnamefont {Y.}~\bibnamefont {Taga}},\ }\href
  {https://doi.org/10.1109/20.539303} {\bibfield  {journal} {\bibinfo
  {journal} {IEEE Transactions on Magnetics}\ }\textbf {\bibinfo {volume}
  {32}},\ \bibinfo {pages} {4965} (\bibinfo {year} {1996})}\BibitemShut
  {NoStop}%
\bibitem [{\citenamefont {Yu}\ \emph {et~al.}(2000)\citenamefont {Yu},
  \citenamefont {Zhou}, \citenamefont {Cai},\ and\ \citenamefont
  {Xu}}]{Yu2000}%
  \BibitemOpen
  \bibfield  {author} {\bibinfo {author} {\bibfnamefont {J.}~\bibnamefont
  {Yu}}, \bibinfo {author} {\bibfnamefont {Y.}~\bibnamefont {Zhou}}, \bibinfo
  {author} {\bibfnamefont {B.}~\bibnamefont {Cai}},\ and\ \bibinfo {author}
  {\bibfnamefont {D.}~\bibnamefont {Xu}},\ }\href@noop {} {\bibfield  {journal}
  {\bibinfo  {journal} {Journal of Magnetism and Magnetic Materials}\ }\textbf
  {\bibinfo {volume} {213}},\ \bibinfo {pages} {32} (\bibinfo {year}
  {2000})}\BibitemShut {NoStop}%
\bibitem [{\citenamefont {Yokoyama}\ \emph {et~al.}(2019)\citenamefont
  {Yokoyama}, \citenamefont {Kusunoki}, \citenamefont {Hayashi}, \citenamefont
  {Hashi},\ and\ \citenamefont {Ishiyama}}]{Yokoyama2019}%
  \BibitemOpen
  \bibfield  {author} {\bibinfo {author} {\bibfnamefont {H.}~\bibnamefont
  {Yokoyama}}, \bibinfo {author} {\bibfnamefont {K.}~\bibnamefont {Kusunoki}},
  \bibinfo {author} {\bibfnamefont {Y.}~\bibnamefont {Hayashi}}, \bibinfo
  {author} {\bibfnamefont {S.}~\bibnamefont {Hashi}},\ and\ \bibinfo {author}
  {\bibfnamefont {K.}~\bibnamefont {Ishiyama}},\ }\href
  {https://doi.org/10.1016/j.jmmm.2019.01.066} {\bibfield  {journal} {\bibinfo
  {journal} {Journal of Magnetism and Magnetic Materials}\ }\textbf {\bibinfo
  {volume} {478}},\ \bibinfo {pages} {38} (\bibinfo {year} {2019})}\BibitemShut
  {NoStop}%
\bibitem [{\citenamefont {García-Arribas}\ \emph {et~al.}(2017)\citenamefont
  {García-Arribas}, \citenamefont {Fernández},\ and\ \citenamefont
  {de~Cos}}]{Garcia-Arribas2017}%
  \BibitemOpen
  \bibfield  {author} {\bibinfo {author} {\bibfnamefont {A.}~\bibnamefont
  {García-Arribas}}, \bibinfo {author} {\bibfnamefont {E.}~\bibnamefont
  {Fernández}},\ and\ \bibinfo {author} {\bibfnamefont {D.}~\bibnamefont
  {de~Cos}},\ }\href {https://doi.org/10.5772/66603} {\emph {\bibinfo {title}
  {Magnetic Sensors - Development Trends and Applications}}}\ (\bibinfo
  {publisher} {InTech},\ \bibinfo {year} {2017})\ pp.\ \bibinfo {pages}
  {39--62}\BibitemShut {NoStop}%
\bibitem [{\citenamefont {Saito}\ \emph {et~al.}(1964)\citenamefont {Saito},
  \citenamefont {Fujiwara},\ and\ \citenamefont {Sugita}}]{Saito1964}%
  \BibitemOpen
  \bibfield  {author} {\bibinfo {author} {\bibfnamefont {N.}~\bibnamefont
  {Saito}}, \bibinfo {author} {\bibfnamefont {H.}~\bibnamefont {Fujiwara}},\
  and\ \bibinfo {author} {\bibfnamefont {Y.}~\bibnamefont {Sugita}},\ }\href
  {https://doi.org/10.1143/JPSJ.19.1116} {\bibfield  {journal} {\bibinfo
  {journal} {Journal of the Physical Society of Japan}\ }\textbf {\bibinfo
  {volume} {19}},\ \bibinfo {pages} {1116} (\bibinfo {year}
  {1964})}\BibitemShut {NoStop}%
\bibitem [{\citenamefont {Silva}\ \emph {et~al.}(2017)\citenamefont {Silva},
  \citenamefont {Corrêa}, \citenamefont {Pace}, \citenamefont {Cid},
  \citenamefont {Kern}, \citenamefont {Carara}, \citenamefont {Chesman},
  \citenamefont {Santos}, \citenamefont {Rodríguez-Suárez}, \citenamefont
  {Azevedo}, \citenamefont {Rezende},\ and\ \citenamefont {Bohn}}]{Silva2017}%
  \BibitemOpen
  \bibfield  {author} {\bibinfo {author} {\bibfnamefont {E.~F.}\ \bibnamefont
  {Silva}}, \bibinfo {author} {\bibfnamefont {M.~A.}\ \bibnamefont {Corrêa}},
  \bibinfo {author} {\bibfnamefont {R.~D.~D.}\ \bibnamefont {Pace}}, \bibinfo
  {author} {\bibfnamefont {C.~C.~P.}\ \bibnamefont {Cid}}, \bibinfo {author}
  {\bibfnamefont {P.~R.}\ \bibnamefont {Kern}}, \bibinfo {author}
  {\bibfnamefont {M.}~\bibnamefont {Carara}}, \bibinfo {author} {\bibfnamefont
  {C.}~\bibnamefont {Chesman}}, \bibinfo {author} {\bibfnamefont {O.~A.}\
  \bibnamefont {Santos}}, \bibinfo {author} {\bibfnamefont {R.~L.}\
  \bibnamefont {Rodríguez-Suárez}}, \bibinfo {author} {\bibfnamefont
  {A.}~\bibnamefont {Azevedo}}, \bibinfo {author} {\bibfnamefont {S.~M.}\
  \bibnamefont {Rezende}},\ and\ \bibinfo {author} {\bibfnamefont
  {F.}~\bibnamefont {Bohn}},\ }\href {https://doi.org/10.1088/1361-6463/aa6665}
  {\bibfield  {journal} {\bibinfo  {journal} {Journal of Physics D: Applied
  Physics}\ }\textbf {\bibinfo {volume} {50}},\ \bibinfo {pages} {0} (\bibinfo
  {year} {2017})}\BibitemShut {NoStop}%
\bibitem [{\citenamefont {de~Cos}\ \emph {et~al.}(2008)\citenamefont {de~Cos},
  \citenamefont {Lepalovskij}, \citenamefont {Kurlyandskaya}, \citenamefont
  {García-Arribas},\ and\ \citenamefont {Barandiarán}}]{DeCos2008}%
  \BibitemOpen
  \bibfield  {author} {\bibinfo {author} {\bibfnamefont {D.}~\bibnamefont
  {de~Cos}}, \bibinfo {author} {\bibfnamefont {V.~N.}\ \bibnamefont
  {Lepalovskij}}, \bibinfo {author} {\bibfnamefont {G.~V.}\ \bibnamefont
  {Kurlyandskaya}}, \bibinfo {author} {\bibfnamefont {A.}~\bibnamefont
  {García-Arribas}},\ and\ \bibinfo {author} {\bibfnamefont {J.~M.}\
  \bibnamefont {Barandiarán}},\ }\href
  {https://doi.org/10.1016/j.jmmm.2008.04.174} {\bibfield  {journal} {\bibinfo
  {journal} {Journal of Magnetism and Magnetic Materials}\ }\textbf {\bibinfo
  {volume} {320}},\ \bibinfo {pages} {954} (\bibinfo {year}
  {2008})}\BibitemShut {NoStop}%
\bibitem [{\citenamefont {Corrêa}\ \emph {et~al.}(2010)\citenamefont
  {Corrêa}, \citenamefont {Bohn}, \citenamefont {Chesman}, \citenamefont
  {da~Silva}, \citenamefont {Viegas},\ and\ \citenamefont
  {Sommer}}]{Correa2010}%
  \BibitemOpen
  \bibfield  {author} {\bibinfo {author} {\bibfnamefont {M.~A.}\ \bibnamefont
  {Corrêa}}, \bibinfo {author} {\bibfnamefont {F.}~\bibnamefont {Bohn}},
  \bibinfo {author} {\bibfnamefont {C.}~\bibnamefont {Chesman}}, \bibinfo
  {author} {\bibfnamefont {R.~B.}\ \bibnamefont {da~Silva}}, \bibinfo {author}
  {\bibfnamefont {A.~D.~C.}\ \bibnamefont {Viegas}},\ and\ \bibinfo {author}
  {\bibfnamefont {R.~L.}\ \bibnamefont {Sommer}},\ }\bibfield  {journal}
  {\bibinfo  {journal} {Journal of Physics D: Applied Physics}\ }\textbf
  {\bibinfo {volume} {43}},\ \href
  {https://doi.org/10.1088/0022-3727/43/29/295004}
  {10.1088/0022-3727/43/29/295004} (\bibinfo {year} {2010})\BibitemShut
  {NoStop}%
\bibitem [{\citenamefont {Kurlyandskaya}\ \emph {et~al.}(2010)\citenamefont
  {Kurlyandskaya}, \citenamefont {Svalov}, \citenamefont {Fernández},
  \citenamefont {García-Arribas},\ and\ \citenamefont
  {Barandiarán}}]{Kurlyandskaya2010}%
  \BibitemOpen
  \bibfield  {author} {\bibinfo {author} {\bibfnamefont {G.~V.}\ \bibnamefont
  {Kurlyandskaya}}, \bibinfo {author} {\bibfnamefont {A.~V.}\ \bibnamefont
  {Svalov}}, \bibinfo {author} {\bibfnamefont {E.}~\bibnamefont {Fernández}},
  \bibinfo {author} {\bibfnamefont {A.}~\bibnamefont {García-Arribas}},\ and\
  \bibinfo {author} {\bibfnamefont {J.~M.}\ \bibnamefont {Barandiarán}},\
  }\bibfield  {journal} {\bibinfo  {journal} {Journal of Applied Physics}\
  }\textbf {\bibinfo {volume} {107}},\ \href
  {https://doi.org/10.1063/1.3355473} {10.1063/1.3355473} (\bibinfo {year}
  {2010})\BibitemShut {NoStop}%
\bibitem [{\citenamefont {Svalov}\ \emph {et~al.}(2012)\citenamefont {Svalov},
  \citenamefont {Fernández}, \citenamefont {García-Arribas}, \citenamefont
  {Alonso}, \citenamefont {Fdez-Gubieda},\ and\ \citenamefont
  {Kurlyandskaya}}]{Svalov2012}%
  \BibitemOpen
  \bibfield  {author} {\bibinfo {author} {\bibfnamefont {A.~V.}\ \bibnamefont
  {Svalov}}, \bibinfo {author} {\bibfnamefont {E.}~\bibnamefont {Fernández}},
  \bibinfo {author} {\bibfnamefont {A.}~\bibnamefont {García-Arribas}},
  \bibinfo {author} {\bibfnamefont {J.}~\bibnamefont {Alonso}}, \bibinfo
  {author} {\bibfnamefont {M.~L.}\ \bibnamefont {Fdez-Gubieda}},\ and\ \bibinfo
  {author} {\bibfnamefont {G.~V.}\ \bibnamefont {Kurlyandskaya}},\ }\href
  {https://doi.org/10.1063/1.4704984} {\bibfield  {journal} {\bibinfo
  {journal} {Applied Physics Letters}\ }\textbf {\bibinfo {volume} {100}},\
  \bibinfo {pages} {1} (\bibinfo {year} {2012})}\BibitemShut {NoStop}%
\bibitem [{\citenamefont {Vas'kovskii}\ \emph {et~al.}(2013)\citenamefont
  {Vas'kovskii}, \citenamefont {Savin}, \citenamefont {Volchkov}, \citenamefont
  {Lepalovskii}, \citenamefont {Bukreev},\ and\ \citenamefont
  {Buchkevich}}]{Vaskovskii2013}%
  \BibitemOpen
  \bibfield  {author} {\bibinfo {author} {\bibfnamefont {V.~O.}\ \bibnamefont
  {Vas'kovskii}}, \bibinfo {author} {\bibfnamefont {P.~A.}\ \bibnamefont
  {Savin}}, \bibinfo {author} {\bibfnamefont {S.~O.}\ \bibnamefont {Volchkov}},
  \bibinfo {author} {\bibfnamefont {V.~N.}\ \bibnamefont {Lepalovskii}},
  \bibinfo {author} {\bibfnamefont {D.~A.}\ \bibnamefont {Bukreev}},\ and\
  \bibinfo {author} {\bibfnamefont {A.~A.}\ \bibnamefont {Buchkevich}},\ }\href
  {https://doi.org/10.1134/S1063784213010222} {\bibfield  {journal} {\bibinfo
  {journal} {Technical Physics}\ }\textbf {\bibinfo {volume} {58}},\ \bibinfo
  {pages} {105} (\bibinfo {year} {2013})}\BibitemShut {NoStop}%
\bibitem [{\citenamefont {Kikuchi}\ \emph {et~al.}(2014)\citenamefont
  {Kikuchi}, \citenamefont {Takahashi}, \citenamefont {Takahashi},
  \citenamefont {Nakai}, \citenamefont {Hashi},\ and\ \citenamefont
  {Ishiyama}}]{Kikuchi2014}%
  \BibitemOpen
  \bibfield  {author} {\bibinfo {author} {\bibfnamefont {H.}~\bibnamefont
  {Kikuchi}}, \bibinfo {author} {\bibfnamefont {Y.}~\bibnamefont {Takahashi}},
  \bibinfo {author} {\bibfnamefont {K.}~\bibnamefont {Takahashi}}, \bibinfo
  {author} {\bibfnamefont {T.}~\bibnamefont {Nakai}}, \bibinfo {author}
  {\bibfnamefont {S.}~\bibnamefont {Hashi}},\ and\ \bibinfo {author}
  {\bibfnamefont {K.}~\bibnamefont {Ishiyama}},\ }\href
  {https://doi.org/10.1063/1.4859095} {\bibfield  {journal} {\bibinfo
  {journal} {Journal of Applied Physics}\ }\textbf {\bibinfo {volume} {115}},\
  \bibinfo {pages} {14} (\bibinfo {year} {2014})}\BibitemShut {NoStop}%
\bibitem [{\citenamefont {Kikuchi}\ \emph
  {et~al.}(2015{\natexlab{a}})\citenamefont {Kikuchi}, \citenamefont {Kamata},
  \citenamefont {Oe}, \citenamefont {Nakai}, \citenamefont {Hashi},\ and\
  \citenamefont {Ishiyama}}]{Kikuchi2015}%
  \BibitemOpen
  \bibfield  {author} {\bibinfo {author} {\bibfnamefont {H.}~\bibnamefont
  {Kikuchi}}, \bibinfo {author} {\bibfnamefont {S.}~\bibnamefont {Kamata}},
  \bibinfo {author} {\bibfnamefont {S.}~\bibnamefont {Oe}}, \bibinfo {author}
  {\bibfnamefont {T.}~\bibnamefont {Nakai}}, \bibinfo {author} {\bibfnamefont
  {S.}~\bibnamefont {Hashi}},\ and\ \bibinfo {author} {\bibfnamefont
  {K.}~\bibnamefont {Ishiyama}},\ }\href
  {https://doi.org/10.1109/TMAG.2015.2441748} {\bibfield  {journal} {\bibinfo
  {journal} {IEEE Transactions on Magnetics}\ }\textbf {\bibinfo {volume}
  {51}},\ \bibinfo {pages} {18} (\bibinfo {year}
  {2015}{\natexlab{a}})}\BibitemShut {NoStop}%
\bibitem [{\citenamefont {Kikuchi}\ \emph
  {et~al.}(2015{\natexlab{b}})\citenamefont {Kikuchi}, \citenamefont {Kamata},
  \citenamefont {Takahashi}, \citenamefont {Nakai}, \citenamefont {Hashi},\
  and\ \citenamefont {Ishiyama}}]{Kikuchi2015a}%
  \BibitemOpen
  \bibfield  {author} {\bibinfo {author} {\bibfnamefont {H.}~\bibnamefont
  {Kikuchi}}, \bibinfo {author} {\bibfnamefont {S.}~\bibnamefont {Kamata}},
  \bibinfo {author} {\bibfnamefont {Y.}~\bibnamefont {Takahashi}}, \bibinfo
  {author} {\bibfnamefont {T.}~\bibnamefont {Nakai}}, \bibinfo {author}
  {\bibfnamefont {S.}~\bibnamefont {Hashi}},\ and\ \bibinfo {author}
  {\bibfnamefont {K.}~\bibnamefont {Ishiyama}},\ }\bibfield  {journal}
  {\bibinfo  {journal} {IEEE Transactions on Magnetics}\ }\textbf {\bibinfo
  {volume} {51}},\ \href {https://doi.org/10.1109/TMAG.2014.2358221}
  {10.1109/TMAG.2014.2358221} (\bibinfo {year}
  {2015}{\natexlab{b}})\BibitemShut {NoStop}%
\bibitem [{\citenamefont {Kikuchi}\ \emph {et~al.}(2020)\citenamefont
  {Kikuchi}, \citenamefont {Tanii},\ and\ \citenamefont
  {Umezaki}}]{Kikuchi2020}%
  \BibitemOpen
  \bibfield  {author} {\bibinfo {author} {\bibfnamefont {H.}~\bibnamefont
  {Kikuchi}}, \bibinfo {author} {\bibfnamefont {M.}~\bibnamefont {Tanii}},\
  and\ \bibinfo {author} {\bibfnamefont {T.}~\bibnamefont {Umezaki}},\
  }\bibfield  {journal} {\bibinfo  {journal} {AIP Advances}\ }\textbf {\bibinfo
  {volume} {10}},\ \href {https://doi.org/10.1063/1.5130410}
  {10.1063/1.5130410} (\bibinfo {year} {2020})\BibitemShut {NoStop}%
\bibitem [{\citenamefont {Kikuchi}\ \emph {et~al.}(2022)\citenamefont
  {Kikuchi}, \citenamefont {Ueno},\ and\ \citenamefont {Tanii}}]{Kikuchi2022}%
  \BibitemOpen
  \bibfield  {author} {\bibinfo {author} {\bibfnamefont {H.}~\bibnamefont
  {Kikuchi}}, \bibinfo {author} {\bibfnamefont {A.}~\bibnamefont {Ueno}},\ and\
  \bibinfo {author} {\bibfnamefont {M.}~\bibnamefont {Tanii}},\ }\bibfield
  {journal} {\bibinfo  {journal} {IEEE Transactions on Magnetics}\ }\textbf
  {\bibinfo {volume} {58}},\ \href {https://doi.org/10.1109/TMAG.2022.3151985}
  {10.1109/TMAG.2022.3151985} (\bibinfo {year} {2022})\BibitemShut {NoStop}%
\bibitem [{\citenamefont {Kikuchi}\ \emph {et~al.}(2023)\citenamefont
  {Kikuchi}, \citenamefont {Ueno},\ and\ \citenamefont {Tanii}}]{Kikuchi2023}%
  \BibitemOpen
  \bibfield  {author} {\bibinfo {author} {\bibfnamefont {H.}~\bibnamefont
  {Kikuchi}}, \bibinfo {author} {\bibfnamefont {A.}~\bibnamefont {Ueno}},\ and\
  \bibinfo {author} {\bibfnamefont {M.}~\bibnamefont {Tanii}},\ }\href
  {https://doi.org/10.1109/TMAG.2023.3280215} {\bibfield  {journal} {\bibinfo
  {journal} {IEEE Transactions on Magnetics}\ }\textbf {\bibinfo {volume}
  {59}},\ \bibinfo {pages} {1} (\bibinfo {year} {2023})}\BibitemShut {NoStop}%
\bibitem [{\citenamefont {Corrêa}\ \emph {et~al.}(2007)\citenamefont
  {Corrêa}, \citenamefont {Viegas}, \citenamefont {da~Silva}, \citenamefont
  {de~Andrade},\ and\ \citenamefont {Sommer}}]{Correa2007}%
  \BibitemOpen
  \bibfield  {author} {\bibinfo {author} {\bibfnamefont {M.~A.}\ \bibnamefont
  {Corrêa}}, \bibinfo {author} {\bibfnamefont {A.~D.~C.}\ \bibnamefont
  {Viegas}}, \bibinfo {author} {\bibfnamefont {R.~B.}\ \bibnamefont
  {da~Silva}}, \bibinfo {author} {\bibfnamefont {A.~M.~H.}\ \bibnamefont
  {de~Andrade}},\ and\ \bibinfo {author} {\bibfnamefont {R.~L.}\ \bibnamefont
  {Sommer}},\ }\bibfield  {journal} {\bibinfo  {journal} {Journal of Applied
  Physics}\ }\textbf {\bibinfo {volume} {101}},\ \href
  {https://doi.org/10.1063/1.2512867} {10.1063/1.2512867} (\bibinfo {year}
  {2007})\BibitemShut {NoStop}%
\bibitem [{\citenamefont {Corrêa}\ \emph {et~al.}(2008)\citenamefont
  {Corrêa}, \citenamefont {Bohn}, \citenamefont {Viegas}, \citenamefont
  {de~Andrade}, \citenamefont {Schelp},\ and\ \citenamefont
  {Sommer}}]{Correa2008}%
  \BibitemOpen
  \bibfield  {author} {\bibinfo {author} {\bibfnamefont {M.~A.}\ \bibnamefont
  {Corrêa}}, \bibinfo {author} {\bibfnamefont {F.}~\bibnamefont {Bohn}},
  \bibinfo {author} {\bibfnamefont {A.~D.~C.}\ \bibnamefont {Viegas}}, \bibinfo
  {author} {\bibfnamefont {A.~M.}\ \bibnamefont {de~Andrade}}, \bibinfo
  {author} {\bibfnamefont {L.~F.}\ \bibnamefont {Schelp}},\ and\ \bibinfo
  {author} {\bibfnamefont {R.~L.}\ \bibnamefont {Sommer}},\ }\bibfield
  {journal} {\bibinfo  {journal} {Journal of Physics D: Applied Physics}\
  }\textbf {\bibinfo {volume} {41}},\ \href
  {https://doi.org/10.1088/0022-3727/41/17/175003}
  {10.1088/0022-3727/41/17/175003} (\bibinfo {year} {2008})\BibitemShut
  {NoStop}%
\bibitem [{\citenamefont {Chen}\ and\ \citenamefont {Munoz}(1999)}]{Chen1999}%
  \BibitemOpen
  \bibfield  {author} {\bibinfo {author} {\bibfnamefont {D.~X.}\ \bibnamefont
  {Chen}}\ and\ \bibinfo {author} {\bibfnamefont {J.~L.}\ \bibnamefont
  {Munoz}},\ }\href {https://doi.org/10.1109/20.764884} {\bibfield  {journal}
  {\bibinfo  {journal} {IEEE Transactions on Magnetics}\ }\textbf {\bibinfo
  {volume} {35}},\ \bibinfo {pages} {1906} (\bibinfo {year}
  {1999})}\BibitemShut {NoStop}%
\bibitem [{\citenamefont {Phan}\ and\ \citenamefont {Peng}(2008)}]{Phan2008}%
  \BibitemOpen
  \bibfield  {author} {\bibinfo {author} {\bibfnamefont {M.}~\bibnamefont
  {Phan}}\ and\ \bibinfo {author} {\bibfnamefont {H.}~\bibnamefont {Peng}},\
  }\href {https://doi.org/10.1016/j.pmatsci.2007.05.003} {\bibfield  {journal}
  {\bibinfo  {journal} {Progress in Materials Science}\ }\textbf {\bibinfo
  {volume} {53}},\ \bibinfo {pages} {323} (\bibinfo {year} {2008})}\BibitemShut
  {NoStop}%
\bibitem [{\citenamefont {Buschow}(2003)}]{Buschow2003}%
  \BibitemOpen
  \bibfield  {author} {\bibinfo {author} {\bibfnamefont {K.}~\bibnamefont
  {Buschow}},\ }\href {https://doi.org/10.1016/0925-8388(93)90087-4} {\emph
  {\bibinfo {title} {Handbook of magnetic materials}}},\ \bibinfo {edition}
  {1st}\ ed.,\ Vol.~\bibinfo {volume} {15}\ (\bibinfo  {publisher} {Elsevier
  B.V.},\ \bibinfo {year} {2003})\BibitemShut {NoStop}%
\bibitem [{\citenamefont {Levenberg}(1944)}]{Levenberg1944}%
  \BibitemOpen
  \bibfield  {author} {\bibinfo {author} {\bibfnamefont {K.}~\bibnamefont
  {Levenberg}},\ }\href {https://doi.org/10.1090/qam/10666} {\bibfield
  {journal} {\bibinfo  {journal} {Quarterly of Applied Mathematics}\ }\textbf
  {\bibinfo {volume} {2}},\ \bibinfo {pages} {164} (\bibinfo {year}
  {1944})}\BibitemShut {NoStop}%
\bibitem [{\citenamefont {Marquardt}(1963)}]{Marquardt1963}%
  \BibitemOpen
  \bibfield  {author} {\bibinfo {author} {\bibfnamefont {D.~W.}\ \bibnamefont
  {Marquardt}},\ }\href {https://doi.org/10.1137/0111030} {\bibfield  {journal}
  {\bibinfo  {journal} {Journal of the Society for Industrial and Applied
  Mathematics}\ }\textbf {\bibinfo {volume} {11}},\ \bibinfo {pages} {431}
  (\bibinfo {year} {1963})}\BibitemShut {NoStop}%
\bibitem [{\citenamefont {Kittel}(1948)}]{Kittel1948}%
  \BibitemOpen
  \bibfield  {author} {\bibinfo {author} {\bibfnamefont {C.}~\bibnamefont
  {Kittel}},\ }\href {https://doi.org/10.1103/PhysRev.73.155} {\bibfield
  {journal} {\bibinfo  {journal} {Physical Review}\ }\textbf {\bibinfo {volume}
  {73}},\ \bibinfo {pages} {155} (\bibinfo {year} {1948})}\BibitemShut
  {NoStop}%
\bibitem [{\citenamefont {Osborn}(1945)}]{Osborn1945}%
  \BibitemOpen
  \bibfield  {author} {\bibinfo {author} {\bibfnamefont {J.~A.}\ \bibnamefont
  {Osborn}},\ }\href {https://doi.org/10.1103/PhysRev.67.351} {\bibinfo {title}
  {Demagnetizing factors of the general ellipsoid}} (\bibinfo {year}
  {1945})\BibitemShut {NoStop}%
\bibitem [{\citenamefont {Aharoni}(1998)}]{Aharoni1998}%
  \BibitemOpen
  \bibfield  {author} {\bibinfo {author} {\bibfnamefont {A.}~\bibnamefont
  {Aharoni}},\ }\href {https://doi.org/10.1063/1.367113} {\bibfield  {journal}
  {\bibinfo  {journal} {Journal of Applied Physics}\ }\textbf {\bibinfo
  {volume} {83}},\ \bibinfo {pages} {3432} (\bibinfo {year}
  {1998})}\BibitemShut {NoStop}%
\bibitem [{\citenamefont {Cullity}\ and\ \citenamefont
  {Graham}(2009)}]{Cullity2009}%
  \BibitemOpen
  \bibfield  {author} {\bibinfo {author} {\bibfnamefont {B.}~\bibnamefont
  {Cullity}}\ and\ \bibinfo {author} {\bibfnamefont {C.~D.}\ \bibnamefont
  {Graham}},\ }\href {https://doi.org/10.1016/s1369-7021(09)70091-4} {\emph
  {\bibinfo {title} {Introduction to magnetic materials}}}\ (\bibinfo
  {publisher} {Wiley},\ \bibinfo {year} {2009})\BibitemShut {NoStop}%
\bibitem [{\citenamefont {Tomus}\ and\ \citenamefont {Ng}(2013)}]{Tomus2013}%
  \BibitemOpen
  \bibfield  {author} {\bibinfo {author} {\bibfnamefont {D.}~\bibnamefont
  {Tomus}}\ and\ \bibinfo {author} {\bibfnamefont {H.~P.}\ \bibnamefont {Ng}},\
  }\href {https://doi.org/10.1016/j.micron.2012.05.006} {\bibfield  {journal}
  {\bibinfo  {journal} {Micron}\ }\textbf {\bibinfo {volume} {44}},\ \bibinfo
  {pages} {115} (\bibinfo {year} {2013})}\BibitemShut {NoStop}%
\bibitem [{\citenamefont {Manuel}\ and\ \citenamefont
  {Quinton}(1963)}]{Manuel1963}%
  \BibitemOpen
  \bibfield  {author} {\bibinfo {author} {\bibfnamefont {A.~J.}\ \bibnamefont
  {Manuel}}\ and\ \bibinfo {author} {\bibfnamefont {J.~M. P.~S.}\ \bibnamefont
  {Quinton}},\ }\href {https://doi.org/10.1098/rspa.1963.0099} {\bibfield
  {journal} {\bibinfo  {journal} {Proceedings of the Royal Society of London.
  Series A. Mathematical and Physical Sciences}\ }\textbf {\bibinfo {volume}
  {273}},\ \bibinfo {pages} {412} (\bibinfo {year} {1963})}\BibitemShut
  {NoStop}%
\bibitem [{\citenamefont {Kolding}(2000)}]{Kolding2000}%
  \BibitemOpen
  \bibfield  {author} {\bibinfo {author} {\bibfnamefont {T.~E.}\ \bibnamefont
  {Kolding}},\ }\href {https://doi.org/10.1109/16.830987} {\bibfield  {journal}
  {\bibinfo  {journal} {IEEE Transactions on Electron Devices}\ }\textbf
  {\bibinfo {volume} {47}},\ \bibinfo {pages} {734} (\bibinfo {year}
  {2000})}\BibitemShut {NoStop}%
\bibitem [{\citenamefont {Kikuchi}\ \emph {et~al.}(2021)\citenamefont
  {Kikuchi}, \citenamefont {Urakawa},\ and\ \citenamefont
  {Tanii}}]{Kikuchi2021}%
  \BibitemOpen
  \bibfield  {author} {\bibinfo {author} {\bibfnamefont {H.}~\bibnamefont
  {Kikuchi}}, \bibinfo {author} {\bibfnamefont {Y.}~\bibnamefont {Urakawa}},\
  and\ \bibinfo {author} {\bibfnamefont {M.}~\bibnamefont {Tanii}},\ }\href
  {https://doi.org/10.1016/j.jmmm.2021.168356} {\bibfield  {journal} {\bibinfo
  {journal} {Journal of Magnetism and Magnetic Materials}\ }\textbf {\bibinfo
  {volume} {539}},\ \bibinfo {pages} {168356} (\bibinfo {year}
  {2021})}\BibitemShut {NoStop}%
\bibitem [{\citenamefont {Ashcroft}\ and\ \citenamefont
  {Mermin}(1976)}]{Ashcroft1976}%
  \BibitemOpen
  \bibfield  {author} {\bibinfo {author} {\bibfnamefont {N.~W.}\ \bibnamefont
  {Ashcroft}}\ and\ \bibinfo {author} {\bibfnamefont {N.~D.}\ \bibnamefont
  {Mermin}},\ }\href@noop {} {\emph {\bibinfo {title} {Solid State Physics}}}\
  (\bibinfo  {publisher} {Saunders College Publishing},\ \bibinfo {year}
  {1976})\BibitemShut {NoStop}%
\bibitem [{\citenamefont {Wakelin}\ and\ \citenamefont
  {Yates}(1953)}]{Wakelin1953}%
  \BibitemOpen
  \bibfield  {author} {\bibinfo {author} {\bibfnamefont {R.~J.}\ \bibnamefont
  {Wakelin}}\ and\ \bibinfo {author} {\bibfnamefont {E.~L.}\ \bibnamefont
  {Yates}},\ }\href@noop {} {\bibfield  {journal} {\bibinfo  {journal}
  {Proceedings of the Physical Society. Section B}\ }\textbf {\bibinfo {volume}
  {66}},\ \bibinfo {pages} {221} (\bibinfo {year} {1953})}\BibitemShut
  {NoStop}%
\bibitem [{\citenamefont {Yamaguchi}\ \emph {et~al.}(2015)\citenamefont
  {Yamaguchi}, \citenamefont {Tanaka}, \citenamefont {Endo}, \citenamefont
  {Muroga},\ and\ \citenamefont {Nagata}}]{Yamaguchi2015}%
  \BibitemOpen
  \bibfield  {author} {\bibinfo {author} {\bibfnamefont {M.}~\bibnamefont
  {Yamaguchi}}, \bibinfo {author} {\bibfnamefont {S.}~\bibnamefont {Tanaka}},
  \bibinfo {author} {\bibfnamefont {Y.}~\bibnamefont {Endo}}, \bibinfo {author}
  {\bibfnamefont {S.}~\bibnamefont {Muroga}},\ and\ \bibinfo {author}
  {\bibfnamefont {M.}~\bibnamefont {Nagata}},\ }\href
  {https://doi.org/10.1109/APEMC.2015.7175409} {\bibfield  {journal} {\bibinfo
  {journal} {2015 Asia-Pacific International Symposium on Electromagnetic
  Compatibility (APEMC)}\ ,\ \bibinfo {pages} {536}} (\bibinfo {year}
  {2015})}\BibitemShut {NoStop}%
\bibitem [{\citenamefont {Klokholm}\ and\ \citenamefont
  {Aboaf}(1981)}]{Klokholm1981}%
  \BibitemOpen
  \bibfield  {author} {\bibinfo {author} {\bibfnamefont {E.}~\bibnamefont
  {Klokholm}}\ and\ \bibinfo {author} {\bibfnamefont {J.~A.}\ \bibnamefont
  {Aboaf}},\ }\href {https://doi.org/10.1063/1.328971} {\bibfield  {journal}
  {\bibinfo  {journal} {Journal of Applied Physics}\ }\textbf {\bibinfo
  {volume} {52}},\ \bibinfo {pages} {2474} (\bibinfo {year}
  {1981})}\BibitemShut {NoStop}%
\bibitem [{\citenamefont {Neidhardt}\ \emph {et~al.}(2008)\citenamefont
  {Neidhardt}, \citenamefont {Mráz}, \citenamefont {Schneider}, \citenamefont
  {Strub}, \citenamefont {Bohne}, \citenamefont {Liedke}, \citenamefont
  {Möller},\ and\ \citenamefont {Mitterer}}]{Neidhardt2008}%
  \BibitemOpen
  \bibfield  {author} {\bibinfo {author} {\bibfnamefont {J.}~\bibnamefont
  {Neidhardt}}, \bibinfo {author} {\bibfnamefont {S.}~\bibnamefont {Mráz}},
  \bibinfo {author} {\bibfnamefont {J.~M.}\ \bibnamefont {Schneider}}, \bibinfo
  {author} {\bibfnamefont {E.}~\bibnamefont {Strub}}, \bibinfo {author}
  {\bibfnamefont {W.}~\bibnamefont {Bohne}}, \bibinfo {author} {\bibfnamefont
  {B.}~\bibnamefont {Liedke}}, \bibinfo {author} {\bibfnamefont
  {W.}~\bibnamefont {Möller}},\ and\ \bibinfo {author} {\bibfnamefont
  {C.}~\bibnamefont {Mitterer}},\ }\bibfield  {journal} {\bibinfo  {journal}
  {Journal of Applied Physics}\ }\textbf {\bibinfo {volume} {104}},\ \href
  {https://doi.org/10.1063/1.2978211} {10.1063/1.2978211} (\bibinfo {year}
  {2008})\BibitemShut {NoStop}%
\bibitem [{\citenamefont {Gall}(2016)}]{Gall2016}%
  \BibitemOpen
  \bibfield  {author} {\bibinfo {author} {\bibfnamefont {D.}~\bibnamefont
  {Gall}},\ }\href {https://doi.org/10.1063/1.4942216} {\bibfield  {journal}
  {\bibinfo  {journal} {Journal of Applied Physics}\ }\textbf {\bibinfo
  {volume} {119}},\ \bibinfo {pages} {1} (\bibinfo {year} {2016})}\BibitemShut
  {NoStop}%
\bibitem [{\citenamefont {Nahrwold}\ \emph {et~al.}(2010)\citenamefont
  {Nahrwold}, \citenamefont {Scholtyssek}, \citenamefont {Motl-Ziegler},
  \citenamefont {Albrecht}, \citenamefont {Merkt},\ and\ \citenamefont
  {Meier}}]{Nahrwold2010}%
  \BibitemOpen
  \bibfield  {author} {\bibinfo {author} {\bibfnamefont {G.}~\bibnamefont
  {Nahrwold}}, \bibinfo {author} {\bibfnamefont {J.~M.}\ \bibnamefont
  {Scholtyssek}}, \bibinfo {author} {\bibfnamefont {S.}~\bibnamefont
  {Motl-Ziegler}}, \bibinfo {author} {\bibfnamefont {O.}~\bibnamefont
  {Albrecht}}, \bibinfo {author} {\bibfnamefont {U.}~\bibnamefont {Merkt}},\
  and\ \bibinfo {author} {\bibfnamefont {G.}~\bibnamefont {Meier}},\ }\bibfield
   {journal} {\bibinfo  {journal} {Journal of Applied Physics}\ }\textbf
  {\bibinfo {volume} {108}},\ \href {https://doi.org/10.1063/1.3431384}
  {10.1063/1.3431384} (\bibinfo {year} {2010})\BibitemShut {NoStop}%
\bibitem [{\citenamefont {Hubert}\ and\ \citenamefont
  {Schäfer}(2009)}]{Hubert2009}%
  \BibitemOpen
  \bibfield  {author} {\bibinfo {author} {\bibfnamefont {A.}~\bibnamefont
  {Hubert}}\ and\ \bibinfo {author} {\bibfnamefont {R.}~\bibnamefont
  {Schäfer}},\ }\href@noop {} {\emph {\bibinfo {title} {Magnetic Domains}}}\
  (\bibinfo  {publisher} {Springer Berlin Heidelberg},\ \bibinfo {year}
  {2009})\BibitemShut {NoStop}%
\bibitem [{\citenamefont {Endo}\ \emph {et~al.}(2015)\citenamefont {Endo},
  \citenamefont {Ito}, \citenamefont {Miyazaki}, \citenamefont {Shimada},\ and\
  \citenamefont {Yamaguchi}}]{Endo2015}%
  \BibitemOpen
  \bibfield  {author} {\bibinfo {author} {\bibfnamefont {Y.}~\bibnamefont
  {Endo}}, \bibinfo {author} {\bibfnamefont {T.}~\bibnamefont {Ito}}, \bibinfo
  {author} {\bibfnamefont {T.}~\bibnamefont {Miyazaki}}, \bibinfo {author}
  {\bibfnamefont {Y.}~\bibnamefont {Shimada}},\ and\ \bibinfo {author}
  {\bibfnamefont {M.}~\bibnamefont {Yamaguchi}},\ }\href
  {https://doi.org/10.1063/1.4917502} {\bibfield  {journal} {\bibinfo
  {journal} {Journal of Applied Physics}\ }\textbf {\bibinfo {volume} {117}},\
  \bibinfo {pages} {1} (\bibinfo {year} {2015})}\BibitemShut {NoStop}%
\bibitem [{\citenamefont {Brömer}\ and\ \citenamefont
  {Huber}(1978)}]{Bromer1978}%
  \BibitemOpen
  \bibfield  {author} {\bibinfo {author} {\bibfnamefont {H.}~\bibnamefont
  {Brömer}}\ and\ \bibinfo {author} {\bibfnamefont {H.}~\bibnamefont
  {Huber}},\ }\href {https://doi.org/10.1016/0304-8853(78)90077-X} {\bibfield
  {journal} {\bibinfo  {journal} {Journal of Magnetism and Magnetic Materials}\
  }\textbf {\bibinfo {volume} {8}},\ \bibinfo {pages} {61} (\bibinfo {year}
  {1978})}\BibitemShut {NoStop}%
\bibitem [{\citenamefont {Barandiarán}\ \emph {et~al.}(2006)\citenamefont
  {Barandiarán}, \citenamefont {García-Arribas},\ and\ \citenamefont
  {de~Cos}}]{Barandiaran2006}%
  \BibitemOpen
  \bibfield  {author} {\bibinfo {author} {\bibfnamefont {J.~M.}\ \bibnamefont
  {Barandiarán}}, \bibinfo {author} {\bibfnamefont {A.}~\bibnamefont
  {García-Arribas}},\ and\ \bibinfo {author} {\bibfnamefont {D.}~\bibnamefont
  {de~Cos}},\ }\href {https://doi.org/10.1063/1.2195898} {\bibfield  {journal}
  {\bibinfo  {journal} {Journal of Applied Physics}\ }\textbf {\bibinfo
  {volume} {99}},\ \bibinfo {pages} {4} (\bibinfo {year} {2006})}\BibitemShut
  {NoStop}%
\bibitem [{\citenamefont {Mohri}\ \emph {et~al.}(1992)\citenamefont {Mohri},
  \citenamefont {Kohzawa}, \citenamefont {Kawashima}, \citenamefont {Yoshida},\
  and\ \citenamefont {Panina}}]{Mohri1992}%
  \BibitemOpen
  \bibfield  {author} {\bibinfo {author} {\bibfnamefont {K.}~\bibnamefont
  {Mohri}}, \bibinfo {author} {\bibfnamefont {T.}~\bibnamefont {Kohzawa}},
  \bibinfo {author} {\bibfnamefont {K.}~\bibnamefont {Kawashima}}, \bibinfo
  {author} {\bibfnamefont {H.}~\bibnamefont {Yoshida}},\ and\ \bibinfo {author}
  {\bibfnamefont {L.~V.}\ \bibnamefont {Panina}},\ }\href@noop {} {\bibfield
  {journal} {\bibinfo  {journal} {IEEE Transactions on Magnetics}\ }\textbf
  {\bibinfo {volume} {28}} (\bibinfo {year} {1992})}\BibitemShut {NoStop}%
\bibitem [{\citenamefont {Hika}\ \emph {et~al.}(1996)\citenamefont {Hika},
  \citenamefont {Panina},\ and\ \citenamefont {Mohri}}]{Hika1996}%
  \BibitemOpen
  \bibfield  {author} {\bibinfo {author} {\bibfnamefont {K.}~\bibnamefont
  {Hika}}, \bibinfo {author} {\bibfnamefont {L.~V.}\ \bibnamefont {Panina}},\
  and\ \bibinfo {author} {\bibfnamefont {K.}~\bibnamefont {Mohri}},\ }\href
  {https://doi.org/10.1109/20.539090} {\bibfield  {journal} {\bibinfo
  {journal} {IEEE Transactions on Magnetics}\ }\textbf {\bibinfo {volume}
  {32}},\ \bibinfo {pages} {4594} (\bibinfo {year} {1996})}\BibitemShut
  {NoStop}%
\bibitem [{\citenamefont {Morikawa}\ \emph {et~al.}(1997)\citenamefont
  {Morikawa}, \citenamefont {Nishibe}, \citenamefont {Yamadera}, \citenamefont
  {Nonomura}, \citenamefont {Takeuchi},\ and\ \citenamefont
  {Taga}}]{Morikawa1997}%
  \BibitemOpen
  \bibfield  {author} {\bibinfo {author} {\bibfnamefont {T.}~\bibnamefont
  {Morikawa}}, \bibinfo {author} {\bibfnamefont {Y.}~\bibnamefont {Nishibe}},
  \bibinfo {author} {\bibfnamefont {H.}~\bibnamefont {Yamadera}}, \bibinfo
  {author} {\bibfnamefont {Y.}~\bibnamefont {Nonomura}}, \bibinfo {author}
  {\bibfnamefont {M.}~\bibnamefont {Takeuchi}},\ and\ \bibinfo {author}
  {\bibfnamefont {Y.}~\bibnamefont {Taga}},\ }\href
  {https://doi.org/10.1109/20.620448} {\bibfield  {journal} {\bibinfo
  {journal} {IEEE Transactions on Magnetics}\ }\textbf {\bibinfo {volume}
  {33}},\ \bibinfo {pages} {4367} (\bibinfo {year} {1997})}\BibitemShut
  {NoStop}%
\bibitem [{\citenamefont {García-Arribas}\ \emph {et~al.}(2008)\citenamefont
  {García-Arribas}, \citenamefont {Barandiarán},\ and\ \citenamefont
  {de~Cos}}]{Garcia-Arribas2008}%
  \BibitemOpen
  \bibfield  {author} {\bibinfo {author} {\bibfnamefont {A.}~\bibnamefont
  {García-Arribas}}, \bibinfo {author} {\bibfnamefont {J.~M.}\ \bibnamefont
  {Barandiarán}},\ and\ \bibinfo {author} {\bibfnamefont {D.}~\bibnamefont
  {de~Cos}},\ }\href {https://doi.org/10.1016/j.jmmm.2008.02.005} {\bibfield
  {journal} {\bibinfo  {journal} {Journal of Magnetism and Magnetic Materials}\
  }\textbf {\bibinfo {volume} {320}},\ \bibinfo {pages} {5} (\bibinfo {year}
  {2008})}\BibitemShut {NoStop}%
\end{thebibliography}

\end{document}